\documentclass[useAMS,usenatbib]{mn2e}
\usepackage{psfig}
\usepackage{url}
\usepackage{color}
\usepackage{graphicx}
\usepackage{amsmath}
\usepackage{breqn}

\def\apj{{\rm ApJ}}
\def\aj{{\rm AJ}}
\def\apjs{{\rm ApJS}}

\def\mnras{{\rm MNRAS}}
\def\cm{\, {\rm cm}}

\def\ergs{\, {\rm erg}\, {\rm s}^{-1}}
\def\angs{\, {\rm \AA }}
\def\asec{\, {\rm arcsec}}
\def\etal{{\rm et~al.\ }}
\def\mpc{\, {\rm Mpc}}
\def\hmpc{h^{-1}{\rm Mpc}}

\def\kms{\, {\rm km}\, {\rm s}^{-1}}

\def\msun{\, M_{\odot}}
\def\lya{Ly$\alpha$ }
\def\lyb{Ly$\beta$ }
\def\xiqa{${\xi}_{{\rm q} \alpha}$ }
\def\xiqar{${\xi}_{{\rm q} \alpha}(r)$ }
\def\simlt{\lower.5ex\hbox{$\; \buildrel < \over \sim \;$}}
\def\simgt{\lower.5ex\hbox{$\; \buildrel > \over \sim \;$}}

\def\bqla{$b_{q}b_{{\rm q-Ly}\alpha e}$}
\def\ampqa{$b_{{\rm q}}b_{\alpha} f_{\beta} \langle \mu \rangle$}

\title[Lyman-$\alpha$ emission intensity]{Large-scale clustering of Lyman-$\alpha$ emission
intensity from SDSS/BOSS
}

\author[R.A.C. Croft et al.]{\parbox{18cm}{
Rupert A.C. Croft$^{1,2}$\thanks{E-mail: rcroft@cmu.edu}
Jordi Miralda-Escud\'{e}$^{3,4}$,
Zheng Zheng$^{5}$,
Adam Bolton$^{5}$,
Kyle S. Dawson$^{5}$,
Jeffrey B. Peterson$^{1}$,
Donald G. York$^{6}$,
Daniel Eisenstein$^{7}$,
Jon Brinkmann$^{24}$,
Joel Brownstein$^{5}$,
Timoth\'ee Delubac$^{9}$, 
Andreu Font-Ribera$^{10}$,
Jean-Christophe Hamilton$^{8}$,
Khee-Gan Lee$^{12}$,
Adam Myers$^{14}$,
Nathalie Palanque-Delabrouille$^{9}$,
Isabelle P\^aris$^{21}$,
Patrick Petitjean$^{16}$,
Matthew M. Pieri$^{17}$,
Nicholas P. Ross$^{10}$,
Graziano Rossi$^{9,25}$,
David J. Schlegel$^{10}$,
Donald P. Schneider$^{18,19}$,
An\v{z}e Slosar$^{20}$,
Jos\'{e} Vazquez$^{20}$,
Matteo Viel$^{21,22}$,
David H. Weinberg$^{23}$,
Christophe Y\`eche$^{9}$
}\vspace{0.3cm}\\
(Author affiliations are listed after the bibliography)
}

\begin{document}


\topmargin=-1.0cm

\maketitle


\begin{abstract}
We detect the large-scale structure of \lya emission in the Universe
at redshifts $z=2-3.5$ by measuring the cross-correlation of \lya surface
brightness with quasars in the Sloan Digital Sky Survey (SDSS/BOSS).
We use nearly a million spectra targeting Luminous Red Galaxies (LRGs) at
$z<0.8$, after subtracting a best fit model galaxy spectrum from
each one, as an estimate of the high-redshift \lya surface brightness.
The quasar-\lya emission cross-correlation we detect has a shape
consistent with a linear $\Lambda$CDM model with
$\Omega_{\rm m} =0.30^{+0.10}_{-0.07}$. The predicted amplitude of this
cross-correlation is proportional to the product of the mean
\lya surface brightness, $\langle \mu_{\alpha} \rangle$, the
amplitude of mass density fluctuations, and the quasar and \lya emission
bias factors. Using published cosmological observations to constrain the
amplitude of mass fluctuations and the quasar bias factor,
we infer the value of the product
$\langle \mu_{\alpha} \rangle \, (b_\alpha/3) = (3.9 \pm 0.9)\times10^{-21}$
$\ergs$ cm$^{-2}$ \AA $^{-1}$ arcsec$^{-2}$, where $b_{\alpha}$ is the
\lya emission linear bias factor.
If the dominant sources of \lya  emission
we measure are star forming galaxies, we infer a 
total mean star formation
rate density 
of ${\rho}_{\rm SFR} = (0.28 \pm 0.07) (3/b_{\alpha}) $ yr$^{-1}$ Mpc$^{-3}$
at $z=2-3.5$.
For $b_{\alpha}=3$, this value is a factor of $21-35$ above previous
estimates relying on individually detected \lya emitters,
although it is consistent with the total star-formation density derived
from dust-corrected, continuum UV surveys.
Our observations therefore imply that $97\%$ of the \lya emission
in the Universe at these redshifts is undetected in previous surveys of
\lya emitters. Our detected \lya emission is also much greater,
by at least an order of magnitude,
than that measured from stacking analyses of faint halos surrounding
previously detected \lya emitters, but
we speculate that it arises from similar low surface brightness \lya 
halos surrounding all luminous star-forming galaxies.
We also detect a redshift space anisotropy of the quasar-\lya emission
cross-correlation, finding evidence at the $3.0 \sigma$ level
that it is radially elongated, contrary to
the prediction for linear gravitational evolution, but consistent with
distortions caused by radiative-transfer effects,
as predicted by Zheng \etal (2011). Our measurements represent
the first application of the intensity mapping technique to 
optical observations.

\end{abstract}


\section{Introduction}
\label{intro}

The \lya emission line
of neutral hydrogen is a strong feature that has been used to detect
galaxies at a wide range of redshifts (e.g., Hu \& McMahon 1996, Keel \etal 
1999, Fujita \etal 2003, Cowie \etal 2010).
Another potentially useful technique is that of intensity mapping
(e.g., Carilli 2011, Peterson \& Suarez
2012, Pullen \etal 2014), which seeks to
map  the large-scale structure using one emission line or more (see e.g.,
Wyithe \& Morales 2007, Visbal \& Loeb 2010),
without resolving individual sources (such as galaxies or
gas clouds). By measuring this structure, one is sensitive to all
clustered emission, without the observational biases which arise from
source detection and luminosity measurement (such as detection
limits, determination of backgrounds and finite aperture size). In this
paper, we seek to perform the first cosmological
measurement of intensity mapping
in the \lya  line, using a large dataset of spectra from the 
Sloan Digital Sky Survey III (SDSS-III, Eisenstein \etal 2011)
 Baryon Oscillation Spectroscopic Survey
(BOSS, Dawson \etal 2013). 
We use spectra that were targeted at massive galaxies at $z<0.8$.
After subtracting best fit model galaxy spectra, we expect that
any high redshift \lya emitters that are within the fiber aperture
result in a residual flux, present also in sky fibers. Even if not
detectable as individual sources, we can search for large-scale structure
in this emission by determining its spatial cross-correlation function
with the positions of BOSS quasars, which are tracers of structure with
a known bias factor (e.g., White \etal 2012) at redshifts $z>2$, where
the \lya emission line is in the optical part of the spectrum.

  Following the early prediction by Partridge \& Peebles (1967) that
galaxies should be detectable at high redshift from their \lya emission
line, many surveys have been 
designed to detect individual galaxies as sources of \lya emission.
These include narrow band imaging (e.g., Ouchi \etal 2003, 
Gronwall \etal 2007),
serendipitous slit spectroscopy (e.g., Cassata \etal 2011) and 
integral field spectroscopy (e.g.,
van Bruekelen \etal 2005, Blanc \etal 2011). These techniques have 
resulted in the  compilation of  catalogs of several hundred to
a few thousand \lya emitting galaxies from redshifts
$z\sim 2.1$ to the redshifts associated with the end of reionization. 
These samples have been used to show that  \lya emitters of 
line luminosity $L_{\alpha}=10^{42} \ergs$
found at $z=3$ have space densities of 
$\sim 10^{-3}$ Mpc$^{-3}$ (e.g., Gawiser \etal 2007, Cassata \etal 2015) and
are therefore expected to
be the progenitors of $L_{*}$ galaxies at redshift $z=0$.
The clustering of these galaxies has been measured on 
scales of up to 10 Mpc by Gauita \etal (2010) and Gawiser \etal (2007), who
find that at redshifts $z=2-3$ they have a bias factor with respect 
to the underlying matter (in CDM models) of $b\sim 1.5-2$.
Integrating the luminosity functions of \lya emitting galaxies,
assuming a power-law extrapolation for the faint end slope, has revealed
that the comoving volume emissivity of \lya photons declines
significantly from  $z\sim6$ to  $z\sim 2$ (Cassata \etal 2011, 
Gronwall \etal 2007, Ouchi \etal 2008).
This behaviour can be compared to the opposite evolution in redshift 
of galaxies measured in optically thin
parts of the rest-frame UV spectrum
(or using the H$\alpha$ line, e.g., \citealt{Hayes11}). 
This comparison has been used to infer (e.g., by \citealt{Hayes11} and 
\citealt{Cassata11})
that the escape fraction of \lya 
photons produced in star forming regions has significantly
decreased from $z=6$ to $z=2$.

Because \lya photons have a high
cross-section for scattering off neutral hydrogen,
extended \lya emission is expected to be common in many environments.
For example, \lya radiation 
from star forming regions in galaxies should undergo hundreds or 
thousands of
scatterings in gas in any circumgalactic medium before finally
escaping or else being absorbed by dust. 
The existence of a general fluorescent emission from the intergalactic
medium was also hypothesised by Hogan \& Weymann (1987) and
Gould \& Weinberg (1996).
Theoretical work applying line radiative 
transfer on gas distribution in cosmological hydrodynamic simulations
has made predictions for \lya emission around galaxies and quasars
(e.g., Cantalupo \etal 2005, Laursen \etal 2007, Kollmeier \etal 2010,
Zheng \etal 2011a), as well as metal line emission
(Bertone \& Schaye 2012). These studies have
resulted in predictions of extended
\lya halos around galaxies with sizes of hundreds of kpc,
with a strong dependence of their properties on environment that can lead
to new effects on galaxy clustering (Zheng \etal 2011a).

  Observational evidence for extended emission includes the discovery
and characterization of the so-called \lya ``blobs'' 
(Steidel \etal 2000).
Deep spectroscopic searches for  diffuse \lya emission
have been completed by Rauch \etal (2008), finding faint \lya emitting
galaxies. Stacking of spectra of damped \lya absorbers in quasars
has also produced measurements of residual \lya emission (Rahmani et al. 
2010, Noterdaeme et al. 2014). 
 Recently, Martin \etal (2014a,b) published the first results
from the Cosmic Web Imager, an integral field spectrograph designed
to map low surface brightness emission, detecting
\lya emission from filamentary structures around a $z=2.8$ 
quasar as long as 250-400 proper kpc. 
Diffuse \lya halos around high redshift galaxies have
been found to be ubiquitous by Steidel \etal (2011) and Matsuda \etal (2012).
Momose \etal (2014) have assembled several samples
of up to 3600 \lya emitters from Subaru narrowband imaging
at a range of redshifts from $z=2.2$ to $z=6.6$ and,
after controlling for atmospheric and instrumental artifacts, they
detect diffuse extended \lya halos with exponential
scale lengths of $\sim 5- 10$ kpc from $z=2.2-5.7$.
The large scale studies in our paper are an alternative, complementary
observational strategy to these earlier studies,
which involve deep integrations over small fields of view. 

All of these sources should be clustered
on large scales and should contribute to the mean \lya emission
intensity in the Universe. This mean emission is detectable if it
cross-correlates as expected with other tracers of large-scale structure
that we can observe at the same redshift. We shall use the quasars found
by BOSS at $z>2$ to correlate with \lya emission in this work. This
clustered \lya emission is extremely faint, but as we shall demonstrate it can
be detected with BOSS thanks to the enormous number of spectra that are
observed. While large-scale clustering measurements cannot
easily allow separation of the signal into various sources, we may
expect faint \lya emitting galaxies to dominate over quasars due to their
much larger number density. If this is the case, then the mean \lya emission
intensity clustered with quasars can be used as a measure of the global star
formation rate (e.g., Cassata \etal 2011), times the mean (luminosity weighted)
bias factor
of the distribution of these galaxies. Our measurement of \lya
emission will therefore be
useful as a probe of star formation which takes into account all sources
of \lya emission, reaching to arbitrarily faint luminosities and surface
brightnesses from extended halos to faint galaxies.

The structure of this paper is as follows. Section 2 describes the
data samples we use in our work, which include the galaxy and sky spectra from
SDSS DR10, along with the quasar catalog from that data release. We present
our measurement of quasar-\lya emission correlations in Section 3,
including the evolution with redshift and clustering parallel and transverse
to the line of sight. Section 4 describes our tests involving
fitting and subtraction of emission lines. In Section 5, we convert
our determination of the \lya surface brightness into a star formation
rate density and compare to other measurements. In Section 6, we summarise our
results and in Section 7
discuss them further. There are also three appendices to the paper,
A-C; in these we measure stray light contamination, determine a 
large-scale surface brightness correction, and perform some sample
tests of our results.

\section{Data samples}

This study makes use of data from the SDSS BOSS survey Data Release 10
(DR10, \citealt{Ahn14}), including quasar 
position and redshift data, galaxy spectra and
sky fiber spectra. The SDSS camera and telescope 
are  described in Gunn et al. (1998) and Gunn et al. (2006), respectively.
Full information on the SDSS/BOSS spectrographs can be found in 
Smee et al. (2013). The wavelength coverage of the spectrograph is from 
$\lambda=$3560 \AA\ to 10400 \AA\, the resolving power is $R\sim1400$
for the range $\lambda= 3800$ \AA$-$ 4900 \AA, and is kept above
$R=1000$ for the remainder of the wavelength range. The
fibers have a diameter of 120 $\mu$m, corresponding to $2$ arcsec. in angle.
We restrict the redshift range of data we use in our analysis to
$2.0 < z < 3.5$, due to the spectrograph cutoff at low redshift and
the limited number of observed quasars at high redshift.

\subsection{Spectra}
The 987482 galaxy spectra in our sample are of targeted LRGs which are within
redshifts $z\sim0.15$ and $z\sim 0.7$. 
The redshift range of the original targets is not
important to our study, as for each spectrum we make use only of the pixels
for which the \lya emission line lies within the redshift range
specified above ( $2.0 < z < 3.5$). In observed
wavelength units this is 3647 \AA\
to 5470 \AA. We also make use of 146065 sky fiber spectra.

The main BOSS LRG program consists of two galaxy target
samples (see Dawson \etal 2013), designated CMASS
(for “constant mass”) and LOWZ (for “low-redshift”). The
LOWZ galaxy sample is composed of massive red galaxies
spanning the redshift range $0.15 \simlt z  \simlt 0.4.$ The CMASS
galaxy sample is composed of massive galaxies spanning the
redshift range $0.4 \simlt z \simlt 0.7$. Both samples are color-selected
to provide near-uniform sampling over the combined volume.
The faintest galaxies are at $r = 19.5$ for LOWZ and $i = 19.9$ for
CMASS. Colors and magnitudes for the galaxy selection cuts
are corrected for Galactic extinction using Schlegel et al. (1998)
dust maps. We do not differentiate between CMASS and  LOWZ samples
in our analysis.   

The spectroscopic measurement pipeline for BOSS is described in detail in 
Bolton \etal (2012). The most important data products that are used in the 
present analysis are:
(a) Wavelength-calibrated, sky-subtracted, flux-calibrated, and
co-added object spectra, which have been rebinned onto a uniform baseline
of $\Delta \log_{10} \lambda = 10^{-4}$ (about 69 km s$^{-1}$ pixel$^{-1}$).
(b) Statistical error-estimate vectors for each spectrum (expressed
as inverse variance) incorporating contributions
from photon noise, CCD read noise, and sky-subtraction
error. (c) Mask vectors for each spectrum.

\subsubsection{Data preparation}
For each of the  LRG spectra, we subtract the best fit model spectrum
provided by the pipeline. 
This template model spectrum (see Bolton \etal 2012 for details)
is computed using least-squares minimization comparison
of each galaxy spectrum to a full range of 
galaxy templates. A range of redshifts is explored,
with trial redshifts spaced every pixel.
At each redshift the spectrum is fit with an error-weighted
least-squares linear combination of redshifted template
“eigenspectra” in combination with a low-order polynomial.
The polynomial terms absorb Galactic extinction, intrinsic
extinction, and residual spectrophotometric calibration errors
(typically at the $10\%$ level) that are not fully spanned by the
eigenspectra; there are three polynomial degrees of freedom for galaxies.
The template basis sets are derived from restframe
principal-component analyses (PCA) of training samples
of galaxies, and have four degrees of freedom (eigenspectra).

\begin{figure}
\centerline{
\psfig{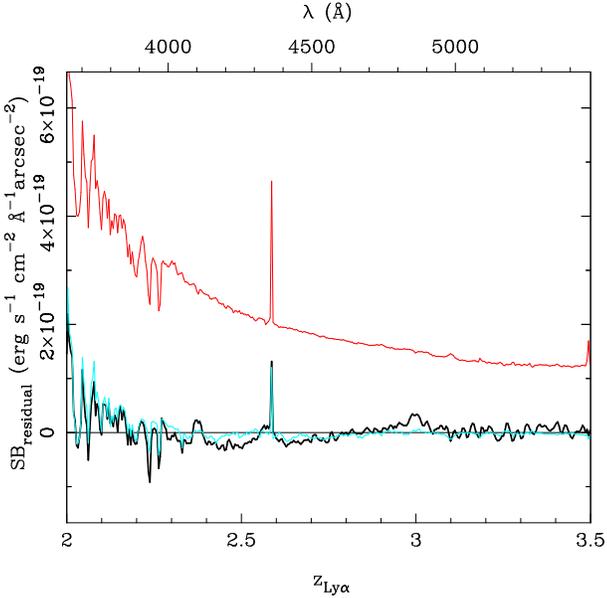}
}
\caption{ Black line: 
the average residual surface brightness in all 987482
 LRG spectra after subtraction 
of the best fit galaxy model. The bottom horizontal axis is in units of the
redshift of the \lya line and the top in units of observed wavelength. 
Red line (top): the average sky fiber surface brightness
in all 146065 sky fiber spectra.
Cyan line: the average sky fiber surface brightness with model sky subtracted
from all sky fiber spectra.
In all curves, the prominent emission line
at wavelength $\lambda=4358$\AA is due to terrestrial airglow from Mercury 
streetlamps (the Hg G-line).
\label{resflux}}
\end{figure}

After subtraction of the best fittting template
spectrum, we compute the average residual spectrum of all galaxies. This 
is displayed in Figure \ref{resflux} where the horizontal axis is labelled
in units of the redshift of the \lya line. 
Figure  \ref{resflux}  also presents the mean sky spectrum and the 
sky-subtracted sky spectrum.

We can see that the residual
surface brightness per unit wavelength (hereafter shortened to
``surface brightness''- we use this term to refer to the quantity measured 
throughout the paper, which is most precisely 
the flux density per unit solid angle per
unit wavelength) in the galaxy fibers is within
 $\pm 10^{-19} \ergs
\cm^{-2} \angs^{-1}\asec^{-2}$ 
for most of the redshift range.
There are however significant excursions corresponding to features including
the zero redshift Calcium H and K lines (at 3969 and 3934 \AA)
and a strong Mercury G line from streetlamps (at 4358 \AA). In 
our analysis we subtract the mean
residual surface brightness, from
all spectra before cross-correlating them as we are only interested in
the fluctuations in the \lya surface brightness. In order to
reduce noise, we also mask two regions
corresponding to large 
features in the residual  surface brightness, 40 \AA\ and 30 \AA\ 
windows centered on wavelengths 3900 \AA\ and 4357 \AA\
respectively (corresponding to redshifts $z=2.21$ and 2.58).
 
Comparing the sky fiber and galaxy fiber residual spectra, we can
see that there are differences at the $\sim 10^{-19}$ $\ergs$
level over much of the spectra. We attribute these to galaxy surface
brightness that was not subtracted perfectly by the galaxy model.
In our cross-correlation technique for measuring
clustering in the \lya emission we necessarily subtract
the mean surface brightness, therefore residual fluctuations seen in 
Figure \ref{resflux} are not problematic except for the noise they
contribute.

\subsection{Quasars}
We use quasars from the SDSS/BOSS DR10 catalogue (Ahn et al. 2014). The
quasar target sample included both color-selected candidates
and known quasars (Bovy et al. 2011; Kirkpatrick et al. 2011;
Ross et al. 2012). The candidate quasar spectra were all visually inspected
and redshift estimates computed using a principal component analysis (see
P\^{a}ris \etal 2012 for the details of the procedure 
as applied to DR9 quasars).
We select the 130812 quasars in the DR10 dataset that have redshifts in the
range $2.0 < z < 3.5$.
Because the galaxy pixels cover this redshift range
uniformly, the central redshift of our measurements 
is the mean redshift of these quasars, $z=2.55$.

\section{Quasar-\lya emission cross-correlation}
\label{xcor}

  Before computing the quasar-\lya emission cross-correlation, we first
split the sample of galaxy spectra into 100 subsamples of
approximately equal sky area based on contiguous groupings of plates.
We then convert the galaxy spectrum pixels and the quasar angular positions
and redshifts into comoving Cartesian coordinates using a flat cosmological
model with matter density $\Omega_{\rm m}=0.315$, consistent with the
Planck, Ade (2014) results (cosmological constant
density $\Omega_{\rm \Lambda}=0.685$). This fiducial model is used throughout
the paper.

We compute the  quasar-\lya emission surface brightness
cross-correlation, \xiqar , using a sum over all
quasar-galaxy spectrum pixel pairs separated by $r$
within a certain bin:
\begin{equation} 
\xi_{{\rm q} \alpha}(r)
 = \frac{1}{\sum_{i=1}^{N(r)}w_{ri}}
\sum^{N(r)}_{i=1} w_{ri}\, \Delta_{\mu,ri},   
                            \label{xieq}          
\end{equation}
where $N(r)$ is the number of pixels in the
bin centered on quasar-pixel distance $r$,
and $\Delta_{\mu,ri}=\mu_{ri}-\langle \mu(z) \rangle$
is the residual surface brightness
in the spectrum at pixel $i$ for the bin $r$.
Note here that we have a different list of pixels labeled as $i$ for
each bin in the separation $r$ between a pixel and a quasar, which has
\lya surface brightness
$\mu_{ri}$. The residual flux at each pixel is obtained by
subtracting the mean at each redshift, $ \langle \mu(z) \rangle$.
We weight each pixel  by $w_{ri} = 1/\sigma^{2}_{ri}$, where 
$\sigma^2_{ri}$ is the pipeline estimate of the inverse variance of the
flux at each pixel. We first present our results
as a function of 
only the modulus of the quasar-pixel separation $r$ in comoving $\hmpc$,
  in
20 bins
logarithmically spaced between $r=0.5 \hmpc$ and $r=150 \hmpc$.
In Section \ref{rparperp} we will also examine redshift space
anisotropies in the
correlation function \xiqa\ by considering bins in the parallel and
perpendicular components of $r$, using the same formulation.

  When evaluating equation \ref{xieq}, a possible 
signficant systematic error is caused by stray light from the
quasars themselves contaminating spectra of nearby galaxies. This
occurs because the 
light from the various fibers is dispersed onto a single CCD, so that 
extraction of each spectrum along one dimension (Bolton \etal 2012) 
may include light from
adjacent fibers of bright sources. We see strong evidence of this
stray light from
quasars in galaxy spectra when the quasar and galaxy spectra are four
fibers apart or fewer, in the list of fibers as they are ordered in
the CCD. The effect is discussed in detail in Appendix A.
When the galaxy and quasar spectra in a pair are more than
4 fibers apart, we see no evidence for this contamination, and the results
are statistically consistent with using only pairs of quasars and galaxies
on different plates (see Appendix A).
In order to safely eliminate this
stray quasar light when computing 
the flux cross-correlation with equation \ref{xieq}, we
therefore apply the constraint that the quasar and galaxy fibers must 
be at least six fibers apart.

There is also the possibility that some clustering in the plane of the sky
is generated by effects (e.g., galactic obscuration) which
modulate both \lya surface 
brightness and quasar target selection. Appendix B presents
 measurements of  \xiqa\
for quasar-pixel pairs which are close together on the sky (i.e., in the
transverse separation)
but widely separated along the line of sight. This measurement enables us
to quantify how much clustering could be caused by effects such as Galactic
obscuration and to compute a \xiqa\ correction term 
to be subtracted from our fiducial clustering result. We also 
measure \xiqa\ for pairs which are close in the line-of-sight separation
but widely separated on the sky. This latter measurement constrains how
much spurious clustering is caused by large-scale variations in the 
line-of-sight direction, for example redshift evolution in the efficiency
of galaxy subtraction, or flux calibration errors with wavelength that
may be associated with sky lines. We apply
the corrections to \xiqa\ from Appendix B to our analysis below and
in the other sections of the paper. We discuss the small-scale anisotropy
of \xiqa\ in Section \ref{rparperp} below. For now we note that 
application of the correction factors described 
above changes the amplitude
and shape of our measured \xiqa\ by less than 
$1 \sigma$ in all cases.

We perform the pairwise computation of equation \ref{xieq}
for each of our 100 subsamples, and then compute the mean
and standard deviation of \xiqar\ using a Jackknife estimator. The 
Jackknife estimator is also used to compute the covariance matrix of 
\xiqar:

\begin{equation}
\label{covm}
C_{ij}=\sum^{M}_{k=1} [\xi_{{\rm q} \alpha,k}(r_{i})-\overline{\xi_{{\rm q} \alpha}}(r_{i})] [\xi_{{\rm q} \alpha,k}(r_{j})-\overline{\xi_{{\rm q} \alpha}}(r_{j})],
\end{equation}
where $\xi_{{\rm q} \alpha,k}(r_{i})$ is the cross-correlation in bin
$i$ for Jackknife sample $k$, $\overline{\xi_{{\rm q} \alpha}}(r_{i})$
is the cross-correlation for bin $i$ for the full dataset, and the
number of Jackknife samples is $M=100$.

\subsection{Fiducial result}
\label{fidresult}

\begin{figure}
\centerline{
\psfig{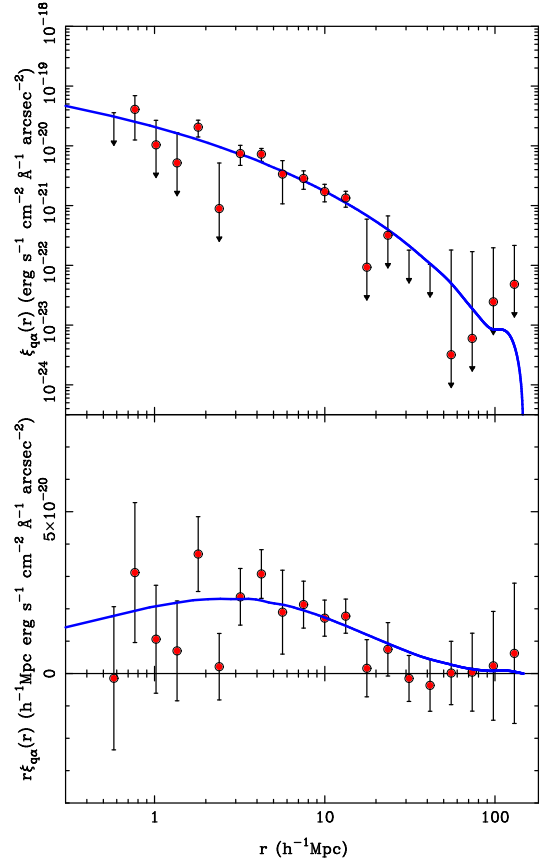}
}
\caption{ The quasar-\lya emission cross-correlation function,
\xiqar\ (see Equation \ref{xieq}). The points represent
results for the fiducial sample that covers redshift range  $2.0 < z< 3.5$.
The error bars have been calculated using a jackknife estimator and
100 subsamples of the data. The smooth curve is a best fit linear CDM
correlation function (see Section \ref{modfit}). The top panel shows the
\xiqar\ results with a log y-axis scale, and the bottom panel displays
$r$\xiqar\ on a linear scale in order to allow points which are negative
to be visible. 
\label{cdmfid}}
\end{figure}

We show \xiqar\ for our fiducial sample (which is the entire
dataset over the redshift range $2.0 < z< 3.5$)
in Figure \ref{cdmfid}. The mean \lya redshift of the galaxy pixels in this
sample is $z=2.71$ and of the quasars $z=2.55$. Because the galaxy pixel
distribution is uniform in redshift, and because quasar-pixel pairs with small
separations contribute most to the clustering signal, we adopt the effective 
mean redshift of our fiducial measurement to be $z=2.55$.
 The cross-correlation
function is in units of the surface brightness of \lya emission,
and its amplitude is directly proportional to that surface brightness.
The \xiqar\ points reveal that there is significant measurable large-scale
structure present in the \lya emission, on scales from $1$ to $\sim 15 \hmpc$. 
Figure \ref{cdmfid} also displays a linear CDM fit to the cross-correlation 
function, which is consistent with
observational results on scales from  $1-100 \hmpc$.
 We turn to this fit in the next subsection. 

\subsection{Model fit}
\label{modfit}

If the \lya emission clustering is due to a linearly biased 
version of the density field, then a model for the isotropically
averaged quasar-\lya cross-correlation \xiqar is as follows:
\begin{equation}
\xi_{{\rm q}\alpha}(r)= b_{q}b_{\alpha} f_{\beta} \langle \mu_{\alpha} \rangle \xi(r)
\label{model}
\end{equation}
where $\langle \mu_{\alpha} \rangle$ is the mean surface 
brightness of \lya emission, $b_{\rm q}$ and $b_{\alpha}$ are the 
quasar and \lya emission linear bias factors, 
and $\xi(r)$ is the linear $\Lambda$CDM correlation function.

It is important to note that $b_{\alpha}$ is the bias factor for
\lya surface brightness fluctuations, and is different in definition from the 
usually quoted bias factor of
\lya emitters, $b_{\rm LAE}$ (e.g., as measured  by 
\citealt{Gawiser07} and \citealt{Guaita10}).
The bias factor $b_{\rm LAE}$ reflects the relation between the fluctuations 
$\delta_n$ in the number density $n$ of \lya emitters and that in the matter
density $\delta$,
\begin{equation}
\delta_n=\frac{n-\langle n \rangle}{\langle n \rangle}=b_{\rm LAE} \delta,
\label{blyae}
\end{equation}
where $\delta = (\rho-\langle \rho \rangle)/\langle \rho \rangle$ and
$\rho$ is the matter density field. 
The factor $b_{\alpha}$ in Equation \ref{model} relates fluctuations 
$\delta_\mu$ in the \lya surface brightness $\mu$ to matter fluctuations 
according to \begin{equation}
\delta_\mu=
\frac{\mu-\langle \mu \rangle}{\langle \mu \rangle}=b_{\alpha} \delta.
\label{blya}
\end{equation}
In the absence of radiative transfer effect (\citealt{Zheng11a}; see 
Section~\ref{sec:RT}), the \lya surface brightness $\mu$ is proportional to 
the \lya luminosity density $\rho_L$ of the {\it underlying} 
star-forming galaxy population. The fluctuations $\delta_L$ of the 
latter can be characterized by the bias factor $b_L$,
\begin{equation}
\delta_L=\frac{\rho_L-\langle \rho_L \rangle}{\langle \rho_L \rangle}=b_L\delta,
\label{blum}
\end{equation}
and we have $b_\alpha=b_L$.
As $b_L$ reflects weighting by luminosity rather than by number, it is 
likely to be significantly higher than $b_{\rm LAE}$, because higher 
luminosity emitters tend to be more strongly clustered. We will return to 
this topic 
in Section \ref{sfrd}. Radiative transfer effect leads to a modification
in the relation $b_\alpha=b_L$. In a simple model, we have 
$b_\alpha=b_L+\alpha_1$ with $\alpha_1$ a positive number (see 
Section~\ref{sec:RT}). Overall, we expect $b_\alpha$ to be substantially 
higher than $b_{\rm LAE}$.

In Equation \ref{model}, $f_{\beta}$ is a constant enhancement to 
the correlation function on linear scales of the form that 
is caused by peculiar velocity redshift-space distortions
(Kaiser 1987). We use the linear CDM transfer function of
Lewis \etal (2000) to compute $\xi(r)$. 
In our computations we choose to vary the shape of the correlation
function by changing $\Omega_{\rm m}$, the matter density,
in the context of the $\Lambda$CDM model, 
 keeping the other
parameters which influence the shape (such as $h$, and the baryon density
$\Omega_{b}$) fixed. Parameters have been reported for the best fit 
$\Lambda$CDM model to the Planck Satellite 
data by Ade \etal 2014. We assume Planck values
for $\Omega_{\rm b}=0.049$ and the spectral index $n_{\rm s}=0.9603$ but set
$h=0.7$.  Note that we are merely using $\Omega_{\rm m}$ to parametrize
the shape of the correlation function to see if it is consistent
with other observations, and are not presenting our
results for $\Omega_{\rm m}$ as properly
marginalized measurements of that parameter.

The other free parameter is 
the amplitude, \ampqa . We assume in all cases that the
underlying amplitude of mass fluctuations $\sigma_{8}(z=0)=0.83$, and
therefore that $\sigma_{8}(z=2.55)=0.294$, again consistent
with Ade \etal 2014.

 We fit our model to the data in Figure \ref{cdmfid} by varying these
two parameters. The $\chi^{2}$ value is given by
\begin{equation}                                                      
\chi^{2} = \sum_{N} 
[\xi^{\rm obs}_{{\rm q}\alpha}(r_{i}) - 
\xi^{\rm mod}_{{\rm q}\alpha}(r_{i})]      
C_{ij}^{-1}  [\xi^{\rm obs}_{{\rm q}\alpha}(r_{j}) - 
\xi^{\rm mod}_{{\rm q}\alpha}(r_{j})]      
\label{chi2}                                                                
\end{equation}
where the sum is over the $N=20$ bins,
$\xi^{\rm obs}_{{\rm q}\alpha}(r_{i})$ is
the observed cross-correlation measured in bin $i$, 
$\xi^{\rm mod}_{{\rm q}\alpha}(r_{i})$
is the model prediction for bin $i$,  and $C_{ij}$
is the covariance matrix  computed using our 100
Jackknife samples  in Equation \ref{covm}.

\begin{figure}
\centerline{
\psfig{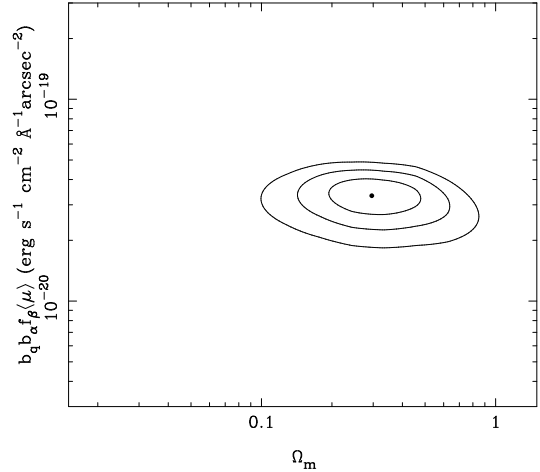}
}
\caption{
Fit parameters for the amplitude
\ampqa\  and shape $\Omega_{\rm m}$ (for fixed $h$ and other parameters)
of a linearly biased CDM model fit to the \lya cross-correlation
function plotted in Figure \ref{cdmfid}. The dot indicates the best fit
parameters and the contours show the 1, 2 and 3 $\sigma$ confidence contours.
\label{gacont}}
\end{figure}

The best fit values and one sigma 
error bars are as follows:
\begin{equation} 
\text{\ampqa}  = 3.33^{+0.41}_{-0.43} \times 10^{-20} \ergs \cm^{-2} \angs^{-1}
\asec^{-2},
\end{equation}
and $\Omega_{m}=0.296^{+0.103}_{-0.071}$.
The shape parameter $\Omega_{\rm m}$ is consistent
with the best fit value from the Planck
satellite results  (Ade \etal 2014, $\Omega_{\rm m}=0.30$).
The 1, 2 and 3 $\sigma$ confidence contours in these parameters
considered together (i.e. $\Delta \chi^{2}$=2.3, 6.17 and 11.8)
are displayed along with the best fit values in Figure \ref{gacont}.

When carrying out our linear fit to the isotropically averaged
correlation function we are neglecting non-linear effects and redshift
measurement errors which are expected to suppress
the correlation function on small scales (Davis and Peebles
1983). We are also not including in our model of Equation \ref{model}
the one-halo term (see e.g., Cooray and Sheth 2002) representing
structure inside virialized halos, which would cause the correlation
function to be increased at small scales. An
estimate of the size of the redshift-space distortion effects is
given below.
There is some evidence for suppression of redshift-space distortions of the
form that is expected from radiative transfer effects \citep{Zheng11a}.
The effect of the suppression terms in the cross-correlation
function will be of opposite
sign to the boost due to a one-halo term, so that the partial
cancellation will further lower the impact of these effects.
Because of these interactions, we leave more detailed nonlinear and halo
modelling of the correlation function to future work.

\subsection{Clustering transverse and parallel to line of sight}
\label{rparperp}

\begin{figure*}
\centerline{
\psfig{file=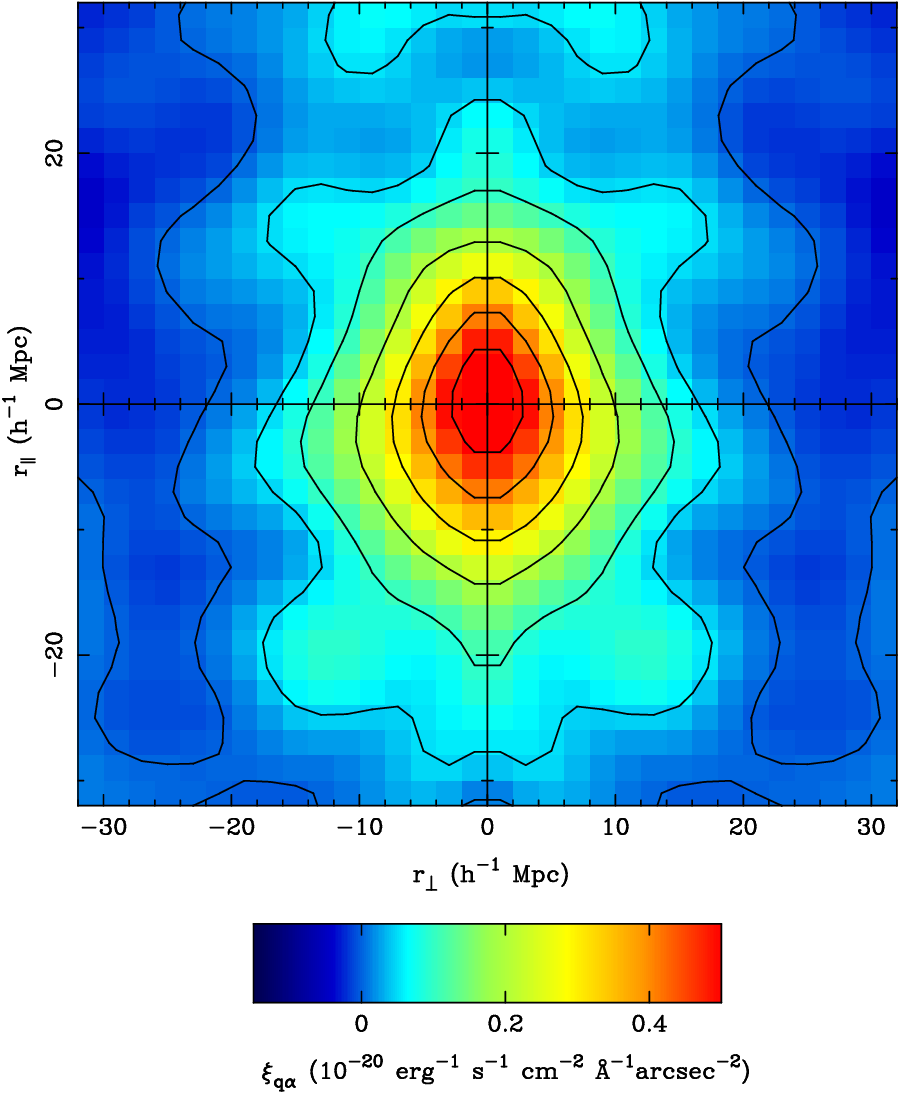,angle=0.,width=8.5truecm}
\psfig{file=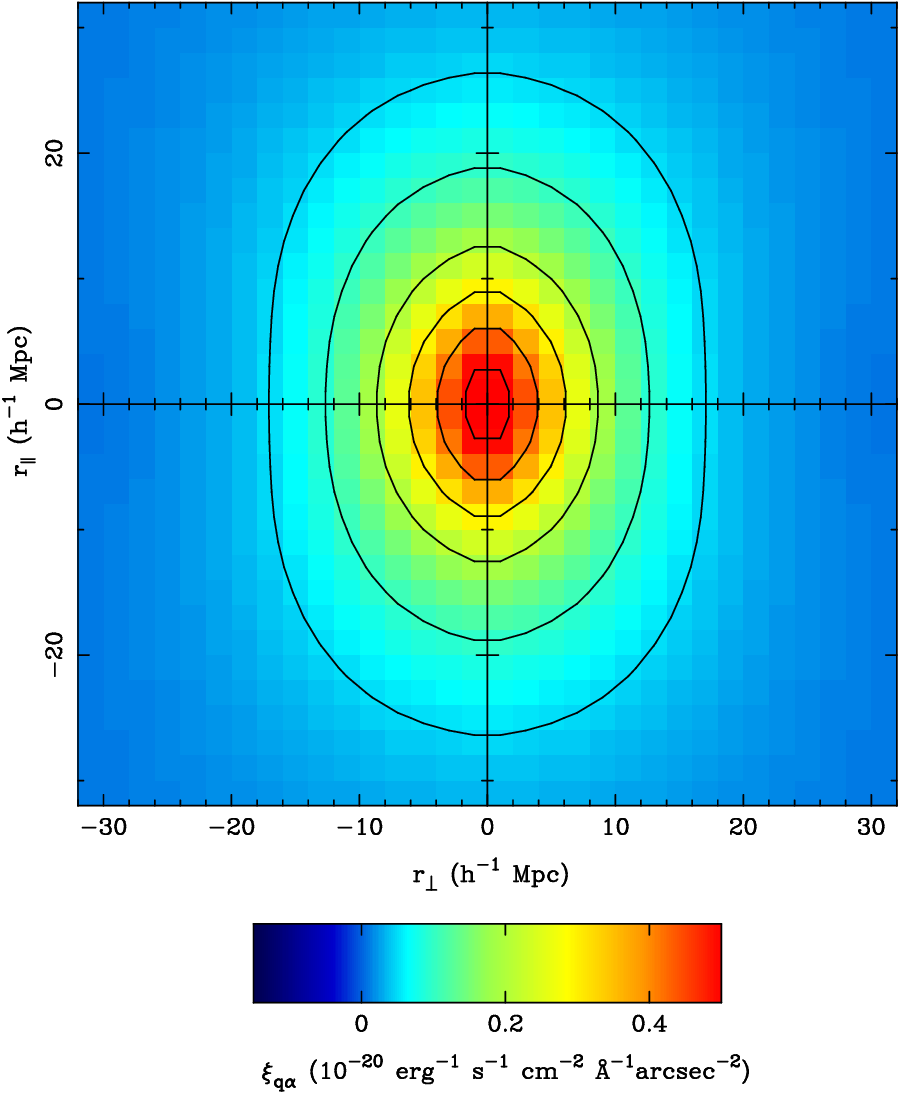,angle=0.,width=8.5truecm}
}
\caption{ 
Left panel: The quasar-\lya cross-correlation \xiqa\ as a function
of $r_{\parallel}$ and $r_{\perp}$. The units (of \lya 
surface brightness) are the same as in Figure 
\ref{cdmfid}. The contours are spaced at values of $10^{-21} \ergs
\cm^{-2} \angs^{-1} \asec^{-2}$. To reduce noise in
the image, the dataset was smoothed with a Gaussian filter with 
$\sigma=4 \hmpc$ (2 cells) before plotting. 
Right panel: The model fit to the quasar-\lya cross-correlation including
redshift-space distortions (see Section \ref{zdistort}).
\label{sigmapi}}
\end{figure*}

  We also compute the quasar-\lya cross-correlation as a function
of $r_{\parallel}$ and $r_{\perp}$, the quasar-pixel pair separation
along and across the line of sight, shown in Figure \ref{sigmapi},
on a linear scale. We can see that the contours are relatively
symmetric about the $r_{\parallel}=0$ axis and somewhat stretched along the
$r_{\parallel}$ direction. Font-Ribera 
et al. (2013) found a redshift offset between
quasars and the \lya forest of $\delta_{z}=-160 \kms$, due to the quasar
catalogue redshifts being on average too small by this amount. This 
correlation resulted
in the quasar-\lya forest cross-correlation being shifted upwards by this 
amount. The precision of the quasar-\lya emission cross-correlation in
our paper is smaller, but visual inspection of Figure \ref{sigmapi}
reveals that the position of the centroid
of the cross-correlation is consistent with a small upward shift of
this magnitude ($1.6 \hmpc$ at this redshift).

On the scales where the cross-correlation is easy to
discern ($r \simlt 20 \hmpc$), there is no sign of compression due to 
linear infall (Kaiser 1987). The redshift-space distortions should be less
prominent compared to the \lya forest because of the expected
higher bias for the \lya emission. In reality there appears to be
stretching along the line of sight (we quantify this below),
which might be due to a combination of quasar redshift errors, the
intrinsic velocity
dispersion of quasars in their host halos, or the intrinsic velocity
dispersion of the sources of \lya emission. 

 Another source of apparent clustering anisotropy could be the
radiative transfer effects predicted by Zheng \etal (2011a). It was shown
by these authors, using
cosmological radiation hydrodynamic simulations,
that \lya radiative transfer has a strong environmental
dependence which can cause the apparent spatial distribution of
\lya emission to become anisotropic with respect to the line-of-sight
direction. Density fluctuations along the line-of-sight direction
are found to preferentially emit the \lya radiation in that direction
in overdense regions, mainly because of the effect of peculiar velocity
gradients on the \lya radiative transfer. This causes a suppression of
the line-of-sight fluctuation, which can be modeled similarly
to the Kaiser effect (also caused by the peculiar velocity gradient),
even though the sign of the effect is opposite.

\subsection{Fitting redshift-space distortions}
\label{zdistort}

In order to approximately quantify the level of distortion in Figure 
\ref{sigmapi} and its statistical significance, we have investigated fitting 
a redshift-space distortion model to the  \xiqa\ $(r_{\perp},r_{\parallel})$
data. To compute the model for \xiqa\ $(r_{\perp},r_{\parallel})$, we first
assume the linear $\Lambda$CDM correlation function shape used in Equation 
\ref{model} and then use a model for peculiar velocities to distort it
in redshift space.
Our pecular velocity model includes  standard linear infall
for large scale flows (Kaiser 1987) and a small scale
random velocity dispersion (e.g., Davis and Peebles 1983). 

The parameterization of the model for linear infall allows for stretching
(outflow) as well as squashing (infall) along the line of sight.
Although gravitational processeses are expected to result in infall,
as mentioned above Zheng \etal (2011b) 
have shown that radiative transfer effects
on the anisotropy of clustering 
can be approximately parametrized with the same model. We do this
here, allowing a net linear outflow measured by the model to be
interpreted as the radiative transfer effect.

The effects of coherent flows
on the correlation function in linear theory were presented
by Hamilton (1992). We use the formulation
of Hawkins (2003), with modifications to make it appropriate for
the case of cross-correlation functions. This modification involves
the use of the two bias factors $b_{\rm q}$ and $b_{\alpha}$
from Equation \ref{model} to compute redshift-space distortion
factors $\beta_{\rm q}=\Omega_{\rm m}(z=2.55)^{0.6}/b_{\rm q}$
and $\beta_{\alpha}=\Omega_{\rm m}(z=2.55)^{0.6}/b_{\alpha}$. The linearly
distorted quasar-\lya emission cross-correlation function is
then given by
\begin{equation}
\xi'_{{\rm q}\alpha}(r_{\perp},r_{\parallel})=
b_{\rm q} b_{\alpha} \langle \mu_{\alpha\ }  \rangle 
\left[ \xi_0(s)P_0(\mu) + \xi_2(s)P_2(\mu) + \xi_4(s)P_4(\mu) \right],
\end{equation}
where $\mu=r_{\parallel}/r$, and
\begin{equation}                  
\xi_0(s) = \left[1 + \frac{1}{3}(\beta_{\rm q}+\beta_{\alpha}) + \frac{1}{5}\beta_{\rm q}\beta_{\alpha}\right] \xi(r)   
\label{xi0}
\end{equation}
\begin{equation}                                                           
\xi_2(s) = \left[\frac{2}{3}(\beta_{\rm q}+\beta_{\alpha}) + 
\frac{4}{7}\beta_{\rm q}\beta_{\alpha}\right][\xi(r)-\overline{\xi}(r)],                  
\end{equation}
\begin{equation}                                                      
\xi_4(s) = \frac{8}{35}\beta_{\rm q}\beta_{\alpha}\left[\xi(r) + \frac{5}{2}\overline{\xi}(r)  
   -\frac{7}{2}\overline{\overline{\xi}}(r)\right],                
\end{equation}
with
\begin{equation}                                                              
\overline{\xi}(r) = \frac{3}{r^3}\int^r_0\xi(r')r'{^2}dr',      
\end{equation}
\begin{equation}                                                             
\overline{\overline{\xi}}(r) = \frac{5}{r^5}\int^r_0\xi(r')r'{^4}dr'.         
\end{equation}
Here $\xi(r)$ is the linear $\Lambda$CDM correlation function 
of equation (\ref{model}).

We use these relations to create a model $\xi'_{{\rm q}\alpha}(r_{\perp},r_{\parallel})$
which we convolve with the distribution function of random
pairwise motions, $f(v)$, to produce the final model
$\xi_{{\rm q}\alpha}(r_{\perp},r_{\parallel})$:
\begin{equation}                                                
\xi_{{\rm q}\alpha}(r_{\perp},r_{\parallel})=
 \int^{\infty}_{-\infty}\xi'_{{\rm q}\alpha}\left(r_{\perp},r_{\parallel} - \frac{(1+z)v}{H(z=2.55)}\right)f(v)dv           
\end{equation}

The random velocity dispersion we use is an exponential model,
thus the distribution function of velocities is
\begin{equation}
f(v)=\frac{1}{\sigma\sqrt{2}}\exp\left(-\frac{\sqrt{2}|v|}{\sigma}
\right),
\label{fv}
\end{equation}
where $\sigma$ is the pairwise velocity dispersion of quasars and
\lya emission, which we assume to be independent of 
pair separation.

 There is a strong degeneracy between $\beta_{\rm q}$ and
$\beta_{\alpha}$, but the bias factor for BOSS quasars
is reasonably
well measured. The first BOSS measurement, of $b_{\rm q}=3.6 \pm 0.6$, 
was made using the quasar autocorrelation function by White \etal (2012).
Font-Ribera \etal (2013), find an even more precise 
value of $b_{q}=3.64^{+0.13}_{-0.15}$
from  the cross-correlation of BOSS quasars with the \lya forest.
Because of this we set
$b_{\rm q}=3.64$, giving $\beta_{\rm q}=0.27$.
The free parameters in our distortion model are therefore 
$\beta_{\alpha}$, $\sigma$ and an amplitude
parameter $b_{\rm q} b_{\alpha} \langle \mu_{\alpha\ }  \rangle$.

\begin{figure}
\centerline{
\psfig{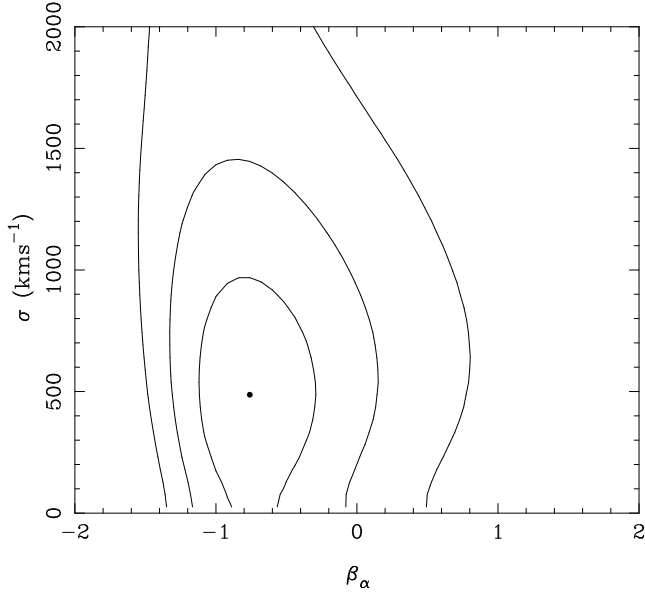}
}
\caption{1, 2 and 3$\sigma$ confidence contours for the redshift-space
distortion parameters $\beta$ and $\sigma$ found from the fit to the 
quasar-\lya emission cross-correlation (Section 3.4). The best fit
values of  $\beta$ and $\sigma$  are shown with a dot. 
\label{betasigmacont}}
\end{figure}

  We set the  parameter governing the shape in our model, $\Omega_{\rm m}$, 
equal to the Planck
value, $\Omega_{\rm m}=0.30$. We compute \xiqa$(r_{\perp},r_{\parallel})$
for our model for a grid of values of varying
$\beta_{\alpha}$, $\sigma$, and $b_{{\rm q}} b_{\alpha}\langle \mu \rangle$.  
We then compare our model to the observed 
\xiqa$(r_{\perp},r_{\parallel})$ from Figure \ref{sigmapi},
performing a $\chi^{2}$  fit to all
points within $r=40 \hmpc$. We again use jackknife error bars
computed from 100 subsamples, but because of difficulties with noisy matrix
inversion, we do not use the off-diagonal ($C_{i\ne j}$) terms when inverting
the covariance matrix.
We marginalize over the amplitude parameter
$b_{{\rm q}} b_{\alpha}\langle \mu_{\alpha\ } \rangle$, and show the confidence
contours (for the two remaining degrees of freedom)
in Figure \ref{betasigmacont}.

We have allowed the $\beta_{\alpha}$ parameter to be negative in our 
fit not because
we believe that outflow of \lya emission is likely around quasars but,
as mentioned above, because this allows quantifying the stretching along 
the line of sight seen in Figure \ref{sigmapi} and its significance.
Our best fit values with 1 $\sigma$ error bars are
$\beta_{\alpha}=-0.76\pm0.36$ and $\sigma=490\pm300 \kms$. From
these values and by observing the contours in Figure \ref{betasigmacont},
 we infer a detection of anisotropies
in the quasar-\lya emission correlation function
that is opposite in sign to that
expected from peculiar velocity flows due to gravitational evolution:
our constraint on $\beta_{\alpha}$ is $2.1 \sigma$  from
$\beta_{\alpha}=0$. We interpret this result as indicating
that there are strong
non-gravitational effects on the \lya emission causing the elongation
of the cross-correlation contours along the line of sight extending
to large separations; the good fit obtained with the redshift-space distortion
model with negative $\beta$ needs to be understood in this case as a
coincidence, since the model is not physically correct.

 If we limit the fit to points
with 20 $\hmpc$, we find central values for  $\beta_{\alpha}$ 
and $\sigma$ that are consistent with those for our fiducial fit,
As might be expected, 
the error bars are larger, however, by a factor of $40\%$ for $\sigma$
and $65\%$ for $\beta_{\alpha}$.

  If we assume a \lya emission bias factor $b_{\alpha}=3$, corresponding
to highly biased star forming galaxies,
(see Section \ref{sfrd} for further discussion of this value), then 
$\beta_{\alpha}=0.32$ at redshift $z=2.55$. For the 
expected value of $\sigma$, we can use as a guide the 
results of Font-Ribera \etal (2013) who constrained $\sigma < 370 \kms$ 
at the 1 $\sigma$ confidence level from the
redshift-space quasar-\lya forest 
cross-correlation function.  
We can see from Figure \ref{betasigmacont}
that although the $\sigma$ measurement 
is consistent with Font-Ribera \etal, $\beta_{\alpha}$ and $\sigma$
considered jointly disagree at the $2.5 \sigma$ level from
our measurement.

We discuss further in Section \ref{disc} how our measurements can
be interpreted as a detection of clustering anisotropies due to
radiative transfer effects (Zheng \etal 2011b).  In this case,
elongation along the line of sight is expected, which can
explain the effective measurement of a negative $\beta_{\alpha}$ 
parameter.

  The cross-correlation contours of the best fit model
are plotted in Figure \ref{sigmapi} (right panel), which  also reveals
stretching along the line-of-sight. 
The $\chi^2$ value for the fit is 610 measured from 400 bins, with 
3 free parameters, a reduced $\chi^2$ of 1.5. The fit is therefore
not good, and the discrepancy arises in large part in the central
region, where the model has lower surface brightness than the
observations. This result may be a sign that adding a one-halo term 
to the correlation function would provide a better
fit, and should be addressed in future work with a larger data sample.


\subsection{Large-scale tests}

\begin{figure}
\centerline{
\psfig{file=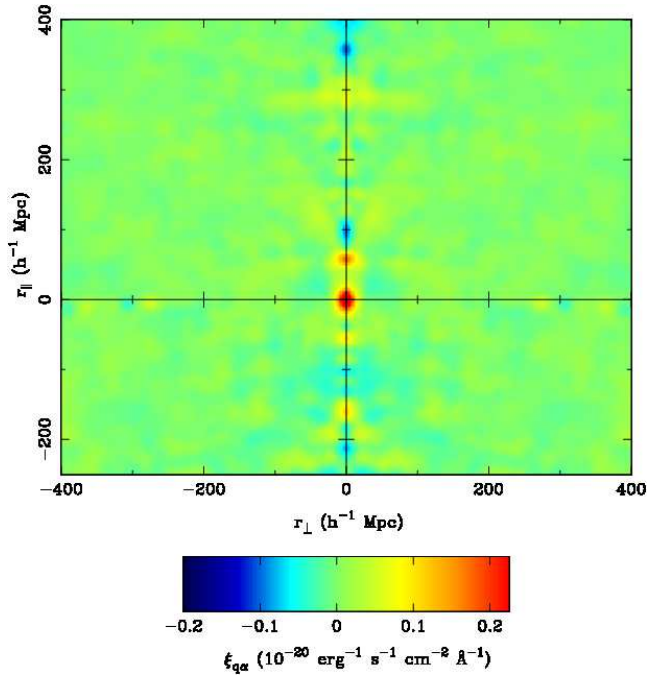,angle=0.,width=8.5truecm}
}
\caption{ 
The quasar-\lya cross-correlation \xiqa\ as a function
of $r_{\parallel}$ and $r_{\perp}$. This figure shows the same information
as Figure \ref{sigmapi}, except over a larger range of scales (see text).
\label{largerange}}
\end{figure}

The cross-correlation across and along the line of sight over the whole
spectrum offers
a way to test whether the detected signal is reasonable. One can
search for any significant cross-correlation signal if a different
wavelength other
than \lya is used, which would indicate either contributions from
other emission lines, or that other effects are causing the cross-correlation
signal meaasured. An equivalent approach is
to extend the line-of-sight range of the cross-correlation 
to large distances.
Figure \ref{largerange}, extends the contours shown
in Figure \ref{sigmapi} to much larger scales.
The positive signal seen in Figure \ref{sigmapi} is the
most significant feature, centered at $r_{\parallel}=0, r_{\perp}=0$. This
is a good indication that \lya emission is the dominant contribution to
our signal. Signal from lines at longer (shorter) wavelengths 
than \lya would appear at positive (negative) values of $r_{\parallel}$.

The second most prominent 
feature, at $r_{\perp}=0, r_{\parallel}\sim 60 \hmpc$,
is significant at the $\sim 1.5 \sigma$ level (the pixel at the center
of the feature is 1.5 $\sigma$ from the zero level, using the 
Jackknife error bars from Section \ref{zdistort}), and so
is consistent with noise. Strong lines that might
be an issue, such as Lyman-$\beta$ and Carbon-IV, are
very far away (at $r_{\parallel}=-490$ and +640  $\hmpc$ respectively)
and so are not a concern. Si-III would appear at $r_{\parallel}=-22 \hmpc$
if it was present.

\subsection{Evolution with redshift}

The redshift coverage of our data sample ($z=2-3.5$) is sufficient that we
can separate it into different bins in redshift and search for evolution.
We do this for four different redshift bins where each bin contains one
quarter of the quasar data. The bin boundaries and mean
quasar redshifts of each bin are given in Table \ref{ztable}. 

\begin{table}
\caption{The amplitude parameter \ampqa\ for different redshift bins}
\label{ztable}
\centering
\begin{tabular}{| c | c | c |c|}
\hline
$\langle z \rangle$ & $z_{\rm min}$ &  $z_{\rm max}$ & \ampqa\ (using $\Omega_{\rm m}=0.30$) \\
& & & ( $10^{-20}$ erg s$^{-1}$ cm$^{-2}$ \AA $^{-1}$ arcsec$^{-2}$)\\ 
\hline
2.20 & 2.0 & 2.29 &  $5.0^{+1.1}_{-1.2}$\\ 
2.37 & 2.29 & 2.46 & $4.0^{+0.8}_{-0.9}$\\ 
2.59 & 2.46 & 2.75 & $3.4^{+0.8}_{-0.9}$\\ 
3.04 & 2.75 & 3.50 & $2.0^{+0.8}_{-0.8}$\\ 
\hline
\end{tabular}
\end{table}

\begin{figure*}
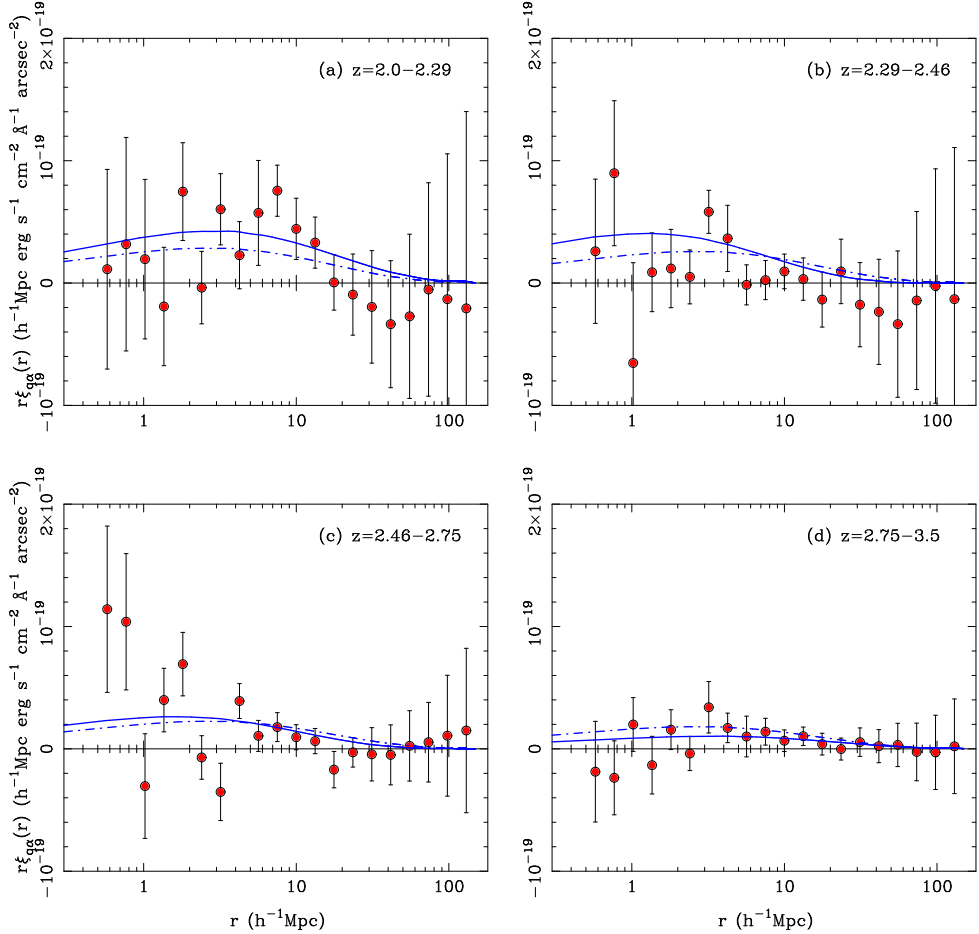
 
\begin{center}
\begin{minipage}{6.5in}
\begin{center}
\includegraphics[width=2.427in,angle=-90]{cdm_z1.ps}
\includegraphics[width=2.427in,angle=-90]{cdm_z2.ps}
\end{center}
\end{minipage}
\begin{minipage}{6.5in}
\begin{center}
\includegraphics[width=2.427in,angle=-90]{cdm_z3.ps}
\includegraphics[width=2.427in,angle=-90]{cdm_z4.ps}
\end{center}
\end{minipage}

\end{center}
\caption{The quasar-\lya emission cross-correlation function
\xiqar\ for
different redshift ranges, which 
correspond to one quarter of the full dataset each.
The solid lines indicate the best fitting CDM model
at each redshift bin and the dash-dot line represents the best fit to the full
dataset (shown in Figure \ref{gacont}). Both the shape and amplitude of the 
model fit are consistent at the $2 \sigma$ level with no evolution over
the redshift range (this is shown in Figures \ref{azbins} and \ref{gzbins}).
\label{cdmz}}
\end{figure*}

  The quasar-\lya emission
cross-correlation results are shown in the four panels of Figure \ref{cdmz}.
The global CDM model fit to the full sample averaged over all redshifts
is indicated as the dash-dot line in every panel.
Although the results for the redshift bins are relatively noisy, as
expected, they are broadly similar to the global result in shape and
amplitude and show no clear evidence for any redshift evolution.

\begin{figure*} 
\begin{center}
\begin{minipage}{6.5in}
\begin{center}
\includegraphics[width=2.427in,angle=-90]{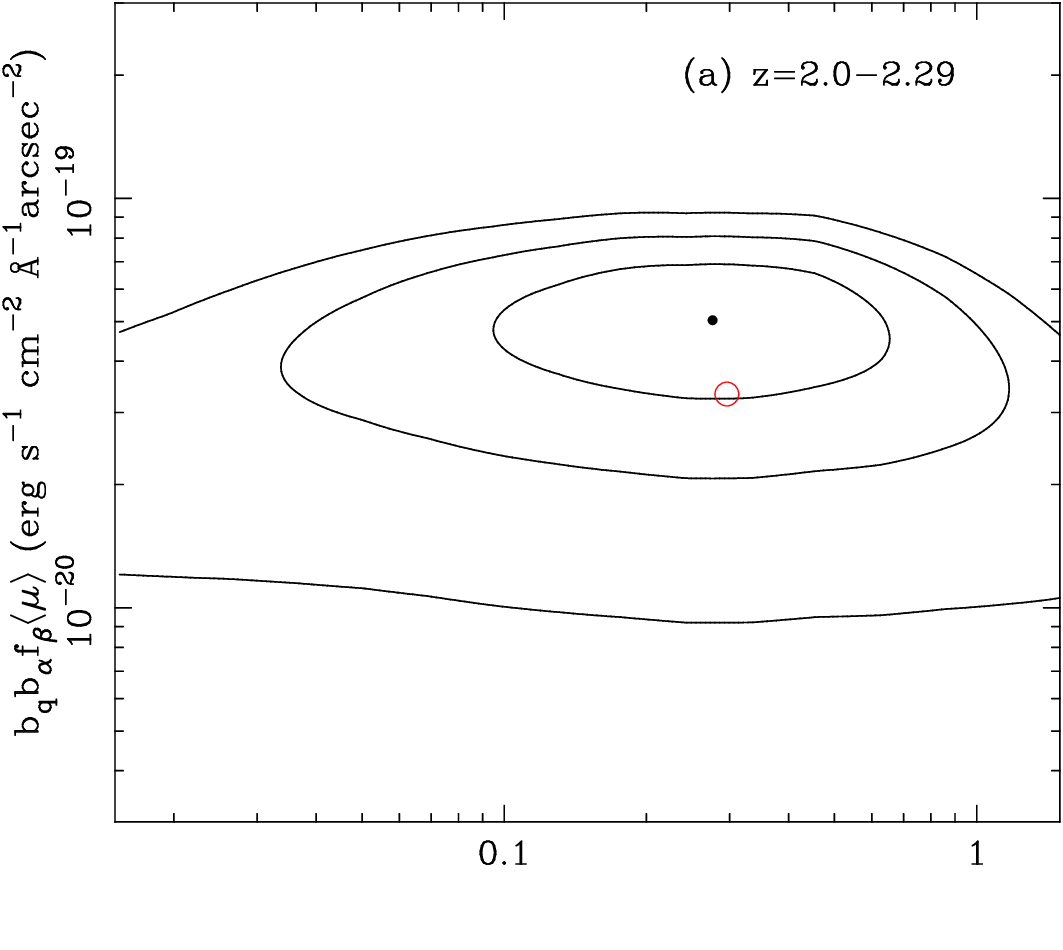}
\includegraphics[width=2.427in,angle=-90]{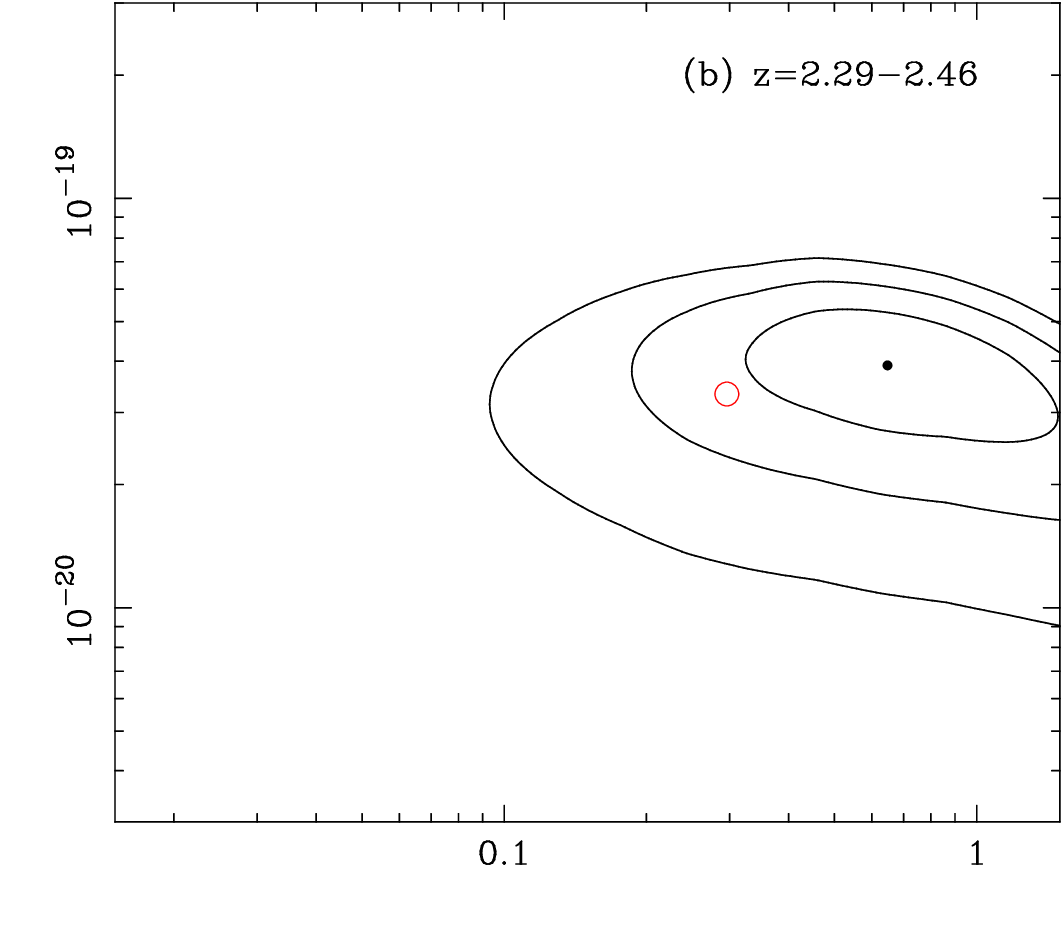}
\end{center}
\end{minipage}
\begin{minipage}{6.5in}
\begin{center}
\includegraphics[width=2.427in,angle=-90]{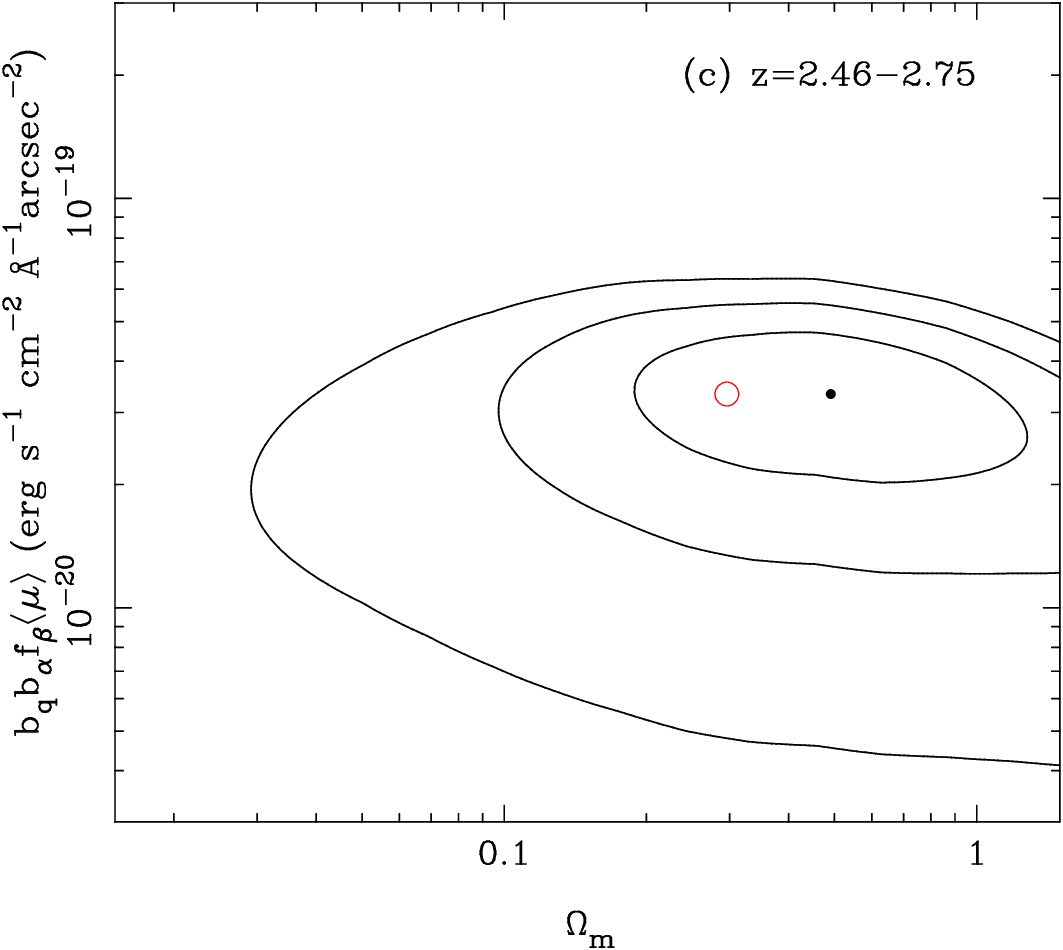}
\includegraphics[width=2.427in,angle=-90]{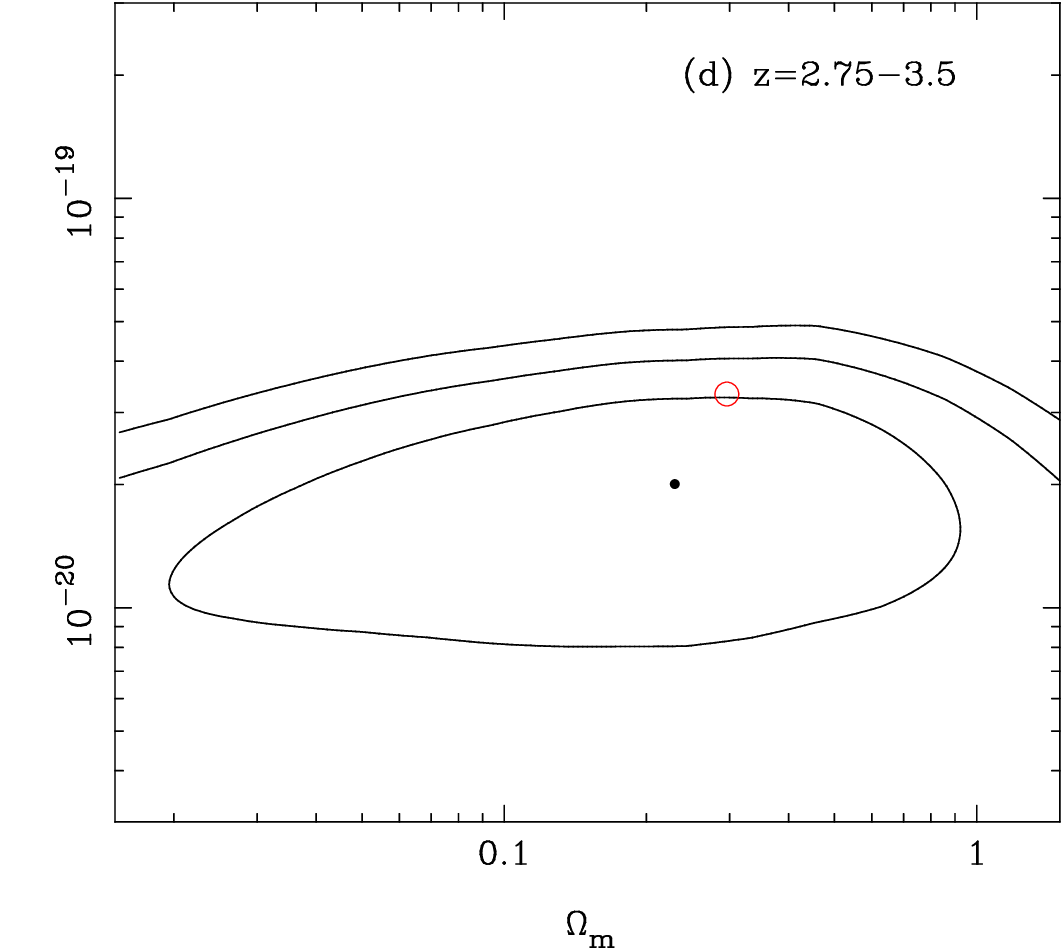}
\end{center}
\end{minipage}

\end{center}
\caption{Likelihood contours for power spectrum parameters
$\Omega_{\rm m}$ (which we are using to
parametrize the CDM shape) and \ampqa\ (the 
amplitude) for different redshift ranges. The filled dot
shows the best fitting values for that redshift range and
the open circles the best fitting values for the full
dataset (shown in Figure \ref{gacont}).
\label{gaz}}
\end{figure*}

  Figure \ref{gaz} presents the best fit CDM shape and amplitude
parameters from Section \ref{modfit}, along with confidence contours.
The fitting was again performed using jackknife error bars.
Figure \ref{gaz} demonstrates that
 the fiducial model results all lie within the
$2 \sigma$ confidence contours for the different redshift bins.

\begin{figure}
\centerline{
\psfig{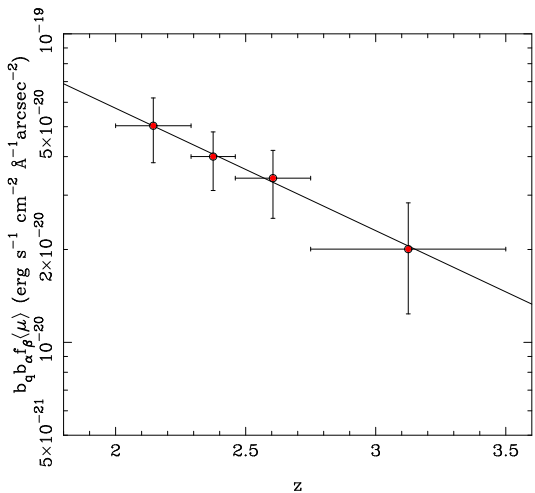}
}
\caption{ 
The evolution of \ampqa\ with redshift.
The results from Figure \ref{cdmz} are used, but here we have fixed the 
CDM shape to one for a $\Lambda$CDM model with $\Omega_{\rm m}=0.30$
(see Section \ref{modfit}) .
The error bars are 
$1 \sigma$
and are estimated using the maximum likelihood fit to the amplitude
of the CDM model. The solid line is a log-linear fit to the data (see text).
\label{azbins}}
\end{figure}

  Assuming that the shape of the cross-correlation function
remains fixed in comoving coordinates (as it would
do if governed by linear biasing), we can search for 
changes in the amplitude of clustering and the mean \lya surface
brightness as a function of redshift. We set $\Omega_{\rm m}=0.30$
(the CDM shape determined by the Planck results; Ade \etal 2014)
and then determine the best fit amplitude parameters, \ampqa,
at each redshift bin. The result, shown in Figure \ref{azbins}, indicates
a decreasing cross-correlation amplitude with redshift, although the
errors are large and a horizontal line would not be an
unreasonable fit ``by eye''. To express this quantitively,
we have carried out a $\chi^{2}$ fit to the function
$\log\ $(\ampqa)$=a+bz$, finding a slope parameter $b=-0.40\pm0.20$,
meaning that the hint of redshift evolution 
is significant at the $2.0 \sigma$ level only.
The values for the fit parameters \ampqa\ for
the different redshift bins (which were used to plot Figure \ref{gaz}
are listed in Table \ref{ztable}.

\begin{figure}
\centerline{
\psfig{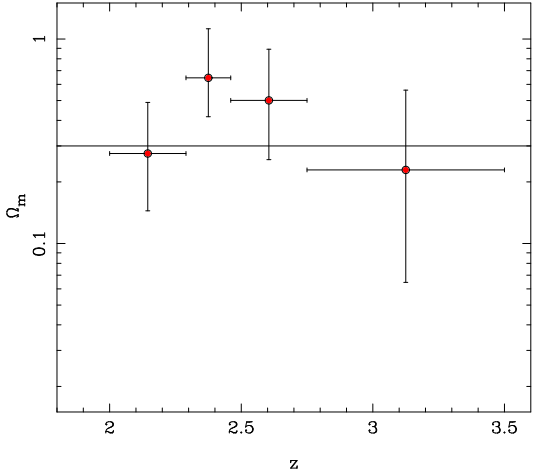}
}
\caption{ 
The value of $\Omega_{\rm m}$ (which we are
using to parametrize the shape of the $\Lambda$CDM correlation function,
holding other parameters fixed) vs redshift.
The results from Figure \ref{cdmz} were used.
The solid line is the best fit to the Planck results (Ade \etal 2014),
$\Omega_{\rm m}=0.30$.
\label{gzbins}}
\end{figure}

We have also looked at redshift evolution of the CDM shape
governed by the parameter $\Omega_{\rm m}$. Within the assumption of 
linear biasing, the shape should remain
constant with redshift. The results are examined in Figure \ref{gzbins},
where the results are indeed consistent with a constant $\Omega_{\rm m}$, 
within the uncertainties.
The Planck value ($\Omega_{\rm m}=0.30$) is also shown and is 
consistent with our results.

\section{Emission lines}
\label{elines}

  Given that the sources of the clustered \lya emission seen in Section 3
could be discrete objects such as star forming galaxies, 
we must investigate see if individual emission lines can be detected in 
our spectra and whether discrete detectable lines can account for the signal.
The BOSS spectra have relatively short integrations on a small (2.5 m)
telescope,
and cannot be expected to compete in individual detections with other
surveys such as that described in Cassata \etal (2011): we can only
detect the most luminous objects. However, our
cross-correlation technique enables us to find the mean total surface
brightness, which includes all emission-line objects no matter how faint
they may be. The difference between the cross-correlation signal 
with and without individually identified lines
therefore enables us to discover what fraction of the \lya surface brightness
lies below our line detection limit.

We note that our line detection procedure is less
sophisticated than that in the BELLS survey (Brownstein \etal 2012), which 
used line detections of galaxies behind LRGs to find gravitational
lenses. In particular, we are not seeking
confirmed detections of objects (which requires multiple emission lines)
and we do not deal with interlopers, except statistically.

\subsection{Line fitting}

  For each LRG spectrum (see  Section 2.1), we subtract the best fit
galaxy spectrum model, as we do in our fiducial
cross-correlation analysis. We then fit lines to this residual 
flux, centering our fitted line profile on the center of each spectrum
pixel, one at a time. 
In this first stage, each spectrum pixel is
therefore the center of a best fit line profile - 
we remove overlapping lines later.
For each of a grid of values of amplitude
$A$ and line width $\sigma$ (between 0 and 20 \AA ) we
compute the $\chi^{2}$ value of a positive
Gaussian emission line profile $G(x)$, where
the profile has the form $G(x)=A \exp -(x^{2}/2\sigma^{2})$ and
$x$ is the separation between the line  center and the pixels we
include in our fit. We use pixels in our fit that are in a 40 \AA\ region 
centered on the line center, excluding
masked regions as in Section 2.1.1. Once the best fitting values of
$A$ and $\sigma$ are found,  we estimate
the significance of each fitted line
from the $\chi^{2}$ difference between the line fit and a
flat interval with zero flux.
After 
fitting to all the pixels we pass through the list and eliminate overlapping
lines, removing the lower significance line when there is an overlap.

We find 5200 lines with a nominal significance
of $5 \sigma$ 
($\Delta \chi^{2} > 30.1$ for the 2 degrees of freedom fitted), 
and $1.6\times10^{6}$ lines with a nominal significance 
of $3 \sigma$ ($\Delta \chi^{2} > 11.8$)
There are $1.3\times10^{9}$ pixels in the search regions of 
the spectra. The detected lines are constrained
to be at least 40 \AA\ (37 pixels) apart,
but for rare lines this should not change the random expectation.
One would therefore expect to find approximately 370 and 
$1.8\times10^{6}$ $5\sigma$ and $3\sigma$ lines
respectively from positive noise fluctuations alone.

For the $3\sigma$ lines, the fact that we have detected fewer lines
than even pure noise fluctuations predict is likely to be a sign that the
fluctuations do not exactly obey a Gaussian noise model.
An additional complication is that the noise estimate from the standard
data pipeline which we have used has been shown to be underestimated by
up to $16\%$ for the relevant wavelengths
(Palanque-Delabrouille \etal 2013).
The detection of more $5\sigma$ lines than randomly expected is likely
to indicate that there are 
false detections arising from unsubtracted features in the galaxy spectra,
sky lines we have not accounted for, and other systematics.
There is also the possibility of interloping [OII] emission
lines from lower redshift galaxies (see Noterdaeme et al. 2010,
 Menard et al. 2011). For our wavelength coverage of 3800-5500 \AA,
this interloper emission would arise from between $z=0.02$ and $z=0.48$.
Without a significant additional
effort to remove false detections
and interlopers, our dataset is not useful for
computing the luminosity function of \lya lines. Instead, we turn 
to the statistical cross-correlation to test for the fraction of
these lines which are really \lya emission lines in our redshift interval.

\subsection{Cross-correlation}
\label{linecrcorr}

\begin{figure}
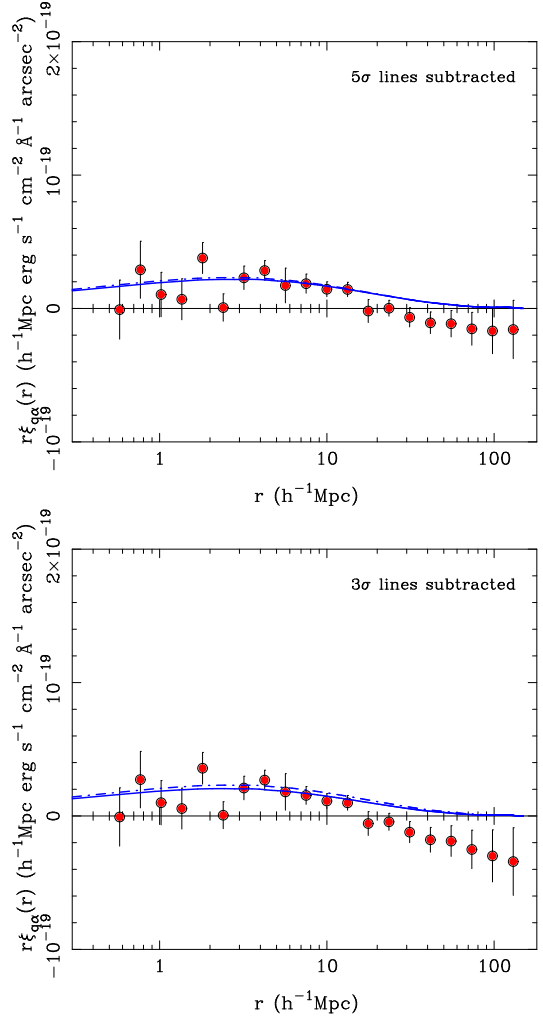

\centerline{
\psfig{file=cdm_5sig.ps,angle=-90.,width=7.0truecm}
}
\centerline{
\psfig{file=cdm_3sig.ps,angle=-90.,width=7.0truecm}
}
\caption{ 
The quasar-\lya emission cross-correlation function \xiqar\
(see Equation \ref{xieq}), as in Figure \ref{cdmfid}, but computed 
after subtracting emission lines that are apparently detected in the
spectra at the $5 \sigma$ significance level in panel (a), and the
$3 \sigma$ level in panel (b).
The smooth curve is the best fit linear CDM
correlation function (see Section \ref{modfit}) and the dash-dotted line
is the best fit CDM curve for the fiducial sample (i.e.,
before subtracting the apparent emission lines). }
\label{cdmsig}
\end{figure}

We subtract the flux in the lines detected in Section 4.1
from each LRG spectrum,
and then recalculate the cross-correlation of quasars and \lya emission
(Equation \ref{xieq}). The results for \xiqar\ (again computing the error
bars using a jackknife estimator and 100 subsamples) are shown in Figure 
\ref{cdmsig},  using our two thresholds on the significance
of the removed lines, $5 \sigma$ and $3 \sigma$.
We can see that in each case, the clustering signal is
still visible and the shape traces that of a CDM curve. This shows
that most of the surface brightness of \lya emission is not
accounted for by these lines. As expected, most of the lines are due to 
noise features. By subtracting these lines, however, we are also subtracting
any possible real lines, and the change in the amplitude of \xiqar\ is a
measure of the fraction of surface brightness that is actually contributed by
strong lines.

\begin{figure}
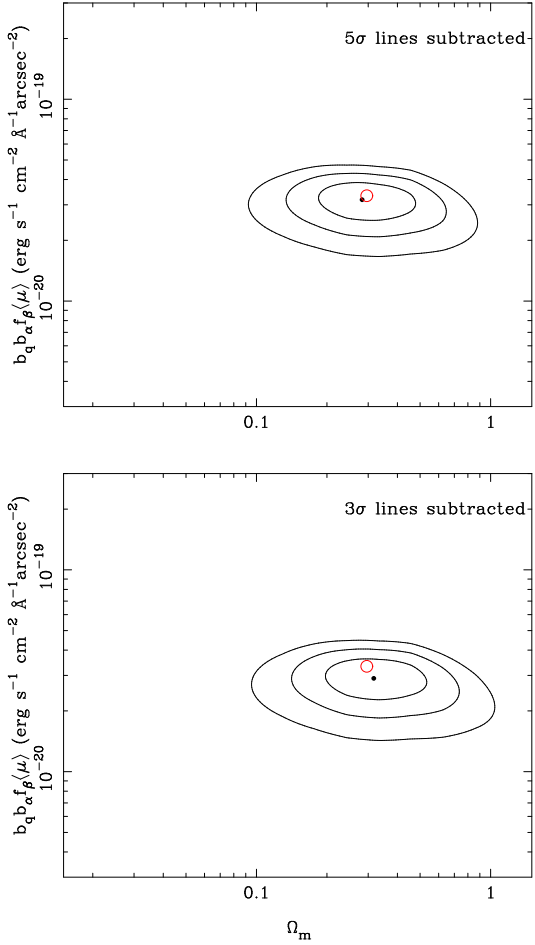

\centerline{
\psfig{file=gamma_amp5sig.ps,angle=-90.,width=7.0truecm}
}
\centerline{
\psfig{file=gamma_amp3sig.ps,angle=-90.,width=7.0truecm}
}
\caption{The effect of removing strong lines on the
shape and amplitude of the quasar-\lya cross-correlation.
We show the fit parameters for the amplitude
\bqla\ $\langle \mu_{Ly\alpha}  \rangle$ and shape $\Omega_{\rm m}$
of a linearly biased CDM model fit to the \lya cross-correlation
function plotted in Figure \ref{cdmsig}. Panel (a) is after removing
$> 5 \sigma$ significance lines and panel (b) 
$> 3 \sigma$ significance lines. The dots indicate the best fit
parameters and the contours show the 1,2 and 3 $\sigma$ confidence contours
on the fit parameters. The open circles show the best fit results
to the fiducial results (from Figure \ref{gacont}).
\label{gasig}}
\end{figure}

The shape and amplitude fitting parameters ($\Omega_{\rm m}$ and \ampqa\ )
for these two cases
($>5 \sigma$ and $>3\sigma$ lines subtracted) are shown in Figure \ref{gasig}.
The shape parameter $\Omega_{\rm m}$ is very similar in the 
two cases and almost
the same as in the fiducial case. The amplitude is lower, as would be 
expected for subtraction of some real lines, but the fiducial result
(with no line subtraction) lies well within the 1 $\sigma$ error contour
of both of the panels in Figure \ref{gasig}, implying that the
contribution to the cross-correlation from emission lines that are
detected and removed is not statistically significant. Quantitatively, this
can be seen by considering that we find the amplitude parameter
for the $>5 \sigma$ case to be
\ampqa\ $= 3.18^{+0.39}_{-0.41} \times 10^{-20} \ergs\cm^{-2}$ \AA $^{-1}$ arcsec$^{-2}$,
and for the $3\sigma$ case to be
\ampqa\ $= 2.89^{+0.43}_{-0.37} \times10^{-20} \ergs\cm^{-2}$ \AA $^{-1}$ arcsec$^{-2}$.
The amplitude parameter is therefore $4\pm12\%$ and 
$13^{+11}_{-13} \%$ lower than the
fiducial case for the $> 5 \sigma$ and $> 3 \sigma$ line removal cases,
but both of these are consistent with zero within the errors.
The analysis is therefore consistent with our line fitting having
found no true \lya\ emission lines at all. 

 In order to relate the significance levels
to line luminosity, we have computed the luminosity
from the surface brightness for each line (bearing in mind that our 
measurements are restricted to a 1 arcsec radius fiber aperture).
We find that the median luminosity of the $>5 \sigma$ lines is 
$L = 9.0\times10^{42}\ergs$ and the $>3 \sigma$ lines
have a median luminosity $L = 1.9\times 10^{42}\ergs$. 
We can compare these luminosities measured  with 
some published values from \lya emitter surveys.
The flux limit of the Guaita \etal (2010) data sample
was $2\times10^{-17} \ergs\cm^{-2}$  (emission line
flux) at z=2.1. This corresponds to a \lya luminosity of $5\times10^{41}\ergs$.
For the Gawiser \etal (2007) sample at z=3.1, 
the line flux limit was $1.5\times 10^{-17}\ergs\cm^{-2}$, corresponding to a
line luminosity of $1.3\times 10^{42}\ergs$.

In our calculation above we have seen that at the 1 $\sigma$
confidence level, $13\%$ of
the \lya cross-correlation signal in our dataset
could be due to lines with a median luminosity 
(in a 1 arcsec radius
aperture) $1.9\times10^{42}$ ergs$^{-1}$.
This is similar to the values of Gawiser \etal 
and Guaita \etal, although given our small aperture, the intrinsic 
luminosity of our fitted line emitters will be even higher.
A small fraction of the  \ampqa\ value we are seeing 
could therefore 
be contributed by emitters similar to those in these two surveys.
As noted above,  the error bar on this fraction
is large, and our result is consistent with zero contribution from such lines.

 There are various possibilities for the nature of the major contributor to
the \lya\ cross-correlation signal. The first is fainter lines than
those seen in \lya\ emitter surveys. This is unlikely because
$b_{\alpha}$ is luminosity weighted, so low
luminosity lines contribute much less to the signal than higher
luminosity lines.
The second is  high luminosity, low surface brightness emission. This
emission is much more difficult to detect, and cannot be seen by
our line search algorithm, which is sensitive to high surface
brightness, narrow lines.
It is also  unlikely to have been seen in previous surveys (we return to this
in Section \ref{disc}). 
This type of emission would also be highly biased, and so if present 
would contribute strongly to the \ampqa\ measurement. 
We discuss the various
possibilities in more detail in Section \ref{disc}.

\section{Star formation rate density}
\label{sfrd}

If the \lya surface brightness we are seeing is produced by 
star forming galaxies, we can convert the mean \lya surface brightness
into a star formation rate density (SFRD) at the mean redshift $z=2.55$
of our observations. Traditionally (e.g., Gronwall \etal 2007) 
narrow band surveys have
been used to detect \lya emitters, compute their luminosity function,
and integrate it to compute a mean \lya luminosity density 
$\epsilon_{\alpha}$ before converting it into a star formation rate,
using a relationship such as:
\begin{equation}
{\rm SFR}/(\msun\,  {\rm yr}^{-1})=L_{\alpha}/(1.1\times10^{42} \ergs )
\label{sflya}
\end{equation}
(Cassata et al. 2011), where $L_{\alpha}$ is the \lya luminosity. The 
conversion factor is based on a stellar population with a Salpeter IMF and
with no correction for effects like dust and escape fraction, which is 
accurate to within a factor of a few for a range of population age, high
mass cutoff of stars, and metallicity \citep{Leitherer99}.

This method assumes that the surveys of \lya emitters are able to capture all
the radiation from young stars. However, these surveys can only detect
the high surface brightness portion of sources within a small angular
aperture, and may be missing much of the \lya line intensity when it is
scattered far out into the galaxy halo. In our case, the statistical
cross-correlation technique we are using should not be affected by any
threshold in \lya surface brightness.
We should therefore be able to compute  the total star formation rate
density from our measurement. One large uncertainty is absorption due to dust,
which is known to significantly affect UV continuum and line 
estimators of star formation. 

\begin{figure}
\centerline{
\psfig{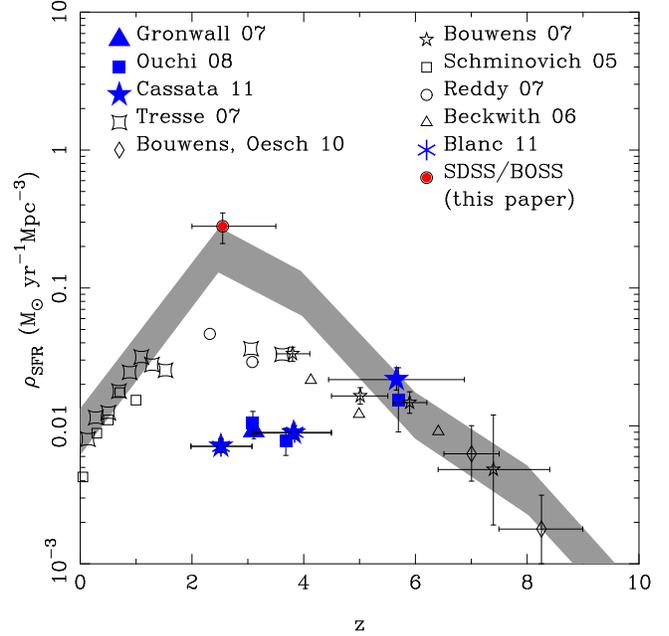}
}
\caption{ 
The star formation rate density (${\rho}_{\rm SFR}$) inferred from 
our measurement
of the mean \lya surface brightness in the Universe between 
$z=2-3.5$ (see Section \ref{sfrd}) is shown as the red point with solid
line error bars, assuming that the linear bias factor
for \lya emission is $b_{\alpha}=3$, a reasonable value for the 
luminosity-weighted clustering of  star forming
forming galaxies (see Section \ref{sfrd}).
The true value of $b_{\alpha}$ is unknown, so this
data point should be scaled by $3/b_{\alpha}$.
Other data values plotted with open (black) symbols 
are from published ${\rho}_{\rm SFR}$ values which
used UV estimators. The solid (blue) 
points show estimates of ${\rho}_{\rm SFR}$
computed from the luminosity functions of surveys
for \lya  emitters. The references are given in Section \ref{sfrd}.
The shaded area represents the range of dust corrected UV estimates
compiled by Bouwens \etal (2010).
\label{sfr}}
\end{figure}

  Recall that our measurement is of the quantity \ampqa, so
to compute the \lya surface brightness
we need to have independent knowledge of $b_{\rm q}$, $b_{\alpha}$
and $f_{\beta}$. For the quasar bias factor
we use the value measured for BOSS quasars by
\citet{Font-Ribera13}, 
$b_{q}=3.64^{+0.13}_{-0.15}$.

  We recall that the bias factor $b_{\alpha}$ is related to a
luminosity-weighted bias factor $b_L$, from the definitions in 
Equations~\ref{blya}--\ref{blum}
($b_\alpha=b_L$ in the absence of radiative transfer effect), 
and $b_L$ is different from
the number weighted bias factor of \lya emitters, $b_{\rm LAE}$.
To understand the difference in the values of the two bias factors,
we start with the following simple model for the \lya emission. 

If we assume that there are $\langle N(M)\rangle$ galaxies per dark matter 
halo of mass $M$ and that 
\lya emission comes from 
galaxies in halos above a mass limit $M_{\rm min}$,
then the spatial bias can be computed as follows \citep[e.g.][]{Berlind02}:
\begin{equation}
b_{\rm LAE}=
\int_{M_{\rm min}}^\infty b_{\rm h}(M) \langle N(M)\rangle \frac{dn}{dM} dM  
\left/ \int_{M_{\rm min}}^{\infty} \langle N(M)\rangle \frac{dn}{dM} dM \right.,
\label{blaemod}
\end{equation}
where $b_{\rm h}(M)$ is the bias factor for halos of mass $M$ and 
$dn/dM$ is the halo mass function. 
For the luminosity weighted bias factor, we have
\begin{equation}
b_L=\int_{M_{\rm min}}^{\infty} b_{h}(M) L(M) \frac{dn}{dM} dM 
\left/ \int_{M_{\rm min}}^{\infty} L(M) \frac{dn}{dM} dM \right.,
\label{bamod}
\end{equation}
where $L(M)$ is the average \lya luminosity in halos of mass $M$.

Observationally, a value of 
$b_{\rm LAE}=1.75 \pm0.23$ results from an average of the values
measured by Gawiser \etal (2007) and Guaita \etal (2010),
who find  $b_{\rm LAE}=1.7^{+0.3}_{-0.4}$ at $z=3.1$
and $b_{\rm LAE}=1.8 \pm0.3$ at $z=2.1$, respectively.
Using Equation~\ref{blaemod}, we find that $b_{\rm LAE}=1.75$
corresponds to $M_{\rm min}=10^{11} \msun$. In the calculation, 
we assume one galaxy per halo ($\langle N(M)\rangle=1$), which overestimates
$M_{\rm min}$ by only a small factor (see the Appendix of \citealt{Zheng07}).
The mass estimation is  consistent with, e.g., the analysis of 
\citet{Gawiser07}.

  To estimate the luminosity-weighted bias factor $b_L$, we need to know the
relation between luminosity and halo mass. Assuming
that $L(M) \propto M^{p}$, for $M_{\rm min}=10^{11} \msun$ we obtain $b_L=$
2.38, 4.40, and 6.84 for the values $p=1, 2,$ and 3, respectively. 
The differences between $b_{\rm LAE}$ and $b_L$
are therefore substantial, the latter being usually much larger.
To proceed further, we need to consider the likely relation between $L$ and 
$M$. The \lya luminosity is related to the star formation rate in galaxies.
Along the star-forming sequence, the star formation rate is inferred to be 
approximately proportional to the stellar mass 
\citep[e.g.][]{Daddi07,Pannella09,Lee11}.
We therefore first make use of the relationship between 
halo mass and stellar mass found with abundance matching by \citet{Moster10}.
It is in the form of a softened broken power law, and at $z\sim 2.55$ the
low-mass-end (high-mass-end) slope of the $L$--$M$ relation is $\sim 2.5$ ($\sim 0.6$) with a transition mass around $10^{12}\msun$. Using this relation in
Equation \ref{bamod} yields $b_L=2.82$ for $M_{\rm min}=10^{11} \msun$. In 
fact, the result is insensitive to $M_{\rm min}$ ($b_L=2.81$ for 
$M_{\rm min}=10^9 \msun$), given the steep $L$--$M$ relation (so low-mass 
halos are weighted less). If we modify the high-mass-end slope to $\sim 1$ to
approximately account for the luminosity contribution from satellite galaxies, 
we obtain $b_L=3.15$.

 There are various uncertainties involved in deriving $b_L$ with this simple
model. 
First, the slope of the star formation rate versus the stellar mass relation
for the star-forming galaxy sequence can be slightly different from unity. We
find that a 10\% deviation from unity in the above slope leads to a $\sim 5\%$
change in the value of $b_L$. Second, we assume that the \lya luminosity 
is proportional to the star formation rate. The way these two quantities 
track each other may vary as a function of star formation rate if, for example,
the escape fraction of \lya photons varies.
Another possibility is that a large fraction of the \lya 
emission comes from previously undetected sources or low surface brightness 
halos around galaxies. These factors will change the $L$--$M$ relation and
therefore the derived value of $b_L$. Even with the above uncertainties, it is
likely that $b_L$ is around 3.

\begin{figure}
\centerline{
\psfig{file=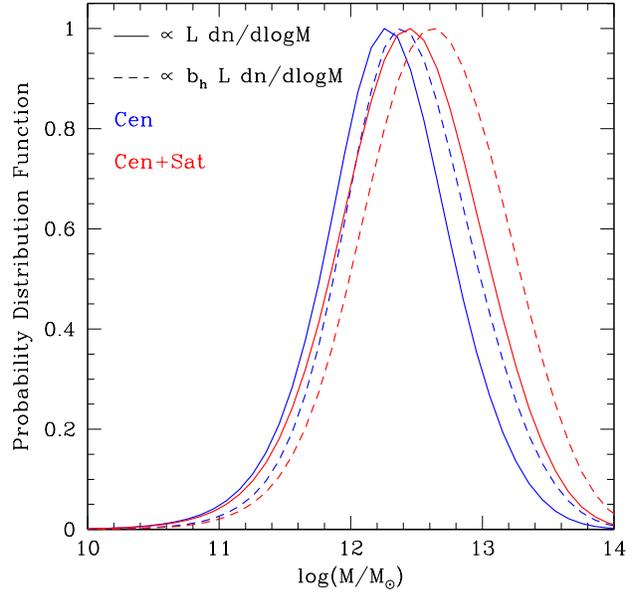,angle=0.,width=8.5truecm}
}
\caption{
\label{fig:MhPDF}
Probability distribution functions (PDFs) of \lya luminosity density (solid
curves) and halo-bias-weighted \lya luminosity density (dashed curves) for
our fiducial model. 
The PDFs are computed as proportional to the average \lya luminosity or 
halo-bias-weighted \lya luminosity in halos of mass $M$ multiplied by the
differential halo mass function $dn/d\log M$. The blue curves use only the 
\lya luminosity of central galaxies, and the red curves include contributions
from satellite galaxies. All curves have been normalized
to unity at their respective maxima. See the text for further details
concerning the model. 
}
\end{figure}

With the above model, we can compute the contribution to the \lya luminosity
density from halos of different masses, which is simply proportional to the 
average \lya luminosity in halos of mass $M$ times the differential halo mass
function $dn/d\log M$. The solid blue curve in Figure~\ref{fig:MhPDF} uses 
central galaxy \lya luminosity only, which peaks around 
$\log (M/M_\odot)=12.25$. Including the contribution from the satellite 
galaxies shifts the peak slightly to a higher mass, around 
$\log (M/M_\odot)=12.45$, as shown by the solid red curve. The curve gives
the probability density of a random \lya photon to come from a halo of 
mass $M$. The \lya emission detected through the
cross-correlation technique probes the halo-bias-weighted luminosity density
distribution. The dashed curves in Figure~\ref{fig:MhPDF} show these 
probability distributions. With the satellite contribution included, the
curve peaks around $\log (M/M_\odot)=12.6$. Taking the full width at half 
maximum of the curve, the fiducial model implies that the signal in the 
cross-correlation should mainly come from \lya emission in halos of mass 
(1--20)$\times 10^{12}M_\odot$.

Finally, as mentioned in Section~\ref{modfit}, $b_L$ is not the same as 
$b_\alpha$ once the radiative transfer effect is taken into account. 
A simple model shows that $b_\alpha=b_L+\alpha_1$ with $\alpha_1$ a 
positive number (see Section~\ref{sec:RT}). The value of $\alpha_1$ is not
readily known without detailed radiative transfer modeling. Overall, we 
expect $b_\alpha$ to be larger than $\sim$3. We choose to parameterize 
derived quantities in terms of $(3/b_{\alpha})$.

An additional uncertain factor to consider is the modification to
clustering caused by
redshift-space distortions. This is embodied in the $f_{\beta}$
parameter. We have seen 
in Section \ref{zdistort} that measurements of anisotropies
in clustering give a measurement of 
$\beta_{\alpha}=\Omega_{\rm m}^{0.6}(z=2.55)/b_{\alpha}=-0.76\pm0.36$.
This negative value of $\beta_{\alpha}$ is of the form
expected to be caused by radiative transfer effects on clustering
\citep{Zheng11a}
and is opposite in sign to the usual \citet{Kaiser87} peculiar velocity 
redshift-space distortions. Neverthless, this redshift-space distortion
model was shown in Section \ref{zdistort} to give a reasonable fit to the data
and one can use this to compute the factor $f_{\beta}$ as
$f_{\beta}=1+\frac{1}{3}(\beta_{\rm q}+\beta_{\alpha})+ \frac{1}{5}
(\beta_{\rm q}\beta_{\alpha})$ (from equation [\ref{xi0}]).
If we do this we find that the value of $f_{\beta}=0.80\pm0.15$,
which
we take as a reasonable estimate of the reduction of the monopole term
due the spreading the correlation along the line of sight. 
Even though gravitational evolution is not the physically correct model for 
interpreting our observations owing to the negative value we obtain for 
$\beta_\alpha$, a model with redshift-space distortion plus radiative transfer 
effect does seem to work reasonably well here (see Section~\ref{sec:RT}).

We use this value and propagate the errors
from the bias measurements and our measurement 
of \ampqa\ (for fixed shape
parametrized by $\Omega_{\rm m}=0.30$), 
to compute the mean \lya surface brightness
at $z=2.55$, finding 
\begin{equation}
\langle \mu_{\alpha} \rangle=(3.9 \pm 0.9) \times10^{-21}
(3/b_{\alpha}) \ergs\cm^{-2} \angs^{-1}\asec^{-2} ~. 
\label{eq:backi}
\end{equation}

  We convert this into a comoving
\lya  luminosity density $\epsilon_{\alpha}$ using
\begin{equation}
\epsilon_{\alpha}=4\pi \langle\mu_\alpha \rangle 
                  \frac{H(z)}{c}\lambda_{\alpha}(1+z)^{2},
\label{eq:backe}
\end{equation}
where $c$ is the speed of light and $\lambda_{\alpha}=1216 \angs $.
We find the value $\epsilon_{\alpha}=3.1\times10^{41}  (3/b_{\alpha})
\ergs \mpc^{-3}$.
We then use Equation \ref{sflya} to convert this into a measurement
of the star formation rate density 
${\rho}_{\rm SFR}(z=2.55)=(0.28\pm0.07) (3/b_{\alpha}) \msun {\rm yr}^{-1} \mpc^{-3}$. As mentioned before, the conversion depends on the assumption 
about the underlying stellar population. A younger population and lower 
metallicity would lead to a lower inferred SFR, which could be an important 
effect for interpreting our results. Keeping this possibility in mind, 
we proceed with the discussion by using the above result from the commonly 
adopted conversion factor.

  We plot this result in Figure \ref{sfr} as the red point, for
the chosen value of $b_{\alpha}=3$. We note that the true value for
the parameter
$b_{\alpha}$ is not well determined, and that the ${\rho}_{\rm SFR}$ datapoint 
should be scaled up and down by the factor $3/b_{\alpha}$. Our discussion
above suggests that $b_{\alpha}$ is likely to be larger than 3,
so that it is more probable for $\rho_{\rm SFR}$
to be scaled downwards than upwards.
Figure \ref{sfr} also shows various estimates of
${\rho}_{\rm SFR}$ from \lya emitter surveys as well as UV continuum
estimates of the star formation rate. We can see that our measurement
is about 30 times  higher (for $b_{\alpha}=3$) than the \lya emitter based 
measurements of Gronwall \etal (2007), Ouchi \etal (2008) 
or Cassata \etal (2011). Note that these \lya emitter based measurements 
result from a direct integration over the observed \lya luminosity function
without corrections for any possible dust effect.

A complication which adds substantial uncertainty is dust absorption, which
may affect \lya and continuum radiation differently. One estimator
of the level of dust extinction in the continuum is the rest-frame UV
continuum slope, $\beta$, which specifies how the flux density of a 
galaxy varies with wavelength (i.e., $f_{\lambda}\propto \lambda^{\beta}$)
in the UV continuum region ($\sim 1300 \AA$ to $\sim 3500 \AA$) of its
spectrum. If an intrinsic $\beta$-dust extinction relationship is assumed,
(usually that measured empirically from $z\sim 0$ galaxies by 
Meurer \etal 1999), one can use observations of $\beta$ for high redshift
galaxies to infer a dust-corrected UV luminosity and then 
star formation rate. This has been done by several
authors, including Bouwens \etal (2009,2010). In Figure \ref{sfr} we show
as a blue band the compilation of dust-corrected UV star formation densities
from Bouwens \etal (2010) computed using this technique. According to 
Bouwens \etal  (2010), the dust correction for a limiting
luminosity of 0.3 $L^{*}_{z=3}$ is $6.0 \pm 2.5$ at redshift $z=2.5$.
Support for the validity of these corrections comes from the agreement
of dust-corrected UV values with ${\rho}_{\rm SFR}$ estimated from infrared
observations (see the recent review by Madau \& Dickinson 2014).

 As for the effect of dust on the \lya radiation, Cassata \etal (2011) 
have speculated that it could be even stronger than proposed by
Bouwens \etal  (2010), as the \lya emitter
inferred ${\rho}_{\rm SFR}$ is less than 20\% of the non-dust corrected
UV continuum value (as can be seen in Figure \ref{sfr}).
One reason which favors this interpretation is
the fact that resonantly scattered radiation does have to
cover a longer path length
than continuum radiation before it leaves a galaxy
so that it could be more vulnerable to dust
extinction. On other hand, there is one well known mechanism
(Neufeld 1991) which could preferentially enhance the escape of \lya
radiation: in a clumpy medium of dusty clouds, continuum (UV) photons are
absorbed as soon as their path crosses an optically thick dust cloud,
whereas \lya photons can bounce off the cloud surfaces and find their way
through the clouds to escape, leading to a lower extinction for \lya
than for continuum photons if the intercloud medium is sufficiently devoid
of dust. The anisotropic escape of \lya radiation \citep{Zheng14}
caused by, for example, a bipolar galactic wind, can also help make
\lya photons follow the path of lower extinction optical depth to escape,
while UV continuum photons are emitted isotropically and on average experience
more extinction. 
From Figure \ref{sfr}, it appears that some mechanism of this sort
is needed if we are to explain our results.
We discuss these issues further in Section \ref{disc}.

In conclusion, the rather surprising result seen in Figure \ref{sfr} is
that the fiducial value of the \lya surface brightness from our
measurement is consistent with all \lya photons produced in stars at $z=2.55$
escaping from their host galaxies and being detected. The dust-corrected
results of Bouwens \etal (2010) imply
${\rho}_{\rm SFR}=0.19^{+0.08}_{-0.06} \msun{\rm yr}^{-1}\mpc^{-3}$ 
at $z=2.55$, and from our measurement,  
${\rho}_{\rm SFR}=(0.28\pm0.07)(3/b_{\alpha})\msun {\rm yr}^{-1}\mpc^{-3}$.
This means that $b_\alpha > 3.0$ is needed for our measurement not to
imply detection of more \lya photons than are actually produced at more than 
the $1-\sigma$ level. Our simple model for $b_\alpha$ does satisfy this
limit.
We note that the intensity mapping technique we use in this paper will
detect \lya photons which are scattered into our sightline from arbitrarily
large distances from the emitting galaxy, and at arbitrarily low
surface brightness. One can characterise our measurement as being 
consistent with a ``total escape fraction'' of \lya photons from star-forming
galaxies of 100 \%. This total escape
fraction could be much higher from the ``detected escape fraction''
measured from traditional surveys of \lya emitters,
which have surface brightness
limitations. We also discuss this in more detail in Section \ref{disc}.

\section{Discussion}
\label{disc}

We can frame further discussion of our results in terms of the following
questions:
\begin{enumerate}
\item Can the observed \lya surface brightness be explained by known 
\lya emitters?
  We have seen in Section \ref{sfrd}
that the simple answer appears to be no.

\item Can the observed \lya surface brightness be explained by faint
\lya emitters below the threshold of published surveys?
Ouchi et al. (2008) have shown that changing the extrapolated luminosity
function faint end slope
$\alpha$ from $-1.0$ to $-2.0$ changes the total integrated \lya luminosity
density they infer by only $20 \%$. The answer to this question
is therefore also no.

\item  Can the observed \lya surface brightness
be explained by extended halos around the known \lya emitters?
We show in Section \ref{halos} below that this is not the case, and
that extended halos around known \lya emitters, while
adding to the mean surface brightness,
fall short of accounting for our results by an order of magnitude .

\item Can the observed \lya surface brightness be explained by star-forming
galaxies, when these are estimated based on the extinction-corrected
UV continuum SFR density? We have seen in Section \ref{sfrd} that
there appears to be just enough star formation per unit volume 
in the Universe at $z=2.5$ that if most of it led to escaped \lya emission,
this could explain what we are seeing. We further discuss
the implications of this below in Sections \ref{budget}, \ref{escape}
and \ref{implications}.

\item What is the contribution to our meaasurement of
sources of \lya emission beyond star-forming galaxies?
We address this in Section \ref{sources} below, showing that other 
contributions are likely to be very small.

\end{enumerate}

\subsection{Potential systematic errors in the measurement}

Our measurement of \lya intensity clustering relies on 
statistical cross-correlation techniques applied to a large
sample of spectra with relatively low signal to noise, all of which were 
targeted at bright foreground galaxies that we have removed in post-processing.
There is therefore ample scope for small instrumental or other
effects to influence the signal we measure. Bearing this in mind,
we have carried out a range
of tests, detailed in Appendices A-C, to make as certain
as possible that the signal is real. Most importantly, these include 
tests of our methods for eliminating
contaminating light from neighboring fibers,
which does have a strong effect.  We have also tested
the effect of eliminating 
quasars clustered with the quasars we are using in our cross-correlation,
and we have checked the dependence of the signal on the luminosity of the
galaxy fiber target and on the quasar luminosity. For these latter tests,
we have found no 
significant effect and therefore conclude that the signal is real,
to the extent we have been able to ascertain.

There remains the possibility that some previously unknown effect,
instrumental or otherwise, is responsible for the cross-correlation
signal. This clustering signal would have to have a shape and amplitude
consistent with $\Lambda$CDM, yield a star formation rate density
consistent with dust-corrected UV estimates, have clustering distortions
of the form predicted by \lya radiative transfer effects, and pass all
the tests mentioned above. This would constitute an unlucky
coincidence; in particular, contamination by quasar light seems very
unlikely once we have eliminated the effect of neighboring fibers in the
BOSS camera and we have tested the absence of a dependence on the quasar
measured flux. Galactic dust absorption or other systematics on the
spectral continua also cannot induce the enhanced emission that is
measured at the inferred \lya wavelength at the redshift of each
observed quasar. However, the reader should bear in mind this possibility
until future independent work is able to confirm our measurement.

We note that the effect of gravitational lensing on our results should be zero,
even though we use spectra of bright galaxies, because our
\lya measurements are of surface brightness, which is
conserved under lensing. If on the other hand
we were detecting \lya\ emitters in the fibers and obtaining their
luminosity function, this would be subject to the well known
magnification bias (e.g., Turner 1980).  In our case, we are computing the
cross-correlation function of the surface brightness
measured from all fibers with quasars, and the
magnification of \lya\ sources cannot change the amplitude of the
cross-correlation. Dust associated with the foreground galaxies might
reduce the \lya emission coming from higher redshift, but this could
only further increase the inferred brightness of the \lya background.

\subsection{Star-forming galaxies and the photon budget}

\label{budget}

  We have found in Section \ref{sfrd} that our detected signal of
cross-correlation of \lya surface brightness with quasars implies a
brightness for the mean \lya photon background given by equation
(\ref{eq:backi}). This at the same time implies an emissivity of
\lya radiation of $\epsilon_\alpha=3.1\times 10^{41} (3/b_\alpha)
\ergs\mpc^{-3}$.
This emissivity can be reexpressed in terms of the rate at which
\lya photons must have been emitted for each baryon in the universe
at the mean redshift of our observation, $z=2.55$. Using the comoving
number density of baryons $n_b = 2.5\times 10^{-7}\cm^{-3}$, and an
expansion rate at $z=2.55$ of $H(z=2.55)=261 \kms\mpc^{-1}$ (using the
parameters $\Omega_b h^2 = 0.0221$, $H_0=68 \kms\mpc^{-1}$, and
$\Omega_m=0.315$, consistent with the most recent determinations
from Planck in Ade et al.\ 2014), we find the following result:
\begin{equation}
 {\epsilon_\alpha \over h\nu_\alpha n_b H(z)} = 306\, {3\over b_\alpha}\,
\frac{\rm photons}{\rm baryon} ~.
\label{eq:epsab}
\end{equation}

  The first, most simple assumption we make is that these photons are
mostly originating from star formation in galaxies. The \lya photons
created by recombinations in the HII regions produced around massive
stars can then be scattered out to gaseous halos surrounding galaxies,
from which they give rise to the background we detect in the
quasar-\lya emission cross-correlation. As discussed in Section 5,
this implies a very large star formation rate at $z=2.55$. Equation
(\ref{sflya}) can also be recast in terms of the number of \lya
photons emitted for each baryon that forms stars, $n_\alpha/n_b$:
\begin{equation}
 {n_\alpha \over n_b } = { m_p (1.1\times 10^{42}\ergs \,{\rm yr})
 \over h\nu_\alpha \, \msun} = 1800\, \frac{\rm photons}{\rm baryon} ~.
\end{equation}
Comparing to equation (\ref{eq:epsab}), we see that this implies that,
for $b_\alpha=3$, about 10\% of all the baryons in the universe would
have to turn into stars if the star formation rate is maintained over
the age of the universe at $z=2.55$, $(2/3)H^{-1}(z=2.55) =
2.5\times 10^9$ years. The difficulty with this very high star
formation rate is twofold: estimates of the {\it total} fraction of
baryons in the form of stars at present are near 6\%
(Fukugita \& Peebles 2014; Shapley 2011), and as described in the previous
section, the total star formation rate at the redshift of our measurement
can reach this value only for the maximum estimates of dust absorption,
which would imply that while the UV continuum has to be absorbed by
factors of $\sim 5$, the \lya photons would have to emerge suffering
little dust absorption.

  A first possible solution to the problem of this extremely high
inferred star formation rate is to modify the Initial Mass Function
(IMF) of the stellar population that is assumed in deriving the
\lya photons emitted per baryon in equation (\ref{eq:epsab}) from population
synthesis models. If the slope of the IMF is flatter in the high
mass range of $10 - 100 \msun$, then stars above
$\sim 20 \msun$, which dominate the production of ionizing photons,
increase their abundance compared to $\sim 10 \msun$ stars, which
dominate the observed UV continuum. Most of the ionizing photons can be
absorbed in HII regions, but most of the \lya photons produced by
subsequent recombinations can escape, and their production rate can
be increased compared to that inferred from the UV continuum
observations. If the IMF stays flat down to lower masses, that can
also greatly reduce the total star formation rate that is implied,
as well as the stellar mass that is derived for the present universe
which is measured from the old stellar population dominating the
present luminosity of galaxies. If this flat IMF occurs particularly in
massive galaxies with high metallicity, then the UV continuum observed
at $z=2.55$ can be further reduced due to the suppression of blue
horizontal branch stars, and the luminosity-weighted bias factor of the
\lya emission can be further increased. A top-heavy IMF during the epoch
when most stars were formed implies a large increase in the production
of heavy elements, but this may be consistent with observations that
show relatively high metallicities in the intracluster medium and in
massive galaxies (Renzini \& Andreon 2014).

  For the rest of this section we assume a standard IMF with the
ratio of \lya photons produced per baryon forming stars in
Equation (\ref{eq:epsab}).
We next enumerate additional sources that might contribute to
this background other than star-forming galaxies, estimating if any
of these contributions might be substantial. Finally, we discuss the
effects of \lya radiative transfer on the quasar-\lya background
cross-correlation we detect.

\subsection{Other sources of \lya emission beyond star-forming galaxies}
\label{sources}

  We shall discuss here four possible contributions to the quasar -- \lya
emission cross-correlation not arising from star-forming galaxies
clustered around the quasars: (1)
Scattering of the quasar \lya broad emission line by the \lya forest.
(2) Fluorescence of the ionizing radiation from the quasar.
(3) Scattering of the cosmic UV background by the \lya forest in the
overdense IGM around quasars.
(4) Fluorescence of the general cosmic ionizing background by the
overdense IGM around quasars.
(5) \lya cooling radiation from radiative dissipation of gas in halos that
are correlated with quasars.

\subsubsection{Scattering of the quasar \lya broad emission line}

 The average observed flux of our sample of BOSS quasars within the
central $\sim 2000 \kms$ of the \lya broad emission line is close to
$f_\alpha\sim 10^{-16} \ergs\cm^{-2}\angs^{-1}$. In general, the
mean fraction of light that is found to be absorbed by the \lya forest
at $z=2.55$ is $1-\bar F \simeq 0.2$ ($\bar F$ is the mean transmitted
fraction; e.g., Faucher-Gigu\`ere \etal 2008). 
At a characteristic impact parameter of $\sim 10\hmpc$ (inside which our
cross-correlation signal is strongest), corresponding to an
angular separation $\theta \sim 500 \asec$,
the surface brightness of the scattered radiation should be
$f_\alpha(1-\bar F)/(\psi \pi\theta^2)$, where $\psi$ is a dimensionless
number that depends on the geometry of the scattering gas around the quasar
and the shape of the \lya emission line, and has a value $\psi\simeq 4$
for a uniform gas density and a flat emission line profile. This yields
a surface brightness $\sim 7\times 10^{-24} \ergs\cm^{-2}\angs^{-1}\asec^{-2}$,
more than two orders of magnitude lower than our measured excess surface
brightness at an impact parameter of $10 \hmpc$ from a quasar, as shown
in Figure \ref{cdmfid}.

  Scattered light from the quasar broad \lya emission line is therefore
a negligible contribution to our detected \lya background. This is
consistent with our test in Appendix C.3 showing no dependence of the
cross-correlation amplitude on the quasar luminosity.

\subsubsection{Fluorescence of the quasar ionizing radiation}

  An excess of \lya emission around the quasar may also arise from
fluorescence of the ionizing radiation from the quasar. The ionizing
radiation is absorbed in intergalactic absorption systems, and about
half of the energy is reemitted as \lya photons when the recombinations
that maintain ionization equilibrium take place. These \lya photons
should be predominantly emitted from systems with column densities
$N_{HI}\sim 10^{17}\cm^{-2}$, for which the Lyman limit optical depth
is of order unity.

  The ratio of this fluorescent emission to the scattered \lya
radiation from the quasar can be estimated as follows. The
characteristic equivalent width of the \lya emission line of quasars
is $c W_\alpha/\lambda_\alpha \sim 2\times 10^4 \kms$ (see, e.g., the
composite spectrum of BOSS quasars in Figure \ref{stackqso}).
Assuming a continuum
slope bluewards of the \lya line $f_\nu \propto \nu^{-1.5}$ for the
average quasar, we find that the ratio of the number of ionizing
photons to \lya photons emitted by a quasar is
$(3/4)^{1.5}\, (2/3) \lambda_\alpha/W_\alpha \simeq 7$. Around two
thirds of these absorptions of ionizing photons result in a \lya
photon (the rest end up producing two-photon emission from the
$2s$ state of the hydrogen atom), whereas only $\sim 20\%$ of the
\lya photons on the blue half of the quasar emission line are scattered
in the \lya forest at $z=2.55$. The ratio of the total number of
fluorescent to scattered \lya photons from a quasar is therefore
$7 (2/3)/0.2/0.5 \sim 40$.

  To estimate the ratio of the contribution to the excess \lya surface
brightness near a quasar from the two processes, we also need to take
into account the different pathlengths over which the photons are
absorbed or scattered. Whereas \lya photons in the blue half of the
quasar emission line will be scattered over a pathlength corresponding
to the typical half-width of the \lya emission line,
$\ell_\alpha \sim 5000 \kms$,
the ionizing photons will be absorbed over the longer mean free path
between Lyman limit absorbers at $z=2.55$, which is $\ell_{LL}\sim
30000 \kms$ (Prochaska \etal 2014). The surface brightness
produced near the quasar is proportional to the total number of \lya
photons produced divided by the pathlength over which they are
emitted. Therefore, we conclude that the fluorescent emission from
quasars should be $40/6\sim 7$ times brighter in surface brightness than
the scattered light from their \lya broad emission line.

  Even this factor of 7, however, raises the estimated surface
brightness contribution from fluorescence at an impact parameter
$r=10 \hmpc$ to $\xi_{q\alpha}(r)\simeq 5\times 10^{-23}
\ergs\cm^{-2}\angs^{-1}\asec^{-2}$, which is still a factor of
$\sim 30$ below our measured value at this impact parameter from
Figure~\ref{cdmfid}.

  We note here that this \lya fluorescent emission from the quasar would
not be of uniform intensity, but would predominantly arise from Lyman
limit system absorbers in the quasar vicinity (see, e.g., the
simulations in \citealt{Kollmeier10}). This does not matter for our
measurement, which is sensitive only to the mean surface brightness. In
any case, our discussion suggests that the main difficulty for detecting
this component of fluorescent emission caused by individual quasars
probably lies in distinguishing it from the \lya emission scattered in
gas halos surrounding star-forming galaxies that are clustered with the
quasar, which is what we believe dominates the production of the \lya
photons in our signal. 

  We also point out here that the quasar fluorescence contribution is
affected by the anisotropy of the quasar emission and the time-delay
between our observation of the quasar and that of the fluorescent light.
Typically, if quasars are highly anisotropic and variable, the quasar
emission in a direction away from the line of sight, or at $\sim 10^7$
years ago corresponding to the time when the quasar was illuminating an
absorber at an impact parameter $r\sim 10 \hmpc$, would be
systematically lower than the one we observe at present because of the
flux-limited selection of quasar catalogues. These effects are likely to
be important in any detailed modelling of the fluorescent emission due
to the quasar.

\subsubsection{Fluorescence of the cosmic ionizing background}

  Fluorescence from the mean cosmic ionizing background can also
contribute to the quasar-\lya cross-correlation we measure. In this
case, if we assume the intensity of ionizing photons from distant
sources is uniform around the quasar, the effect on the
cross-correlation would arise only from the {\it overdensity} of
Lyman limit absorbers near the quasar, because a uniform \lya
brightness does not contribute to our detected signal.
We consider a value of the photoionization rate at $z=2.5$ of
$\Gamma = 10^{-12}\, {\rm s}^{-1}$, which corresponds to a proper
intensity per unit wavelength of the ionizing background of
$i_\lambda \simeq 3\times 10^{-20} \ergs\cm^{-2}\angs\asec^{-2}$
at the ionization edge $\lambda=912\angs$
(e.g., Faucher-Gigu\`ere et al.\ 2008). This ionizing background
is being absorbed over a mean free path $\lambda_i \simeq
cH^{-1}/10$ at $z=2.5$ (Prochaska et al. 2014). We assume that the
ionizing background has a spectral slope $-\beta$ in frequency
(i.e., $i_\lambda \propto \lambda^{\beta-2}$ below $\lambda=912 \angs$).
Each ionizing photon produces a \lya photon with probability close to
$2/3$, so the intensity of the \lya background that is produced is
\begin{equation}
 i_\alpha = {2\over 3}\left({3\over 4}\right)^2 {i_\lambda\over \beta}
{cH^{-1}\over \lambda_i} ~.
\end{equation}
The factor $(3/4)^2$ arises from the ratio of the \lya to the Lyman
limit wavelength (one factor for the energy of each photon, and one
for the fact that the intensity is per unit of wavelength).
This approximates $\lambda_i$ as being independent of frequency,
which is valid for optically thick absorbing systems (for optically
thin systems, $\beta$ in the denominator needs to be replaced by
$\beta+3$, and the reality should be intermediate). Using $\beta=2$, we
obtain $i_\alpha \simeq 5\times 10^{-20} \ergs\cm^{-2}\angs^{-1}\asec^{-2}$,
at $z=2.55$.
When observed at the present time, the intensity of this \lya background
is reduced to
\begin{equation}
 i_{\alpha 0} = i_\alpha /(1+z)^5 \simeq 10^{-22}
\ergs\cm^{-2}\angs^{-1}\asec^{-2} ~.
\end{equation}
This mean intensity now needs to be multiplied by the cross-correlation
function of Lyman limit absorbers and quasars, to obtain the
contribution to our measured cross-correlation. The value of the mass
autocorrelation at impact parameter $r=10 \hmpc$ and $z=2.55$ is
$\xi_m\simeq 0.05$, which needs to be multiplied by the bias factor
of quasars and the bias factor of Lyman limit absorbers. The bias factor
for the BOSS quasars is known, $b_q\simeq 3.5$, and that of the Lyman
limit systems is unlikely to be larger than the value for DLAs,
$b_D\simeq 2$ (Font-Ribera \etal 2012), implying that $f_\beta\sim 1.3$ and 
$b_q b_{LL} f_\beta
\xi_m(r=10 \hmpc)$ is $\sim 0.45$. This yields a value of the
contribution to the quasar-\lya cross-correlation of $\sim 4.5\times 10^{-23}
\ergs\cm^{-2}\angs^{-1}\asec^{-2}$, a factor of $\sim 30$ below
our measured value from Figure~\ref{cdmfid}.

We note that this estimate is consistent with a 
cruder calculation which uses the  
ionizing background observationally inferred
by Fontanot et al. (2014) from the comoving space density of quasars and
star forming galaxies. After exploring the likely parameter space
of limiting magnitudes and escape fractions, Fontanot  et al. (2014)
find that a central value for the ionizing background
comoving emissivity is about
$3\times 10^{50}
{\rm photons\,} {\rm s}^{-1} {\rm Mpc}^{-3}$ at $z\sim 2.55$. 
 If the emissivity of the ionizing photons that 
are converted to \lya photons is at a similar level, with a conversion 
efficiency of 2/3, the fluorescent \lya emissivity is then $2\times 10^{50}
{\rm photons\,} {\rm s}^{-1} {\rm Mpc}^{-3}$. This is $3.2\times 10^{39}
{\rm erg\, s^{-1} Mpc^{-3}}$, $\sim 100$ times lower than the value inferred
from our results.

  We therefore conclude that fluorescence from the cosmic ionizing
background also cannot be responsible for the large \lya brightness
that we detect correlated with quasars.

\subsubsection{Scattered \lya radiation from the radiation background}

  Just as scattering from the quasar by the \lya forest can in principle
produce some contribution to this \lya light around quasars, one may
think that the general ultraviolet continuum background from distant
galaxies and quasars could also give rise to an excess of \lya photons
near quasars because of the overdensity of the \lya forest that scatters
this background radiation. However, this effect actually cancels out
for our observation. High redshift galaxies that are behind the quasar
should show the \lya forest absorption, reducing their ultraviolet
flux, and this \lya absorption will be enhanced because of the
overdensity surrounding the quasar. The background sources (which
are of course too faint to be detected individually, but always
contribute to our total background) are therefore fainter near quasars
compared to any random fields. This is exactly compensated by the
scattered \lya background from these same sources. For this reason,
this contribution to the \lya background could only be detected if the
individual sources behind the quasar were individually detected and
subtracted out, before evaluating the \lya background intensity.

  In addition to \lya photons scattered by the \lya forest, the
background would also have a contribution from background photons
reaching the Ly$\gamma$ wavelength when passing near the quasar and
being downscattered to \lya. This contribution is also very small, and
is also nearly cancelled unless the population of ultraviolet sources
creating the background evolves very fast over the redshift interval
corresponding to the ratio of the Ly$\gamma$ to the \lya wavelength.
(note that the Ly$\beta$ forest does not contribute to this diffuse
emission by intergalactic gas because Ly$\beta$ absorptions can only
end in a 2-photon emission from the $2s$ atomic state of hydrogen.)

\subsubsection{Low surface brightness \lya halos around galaxies}

\label{halos}

We have described in Section \ref{sfrd} how \lya emission from 
galaxies could account for our measured result. 
Due to the magnitude of our signal, this \lya emission
does not appear to originate in a straightforward way from \lya emitters
seen in narrow band surveys at these redshift.

The \lya luminosity from observed \lya emitters is usually defined by
pixels above certain surface brightness threshold or measured within an 
aperture (e.g., $2\arcsec$; \citealt{Ouchi08}). In any case, the surface 
brightness threshold is well above the sky noise and higher than the level 
associated with our measurement in this paper. An extended \lya emitting
halo below sky noise level, resulting from \lya photons scattered by neutral 
gas in the circumgalactic and intergalactic media and clustered unresolved 
\lya sources, is predicted to exist around
a star-forming galaxy based on radiative transfer modeling 
\citep[][]{Zheng11b}. Such diffuse \lya emitting
halos are detected from stacking 
analysis \citep[][]{Steidel11,Matsuda12,Momose14}.

We use the stacked \lya surface brightness profiles and
the fits to the diffuse \lya halos in \citet{Momose14} to estimate the 
contribution of these halos to \lya emission in our detection.
The luminosity inside the aperture of radius $2\arcsec$ roughly corresponds
to the \lya luminosity from the \lya emitter survey. We find that at both
$z=2.2$ and $z=3.1$, the diffuse \lya emission outside of $2\arcsec$ is
about one 
third of the luminosity inside the aperture. Therefore, the observed diffuse
\lya halos, regardless of their origin, may only contribute an additional
one third to the \lya emission from \lya emitter surveys. 

  This implies that our hypothesized extended emission from scattering
halos that may account for our measurement would have to arise in many
more galaxies than those detected in \lya-emission surveys.

\subsubsection{Cooling radiation}


Cooling radiation from gas in galactic halos can produce \lya emission.
A rough estimate of the cooling radiation
can be made if we assume that cooling and star formation reach a steady 
state on average in a galaxy (i.e., $1\msun$ of gas cools per year to feed
1$\msun$ yr$^{-1}$ star formation).
 For gas initially at virial temperature
$T$ dissipating the energy through cooling
(as suggested by cosmological simulations; e.g., 
\citealt{Fardal01,Faucher10,Goerdt10}), we can estimate the 
 corresponding cooling luminosity
as follows:

Let us assume that a fraction $f_{c}$ of all baryons in the Universe fall 
into halos of virial temperature $T$, and they dissipate all their energy 
by emitting \lya photons. Initially the
baryons fall into the halo and are shock heated to temperature $T$ at a 
radius $r_{v}\simeq \sigma t/6$ in the halo, where $t$ is the age
of the Universe, and in the end they have to reach a radius $r_{g}$ by 
dissipating their energy through \lya emission, 
where $r_{g}$ is the half-radius of the galaxy.

The circular velocity $v_{c} = \sqrt{2} \sigma$ is assumed to be 
flat, independent of radius. Then, the
potential difference from $r_{v}$ to $r_{g}$ is $\phi=v^{2}_{c}
\log(r_{v}/r_{c})$. The energy to be dissipated per baryon is therefore
$m_{p}\phi$. The relation between $v_{c}^{2}$ and the temperature
$T$ is $\sigma^{2}=kT/\mu$, where $\mu=0.6 m_{p}$ for the 
fully ionized mixture of hydrogen and helium from Big 
Bang nucleosynthesis. So, the total
energy dissipated per baryon in the universe is
\begin{equation}
\epsilon_{b}=f_{c}m_{p}\phi=\frac{2f_{c}}{0.6}kT\log (r_{v}/r_{c}),
\end{equation}
and the number of \lya photons emitted per baryon is 
\begin{equation}
\epsilon_{\alpha}=\epsilon_{\beta}/(10.2{\rm eV})=
3.3f_{c}\frac{kT}{1.2\times10^{5}{\rm K}}\log(r_{v}/r_{c}).
\end{equation}
Using $T=3\times10^{6}$ K and $r_{v}/r_{c}=20$ we get 
$\epsilon_{\alpha}=240 f_{c}$. With plausible values of
$f_{c}\sim0.2$, this could amount to nearly 15\% of the emission we
observed, and more if the bias of this \lya emission is high.

Cooling radiation is therefore 
the alternative source of \lya emission
which  comes closest to explaining our results, being plausibly less
than an order of magnitude below our measurements.
Cooling radiation could contribute in a substantial way to
the observed \lya emission.


\subsection{Radiative transfer effect}

\label{sec:RT}

The above estimates show that the \lya emission relevant to our clustering
measurements is likely to be
dominated by contributions from star-forming galaxies. 
Regardless of the origin, as long as \lya photons are scattered by neutral 
hydrogen, we expect them to be affected by a radiative transfer effect.
The quasar-\lya cross-correlation function we measure suggests that this
effect is detected, shown as elongated contours along the line of sight 
on scales as large as $\sim$ 20$\hmpc$. This feature cannot solely originate
from the dispersion of the relative peculiar veclocity between quasars and 
galaxies (including redshift errors),
but is consistent with the predicted main radiative effect on the
clustering of \lya emitters in \citet{Zheng11a}.

The radiative transfer effect predicted in \citet{Zheng11a} is a result of the 
anisotropic emission of scattered \lya photons from the anisotropic 
distribution of neutral gas density and velocity (hence anisotropic 
distribution of \lya scattering optical depth). In particular, \lya photons
preferentially escape along the direction for which the neutral gas has the
largest peculiar velocity gradient. Therefore, the \lya emission from less
overdense regions along the 
line-of-sight direction would appear enhanced, suppresing the line-of-sight 
fluctuation. This dependece on the line-of-sight peculiar velocity gradient 
is analogous to that in the Kaiser effect but of opposite sign, leading to
the elongated correlation contours. For all the possible origins of \lya 
emission discussed above, this radiative transfer effect should be at work
as long as \lya photons interact with neutral gas on large scales. 

Our measurement can be used to constrain the parameters relevant to the 
radiative transfer effect. Following the simple model presented in 
\citet{Zheng11a}, the real-space overdensity $\delta_\alpha$ (the 
\lya surface brightness fluctuations here) can be related to 
the matter linear overdensty field $\delta$ and peculiar velocity field as 
\begin{equation}
1+\delta_\alpha \propto (1+b_L\delta)\left[ 1 + \alpha_1\delta 
+ \alpha_2 \frac{1}{Ha}\frac{\partial v_z}{\partial z}
\right],
\end{equation}
where $b_L$ is the \lya luminosity weighted bias of the underlying galaxy 
population and $v_z$ the line-of-sight peculiar velocity. The coefficient 
$\alpha_1$ represents a combined effect of the dependence of the \lya radiative
transfer (i.e., the \lya escape emission) on the density and the transverse 
peculiar velocity gradient, and 
$\alpha_2$ specifies the impact of the line-of-sight peculiar velocity 
gradient (see \citealt{Zheng11a}, in particular their Appendix A). Both 
coefficients are expected to be positive. In redshift space,
we also need to add the Kaiser effect, and in terms of the Fourier component
of the \lya surface brightness fluctuations, we have
\begin{equation}
\delta^s_{\alpha,\mathbf{k}}=(b_L+\alpha_1)[1+\beta_\alpha\mu^2]\delta_{\mathbf k}.
\end{equation}
The $\beta_\alpha$ parameter (as constrained in \S~3.4) is 
\begin{equation}
\beta_\alpha=(1-\alpha_2)\frac{\Omega_{\rm m}(z=2.55)^{0.6}}{b_L+\alpha_1}.
\end{equation}
The factor ``1'' comes from the Kaiser effect. The bias factor $b_\alpha$ 
in Equation~\ref{model} is $b_L+\alpha_1$.
With $\beta_\alpha=-0.76\pm 0.36$, we have $\alpha_2=1+(2.35\pm 1.11)(b_\alpha/3)$ (N.B. this is the $\mathcal{C}_v$ coefficient defined in 
\citealt{Wyithe2011}). That is, the radiative transfer effect (indicated by
a positive $\alpha_2$) shows up or has been detected at a level of 
$\sim 3.0\sigma$ for the fiducial value of $b_\alpha=3$. 

Overall, the quasar-\lya cross-correlation provides tentative evidence to a 
new clustering effect caused by \lya radiative transfer. A better measurement
with a larger data set and a more detailed modeling will help understand this
effect and use it to constrain the neutral gas distribution.

\subsection{Escape fraction and detected fraction}

\label{escape}

Assuming now that our fiducial value of the star formation rate density is
correctly inferred from the quasar-\lya cross-correlation, we conclude that
it is consistent with the whole dust corrected star formation rate in 
\citet{Bouwens10}. At face value, this indicates that dust has little effect 
in reducing the \lya emission, i.e., almost 100\% of the \lya photons
produced from star formation escape. Clumpy dust clouds \citep{Neufeld91} and
anisotropic \lya escape \citep{Zheng14} may be possible mechanisms for
explaining the high escape fraction of \lya photons, together with the much
lower one for continuum UV photons.

We emphasize that the escape fraction is not the same as the fraction that is
detected in \lya emitter surveys. The latter, the detected fraction, comes
only from the central, high surface brightness part of the \lya emission from
galaxies, and not from the extended \lya halos of low surface brightness
\citep{Zheng11b,Steidel11, Matsuda12,Momose14}.
The detected fraction for \lya emission at $z=2$ to 3, 
inferred from comparing the \lya luminosity density and H$_\alpha$ or 
H$_\beta$ luminosity density, is about 5\% \citep[e.g.][]{Hayes11,Ciardullo14}.
This fraction is consistent with the ratio between the star 
formation rate density inferred from \lya emitters and that from quasar-\lya 
cross-correlation in this work (see the blue and red points in 
Figure~\ref{sfr}).

\section{Summary and Conclusions}

\subsection{Summary}

We have carried out a cross-correlation analysis of residual light
in SDSS/BOSS galaxy spectra and SDSS/BOSS quasars at redshifts from 
$z=2.0-3.5$. We have used the \lya emission line (which 
is at wavelengths $\lambda=3647-5471$ \AA\ at these
redshifts) to trace structure
in the cross-correlation.  Our main findings are as follows:\\

\noindent(1) We measure large-scale structure in the quasar-\lya emission
cross-correlation at a mean redshift $z=2.55$
at the 8 $\sigma$ level.
The cross-correlation function shape is
consistent with the linear $\Lambda$CDM correlation function.\\
\noindent(2) Looking at the clustering as a function of separation
across and along the line of sight we see evidence at the $3.0 \sigma$
level for distortions of
clustering of the form  expected to be caused by radiative 
transfer effects.\\
\noindent(3) We detect clustering independently in 4 subsamples
at different redshifts,
finding that the shape of the cross-correlation function is consistent
with the fiducial sample. The amplitude of the cross-correlation 
increases by a factor of 3 between $z=3.5$ and $z=2.0$, although this
detection of evolution is marginal, being consistent with no evolution
at the $2.0\sigma$ level.\\
\noindent(4) Although the spectra are too shallow to allow making a good
catalogue of emission lines, we are able to weakly constrain the 
contribution of
emission lines to our signal statistically 
by fitting lines, subtracting them and remeasuring the 
quasar-\lya cross-correlation
function. We find that lines with luminosities (measured in our 1 arcsec
radius aperture) of $L_{{\rm Ly}\alpha } > 8\times10^{41}$ erg s$^{-1}$ 
may contribute $13^{+11}_{-13} \%$ of the quasar-\lya cross-correlation
amplitude at the relevant redshifts, but this contribution is still consistent
with zero.\\
\noindent(5) In one of our sample tests, we measure the cross-correlation
to be independent of quasar luminosity. This is evidence that the 
large-scale clustering of \lya surface brightness
we measure arises mostly from \lya emission associated with star formation,
and not from any systematic error associated with the quasar light. \\
\noindent(6) We estimate the plausible contribution to the quasar-\lya
surface brightness we measure from a variety of physical processes alternative
to star formation galaxies that follow the same large-scale structure as
quasars, such as fluorescence or scattering of the quasar emission or the
ionizing background, and we find these contributions to be almost all
negligible. Only cooling radiation may contribute in a substantial
fashion (perhaps at the $15\%$ level) to the \lya surface brightness.\\
\noindent(7) Using measurements of clustering from 
SDSS/BOSS quasars, we convert the measured
amplitude of  the cross-correlation function \ampqa\ to the 
product of mean
\lya sky brightness at $z=2.55$ and linear bias factor of \lya emission,
 finding  
$\langle \mu_{\alpha} \rangle(b_{\alpha}/4)=(3.9\pm0.9)\times10^{-21}$
erg s$^{-1}$
cm$^{-2}$ \AA $^{-1}$ arcsec$^{-2}$.
Assuming that this \lya surface brightness is due to star 
formation (see points (5) and (6) above), we convert our measured value to 
the mean star formation rate density in the Universe at redshift $z=2.55$,
finding 
${\rho}_{\rm SFR} = (0.28 \pm 0.07) (3/b_{\alpha}) \msun {\rm yr}^{-1}
{\rm Mpc}^{-3}$.
This is consistent with dust-corrected UV continuum based estimates of 
star formation, but more than an order of magnitude higher than previous
estimates of the SFRD from surveys of individual \lya emitters.

\subsection{Implications}
\label{implications}

We conclude that the high intensity of the \lya background
at $z\simeq 2.5$ that is derived from our measurement can only be reasonably
produced by \lya-emitting galaxies that are clustered with quasars.
If our measurement is confirmed, the consequences for our understanding
of galaxy formation and evolution are dramatic. The \lya
emission directly observed in \lya emitting galaxies so far at these
redshifts contributes only
$ 0.01 \msun\, {\rm yr}^{-1}\mpc^{-3}$ to the mean star formation rate
density. Extended halos that have been seen around these \lya-emitting
galaxies detected in surveys can only contribute an additional $\sim 30\%$
to this value.
We have detected $21-35$ ($\pm 1\sigma$ range) times more
\lya photons than the \lya emitter surveys (with the uncertainty due to the
factor $(3/b_{\alpha})$, and we have argued that most
of these photons also arise from star formation. This
amount of star formation represents most of the massive stars that are
estimated to have formed in the universe and to have generated the
present heavy elements. 

To contribute to our
measurement, these \lya photons cannot be
absorbed by dust before escaping the galaxies. The question that remains
then is how this \lya radiation could have been missed by previous 
observations. Putting forward and testing a detailed scenario
is beyond the scope of this paper, but we speculate that all
star forming galaxies at these redshifts, even if they do not have any
detectable \lya emission line in their central parts, are surrounded by 
low surface brightness halos that nevertheless have a high total
integrated \lya luminosity. These low surface brightness halos could allow
the  bulk of 
\lya photons to be below the detectable levels in narrow band \lya emitter
surveys. As much of the star formation in the Universe at redshifts $z=2-4$
occurs in massive galaxies, the implication is that a large
fraction of the \lya emission we detect 
is from giant low surface brightness halos
around massive, bright, star-forming galaxies, which absorb most of their
continuum photons to reradiate the energy in the infrared, and yet let their
\lya photons escape.

\subsection{The future}
We note that the sky area covered by the million SDSS/BOSS (for our
purposes randomly placed)
fibers in the current study is  $\sim 3\times10^{6}$ square arc seconds.
This is approximately $1/200,000$ of the entire sky area, showing that \lya
intensity mapping holds an enormous promise as a probe of structure in the 
Universe. In addition to the quasar-\lya emission cross-correlation
employed in the present paper, one can imagine carrying out \lya forest-\lya
emission cross-correlations and
\lya emission autocorrelation measurements. Correlations of \lyb absorption
 and \lya emission, and vice-versa would avoid common mode systematics
 in the fluxing
of spectrographs and may reduce the possibility of related error. 
As the \lya signal is distributed on large-scales, a way
 to efficiently carry out intensity mapping surveys (for example for baryonic
acoustic oscillation experiments)  may  be to use integral
field spectroscopy with relatively low angular ($\sim$ 10 arcsec) resolution 
on a wide-field ($\sim$ few deg.) telescope. If bright point sources could be 
masked such an instrument could in principle capture
a dataset containing 
5 orders of magnitude more information on large-scale
clustering in the Universe than our present study.

\section*{Acknowledgements}
RACC was supported by NSF Awards AST-1009781, 
AST-1109730, OCI-0749212, the Moore Foundation
and by the Leverhulme Trust's award of a Visiting Professorship at 
the University of Oxford. RACC would like to acknowledge
the hospitality of the Astrophysics Subdepartment in Oxford
where the intial stage of this work was carried out.  RACC would
also like to thank Matt McQuinn for useful discussions.
JM is partially supported by Spanish grant AYA2012-33938.
ZZ is partially supported by NSF grant AST-1208891 and NASA grant 
NNX14AC89G.

Funding for SDSS-III has been provided by the Alfred P. Sloan
Foundation, the Participating Institutions, the National Science
Foundation, and the U.S. Department of Energy Office of Science.
The SDSS-III web site is http://www.sdss3.org/. SDSS-III is managed
by the Astrophysical Research Consortium for the Participating
Institutions of the SDSS-III Collaboration including the University
of Arizona, the Brazilian Participation Group, Brookhaven
National Laboratory, University of Cambridge, Carnegie Mellon
University, University of Florida, the French Participation Group,
the German Participation Group, Harvard University, the Instituto
de Astrofisica de Canarias, the Michigan State/Notre Dame/JINA
Participation Group, Johns Hopkins University, Lawrence Berkeley
National Laboratory, Max Planck Institute for Astrophysics, Max
Planck Institute for Extraterrestrial Physics, NewMexico State University,
New York University, Ohio State University, Pennsylvania
State University, University of Portsmouth, Princeton University,
the Spanish Participation Group, University of Tokyo, University
of Utah, Vanderbilt University, University of Virginia University of
Washington, and Yale University.

\section*{Author affiliations}
\begin{small}
$^{1}$ McWilliams Center for Cosmology, Dept. of Physics, 
Carnegie   Mellon  University, Pittsburgh, PA 15213, USA\\
$^{2}$ Astrophysics, University of Oxford,
Keble Road, Oxford OX1 3RH, UK\\
$^{3}$ Instituci\'{o} Catalana de Recerca i Estudis  Avan\c{c}ats, 
Barcelona, Catalonia\\
$^{4}$ Institut de Ci\`{e}ncies del Cosmos, Universitat de 
Barcelona/IEEC, Barcelona 08028, Catalonia\\
$^{5}$ Department of Physics and Astronomy, University of Utah, 115 S 1400 E, 
Salt Lake City, UT 84112, USA\\
$^{6}$ Dept. of Astronomy and Astrophysics, The University of Chicago, 
5640 South Ellis Avenue,
Chicago, IL 60615, USA\\
$^{7}$ Department of Astronomy, Harvard University, 60 Garden St.,
Cambridge MA 02138, USA\\
$^{8}$ APC, Universit\'{e} Paris Diderot-Paris 7, CNRS/IN2P3, CEA,
       Observatoire de Paris, 10, rueA. Domon \& L. Duquet,  Paris, France\\
$^{9}$ CEA, Centre de Saclay, IRFU,  F-91191 Gif-sur-Yvette, France\\
$^{10}$ Lawrence Berkeley National Laboratory, 1 Cyclotron Road, Berkeley, 
CA 94720, USA\\
$^{11}$ Department of Physics and Astronomy, University of California, Irvine, CA 92697, USA\\
$^{12}$ Max-Planck-Institut f\"ur Astronomie, K\"onigstuhl 17, 
D69117 Heidelberg, Germany\\
$^{13}$Department of Astronomy, University of Wisconsin, 475 
North Charter Street, Madison, WI 53706, USA\\
$^{14}$Department of Physics and Astronomy, University of Wyoming, Laramie, 
WY 82071, USA\\
$^{15}$ Departamento de Astronomia, Universidad de Chile, Casilla 36-D,
Santiago, Chile\\
$^{16}$Universit\'e Paris 6 et CNRS, Institut d'Astrophysique de Paris,
 98bis blvd. Arago, 75014 Paris, France\\
$^{17}$Institute of Cosmology and Gravitation, University of Portsmouth,
Dennis Sciama building, Portsmouth P01 3FX, UK\\
$^{18}$Department of Astronomy and Astrophysics, The Pennsylvania
State University, University Park, PA 16802, USA\\
$^{19}$ Institute for Gravitation and the Cosmos, The Pennsylvania State
University, University Park, PA 16802, USA\\
$^{20}$ Bldg 510 Brookhaven National Laboratory, Upton, NY 11973, USA\\
$^{21}$ INAF, Osservatorio Astronomico di Trieste, Via G. B. Tiepolo 11,
 34131 Trieste, Italy\\
$^{22}$ INFN/National Institute for Nuclear Physics,
Via Valerio 2, I-34127 Trieste, Italy.\\
$^{23}$ Department of Astronomy, Ohio State University, 
140 West 18th Avenue, Columbus, OH 43210, USA\\
$^{24}$ Apache Point Observatory, PO Box 59, Sunspot, NM 88349, USA\\
$^{25}$ Department of Astronomy and Space Science, Sejong
University, Seoul, 143-747, Korea\\
\end{small}

\section*{Appendix A: Stray light contamination}

As mentioned in Section \ref{xcor} a potentially very important systematic is
the contamination of the galaxy spectra by quasar light. This is particularly
relevant for our cross-correlation measurement because we are searching
for light from sources that are clustered with quasars, and light from 
the quasar itself could mimic this if not treated carefully. The issue 
arises because the 1000 fibers that orginate from the plate in the
telescope focal plane are fed into (each red and blue)
 spectrograph and dispersed onto a single CCD. 1000 spectra are therefore
extracted from a 
4k$\times$4K chip and some light from neighbouring fibers can end
up in CCD columns that are centered on other fibers. 

\subsection*{A.1: Measured Quasar-Lyman-$\alpha$ surface brightness 
correlation}

  We can test for this contaminating light by use of two facts about the
observational setup. The fibers are arranged by fiber number (from 1-1000)
on the CCD, so that for each spectrum we can measure light from 
a certain number of fibers away. Quasars and galaxies that we use
in our cross-correlation will also often be measured on
different plates, so that there is no possibility for this contamination.
We can therefore also test our result by restricting the calculation
of the quasar-\lya cross-correlation to pairs of quasars and galaxies
lying on different plates.

\begin{figure}
\centerline{
\psfig{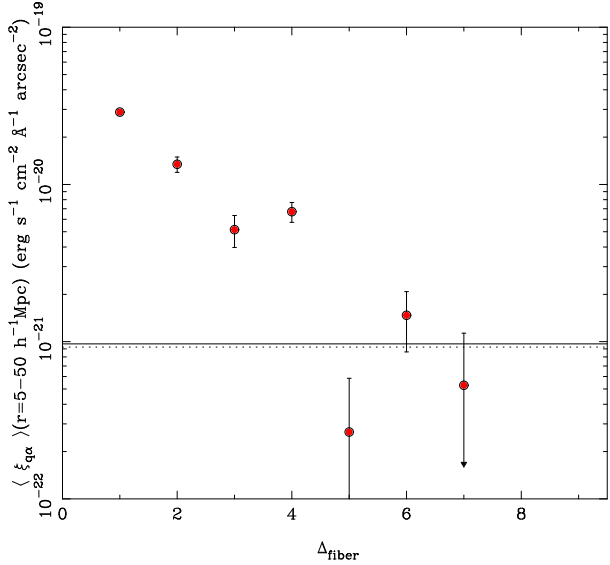}
}
\caption{ A test of stray light from quasars in nearby fibers contaminating
the galaxy spectra. We show the quasar-\lya emission
surface brightness cross-correlation averaged over quasar-pixel pair
separations of $r=5-50 \hmpc$ plotted against $\Delta_{\rm fiber}$, the
difference in fiber number between the quasar and the galaxy spectra.
A value of  $\Delta_{\rm fiber}=1$ means that the quasar and galaxy were in
adjacent fibers and their spectra where dispersed next to each other in the 
CCD. The error bars are jackknife errors computed using 100 subsamples
of the data in each case. The horizontal solid line is the 
quasar-\lya emission surface brightness cross-correlation averaged 
over quasar-pixel pair separations of $r=5-50 \hmpc$ for our fiducial
computation (see Section \ref{fidresult}),
which uses information from all quasar-pixel
pairs 6 fibers apart or greater. The dotted line (which
lies close to the solid line) is the equivalent result but only using
quasar-pixel pairs which are on different plates.
\label{straylight}}
\end{figure}

In the first test, we have computed the quasar-\lya cross-correlation of 
Equation \ref{xieq} restricting ourselves to only quasar-galaxy spectrum pairs 
separated by specific values of $\Delta_{\rm fiber}$, the difference
in fiber number of the two spectra. Spectra with $\Delta_{\rm fiber}=1$,
are adjacent on the CCD, for example, and have the highest potential
for cross-contamination of light. We reduce the quasar-\lya cross-correlation
for each value of $\Delta_{\rm fiber}$ by averaging the \xiqar\
results over a range from $r=5-50 \hmpc$. Our conclusions about 
stray light contamination are not dependent on the range picked.

The results are shown in Figure \ref{straylight}, where we show
\xiqa\ $(r=5-50\hmpc)$ against $\Delta_{\rm fiber}$. In the plot the value
of \xiqa\ $(r=5-50\hmpc)$ for our fiducial sample 
(Section \ref{fidresult}), which uses
all pairs with $\Delta_{\rm fiber} \ge 6$ is shown as a horizontal line.
We can see that for $\Delta_{\rm fiber}=1$, the mean surface brightness
inferred from the cross-correlation is over 20 times the fiducial value.
It remains significantly higher for all $\Delta_{\rm fiber} \le 4$. This
is an indication that even when quasar and galaxy  spectra are separated by 3
other spectra that light from the quasar is able to leak into
the galaxy spectrum and contaminate it. Of course the particular region
of the spectrum we are looking at (close to the \lya emission line
at the redshift of the quasar) is 
the one in which the quasar is extremely bright, and one would not expect
it to contaminate other parts of galaxy spectrum as much. Nevertheless, 
for our project, this is the important region of the spectrum, and we 
therefore must apply a cut on $\Delta_{\rm fiber}$. Based on Figure 
\ref{straylight}, we have chosen that $\Delta_{\rm fiber} \ge 6$ 
in our analysis. 

A related test is to compute \xiqa$(r=5-50\hmpc)$ for quasars and 
galaxies
that are on different plates (and therefore not dispersed at the
same time onto the CCD). We show the results as a dashed line in Figure 
\ref{straylight}, which is indistinguishable from the results from our
fiducial analysis. This is good evidence that our cut 
$\Delta_{\rm fiber} \ge 6$ is sufficient to eliminate stray quasar light.

Yet another related test is to eliminate all galaxy fibers which have a
quasar within a certain number of fibers. This is different from 
eliminating quasar-pixel pairs on a case-by-case basis because
it will also eliminate any potential contamination which could come from
quasars being clustered in space with other quasars. In the fiducial
case, the contaminating quasars will have been eliminated directly, but
those which are clustered with them could still contaminate the neigbouring
galaxy spectra. One good reason to believe that this is not occuring 
to a detectable degree is
that any contamination of this type would be a convolution of the redshift
space autocorrelation function of quasars and the contaminating surface 
brightness
from the quasar lya line, which is highly asymmetric and extremely
elongated along the line of sight (see below). This does not appear to describe
the measured signal (e.g., Figure \ref{sigmapi}).

 Nevertheless, we have
carried out the test, which eliminates $50$\% of the galaxy
fibers from the data sample. We find 
an amplitude 
\ampqa\ $=
1.98^{+0.66}_{-0.65} \times10^{-20}$ 
 erg s$^{-1}$
cm$^{-2}$ \AA $^{-1}$ arcsec$^{-2}$. 
and shape $\Omega_{\rm m}=0.69^{+0.71}_{-0.36}$.
These are consistent at about the 2 $\sigma$ 
 with the measurement from the fiducial
sample.
We note that the detection level of the signal is only about 
3 $\sigma$ overall,
compared to $\sim 9 \sigma$ for the fiducial measurement. This can be explained
by the fact that eliminating all galaxies which have a quasar within
5 fibers will disproportionately affect the number of close quasar-pixel pairs.
By directly counting, we find the number of quasar-pixel 
pairs with separations below 40 $\hmpc$ has fallen by a factor of 3.5
rather than the factor of 2 expected if pairs were drawn uniformly from
all fiber separations. 

We have also carried out another,
similar test, which eliminates a smaller fraction
of the dataset, but which should have the same effect. For this test
we remove from our list of galaxy pixels all pixels which have
more than 1 quasar
within $r=50 \hmpc$. In this way, we eliminate all potential 
cross-contamination from quasars clustered with the target quasars in
the cross-correlation, but at a much reduced cost (doing this only eliminates
0.3 \% of the galaxy spectrum pixels). After fitting the 
cross-correlation of this sample, we find
an amplitude
\ampqa\ $=                                                                     
3.04^{+0.37}_{-0.45} \times10^{-20}$
 erg s$^{-1}$
cm$^{-2}$ \AA $^{-1}$ arcsec$^{-2}$.
and shape $\Omega_{\rm m}=0.33^{+0.13}_{-0.10}$. This is consistent 
with the fiducial 
result within the error bars. It is slightly lower (by less than 1 $\sigma$), 
which should not be surprising, as we are presumably preferentially 
removing some  pixels in overdense regions. 
Overall these two test results are a good sign that significant
cross-contamination from
quasars clustered with the target quasar is not ocurring.

\begin{figure}
\centerline{
\psfig{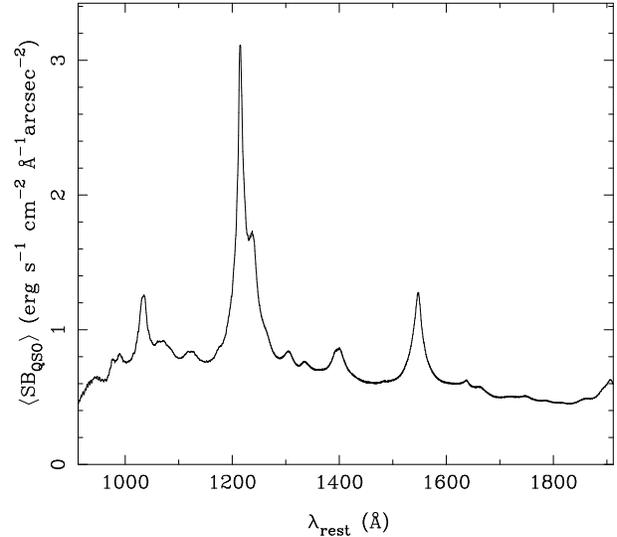}
}
\caption{ 
The mean surface brightness from all DR10 quasar spectra 
as a function of wavelength. We shift all spectra 
to their rest wavelength and stack them with equal weight to make the curve.
\label{stackqso}}
\end{figure}

\subsection*{A.2 Modelling the stray light contamination}
If the excess surface brightness seen above 
in galaxy spectra which are close on  the CCD to quasar spectra is indeed
due to cross-talk from quasar light, we would expect the contamination
to have a quasar-like spectrum. 
To examine this, we first make a stacked spectrum
of the DR10 quasar sample, by averaging all the spectra together in the 
quasar rest frame with unit
weight. This is shown in Figure \ref{stackqso}. We can see from that figure 
that the quasar \lya line is extremely broad, with a FWHM of $\sim 50$ \AA\ 
in the rest frame. There is also the noticeable emission feature
due to NV on the red wing of the line.

\begin{figure}
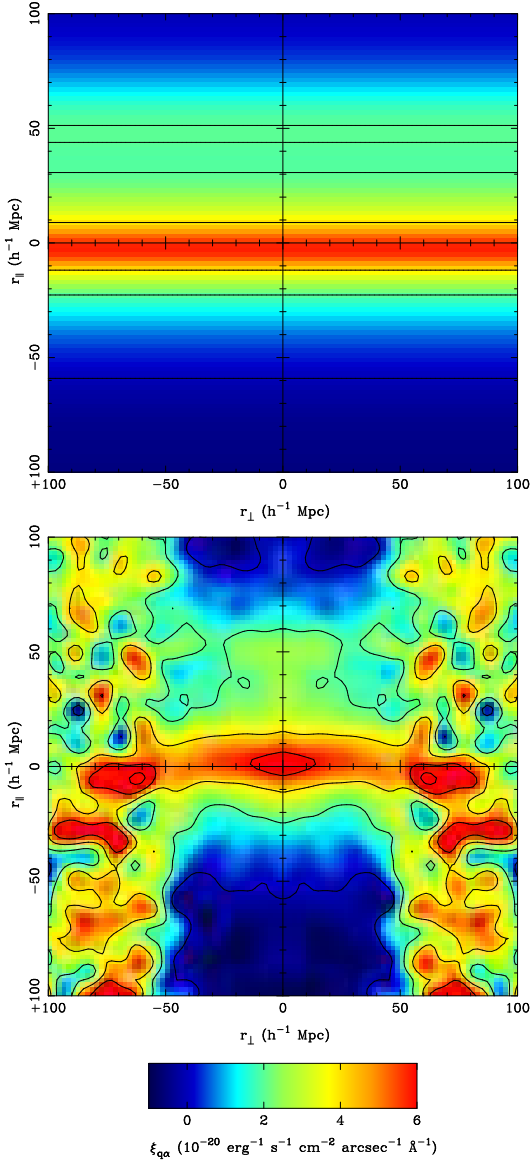

\centerline{
\psfig{file=stackqsosigpi.ps,angle=0.,width=7.0truecm}
}
\centerline{
\psfig{file=deltafiber1_sigpi.ps,angle=0.,width=7.0truecm}
}
\caption{ 
Top panel: The stacked quasar surface brightness from 
Figure \ref{stackqso} multiplied by $2.0\times10^{-3}$.
The stack is centered on the redshift of the \lya line
at redshift $z=2.55$ and the fiducial cosmology has been
used to convert \AA\ to comoving $\hmpc$. The stack is plotted as a function
of $r_{\parallel}$ and is independent of $r_{\perp}$.  
Bottom panel: The measured quasar-\lya emission
correlation for quasar-galaxy pixel pairs separated by 
1 fiber ($\Delta_{\rm fiber}=1$)  as a function
of $r_{\parallel}$ and $r_{\perp}$. 
 In both panels a Gaussian
filter with $\sigma=4 \hmpc$ was used to smooth the image 
(as in Figure \ref{sigmapi}).
\label{stacksigpi}}
\end{figure}

We next take this stacked spectrum, and convert the wavelength 
units into comoving $\hmpc$ at $z=2.55$, the mean redshift of our measurements.
If the quasar light is truly a contaminant, we expect the strength of the
contamination to depend on the difference between fiber numbers 
($\Delta_{\rm fiber}$, as defined above) and not on the actual physical
separation between quasar and pixel across the line of sight ($r_{\perp}$).
Plotting the quasar-\lya cross-correlation for particular $\Delta_{\rm fiber}$
values as a function of $r_{\perp}$ is therefore a useful check that
 light contamination is occuring as we believe. The prediction for the 
contamination, taken directly from the quasar spectrum is 
shown in the top panel of Figure \ref{stacksigpi}. In that Figure we have 
scaled the overall amplitude by a factor of $2\times10^{-3}$ (see below).
We can see that 
the large width of the \lya line, and its asymmetry due to the NV
feature are both apparent.

The data with the strongest contamination should be from galaxy 
pixel-quasar pairs which are 1 fiber apart ($\Delta_{\rm fiber}$=1). We
therefore plot the quasar-\lya correlation for just those 
pixel-quasar pairs 
in the bottom panel of Figure \ref{stacksigpi}. It is immediately
apparent that there is a stripe across the middle of the plot
corresponding to the contaminating quasar \lya emission line. The 
contamination is clearly asymmetric, extending further to positive
values of $r_{\parallel}$ than negative,
and overall is visually consistent with the 
prediction based on the stacked quasar spectrum (top panel of Figure
\ref{stacksigpi}). Because the quasar-pixel pairs that are exactly
1 fiber apart only comprise a very small fraction of the data, we
expect there to be lots of noise in the plot, particularly for large
values of $r_{\perp}$. This latter is because close pairs of quasars
and galaxies on the sky are more likely to be close in $\Delta_{\rm fiber}$.
There is neverthless enough range that we can examine by eye whether the 
contamination depends on $r_{\perp}$, with the answer appearing to 
be no. We examine this more quantitively below.

\begin{figure}
\psfig{file=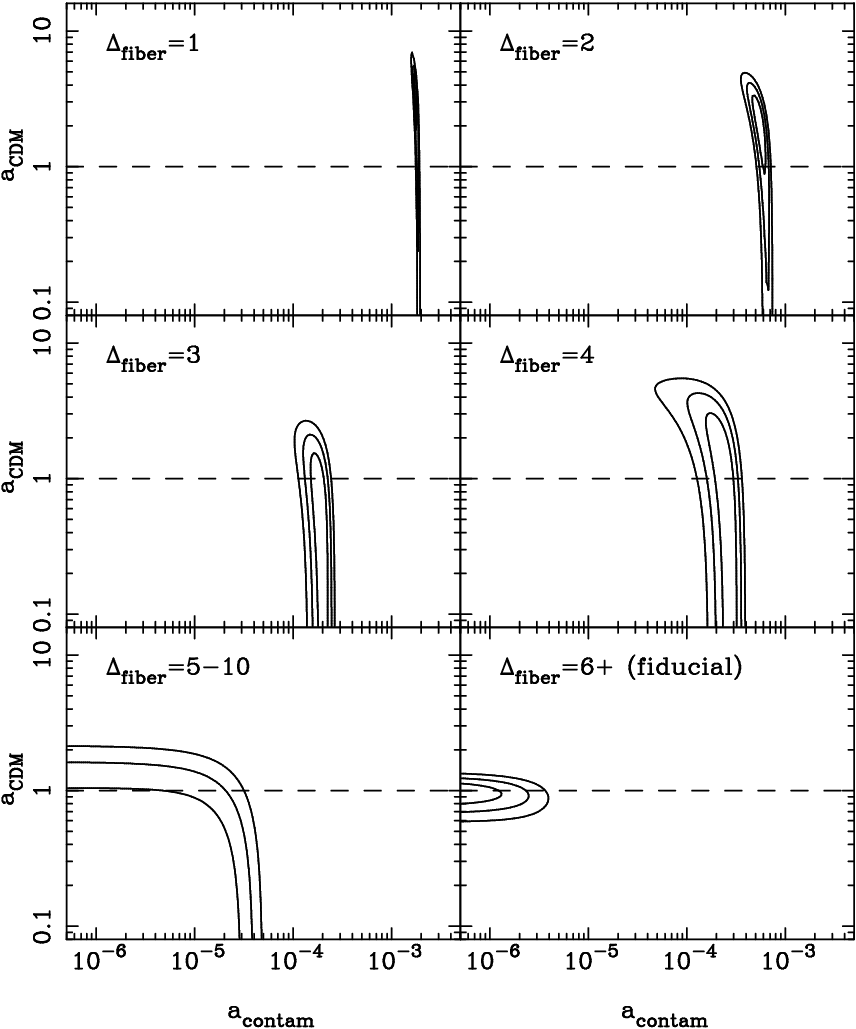,angle=-90.,width=8.5truecm}
\caption{ 
A joint fit of quasar scattered light contamination and 
cosmological model for \xiqar\ for different values of $\Delta_{\rm fiber}$.
We show 1,2 and 3 $\sigma$
 confidence limits on the joint parameters
$a_{\rm contam}$ and $a_{\rm CDM}$  from Equation \ref{contamfit}.
\label{ampcontam}}
\end{figure}

When comparing the top and bottom panels
of Figure \ref{stacksigpi} we note that the amplitude of the stacked quasar
spectrum is scaled by a factor of $2\times10^{-3}$, which is the 
level required to match the observations in a $\chi^{2}$ fit (see below).
This means that 0.2 \% of the quasar light is scattered into neighboring 
spectra (i.e. spectra with $\Delta_{\rm fiber}=1$).

\begin{figure*}
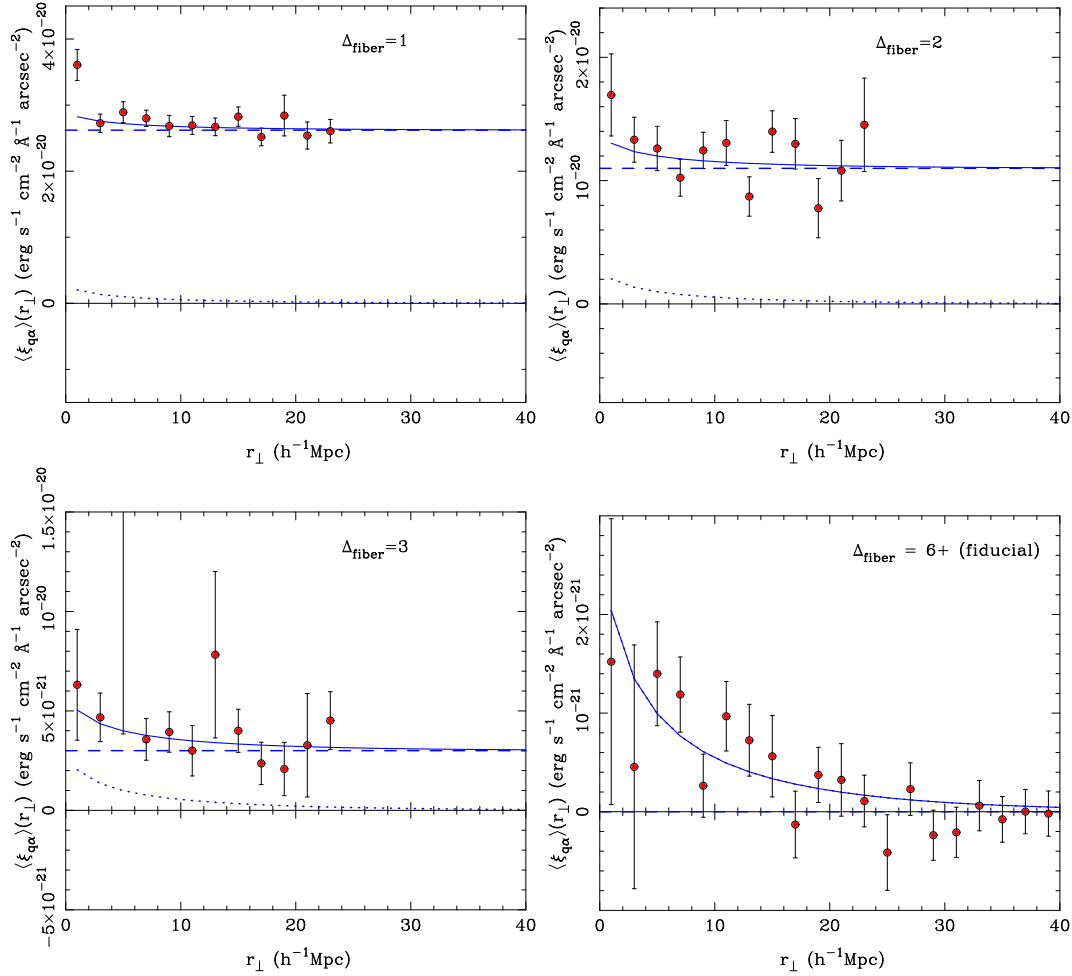

\centerline{
\psfig{file=vsrperp_df1.ps,angle=-90.,width=7.0truecm}
\psfig{file=vsrperp_df2.ps,angle=-90.,width=7.0truecm}
}
\centerline{
\psfig{file=vsrperp_df3.ps,angle=-90.,width=7.0truecm}
\psfig{file=vsrperp_dffid.ps,angle=-90.,width=7.0truecm}
}
\caption{
The measured \xiqa\ from BOSS data averaged between 
$r_{\parallel}=\pm 50 \hmpc$
and plotted as function of $r_{\perp}$. We show results for
different values of $\Delta_{\rm fiber}$ as indicated in each panel.
The points with error bars show the observational results. The prediction
of the theoretical model from Equation 11 is shown as a dotted line. A
dashed line indicates scattered light contamination of the
sort discussed in Section 3.2 which is independent of $r_{\perp}$
and the solid line is the sum of the two components.
\label{vsrperp}}
\end{figure*}

If there is a signal that is actually coming from quasar-\lya emission
correlations in the Universe, one would expect that to be superimposed on
top of any contamination from stray light. We have therefore carried out
a simple joint fit of \xiqa\ with a sum of
the fiducial theoretical model from Figure \ref{sigmapi} (right panel)
scaled by an amplitude
factor $a_{\rm CDM}$ and the stacked quasar contamination 
scaled by a factor $a_{\rm contam}$. The fit is therefore as follows:
\begin{equation}
\xi_{\rm fit}(r_{\perp},r_{\parallel})=a_{\rm CDM}\xi_{{\rm q}\alpha,{CDM}}
(r_{\perp},r_{\parallel})+a_{\rm contam}\xi_{\rm stack}(r_{\perp},r_{\parallel}),
\label{contamfit}
\end{equation}
where $\xi_{\rm stack}(r_{\perp},r_{\parallel})$ 
(multiplied by $2\times10^{-3}$) is shown in Figure \ref{stackqso} 
(top panel). We separate the  observational \xiqa\ 
results by $\Delta_{\rm fiber}$ and then carrying out a $\chi^{2}$ fit
of the form given by Equation \ref{contamfit} to determine the best amplitude
parameters $a_{\rm contam}$ and $a_{\rm CDM}$ as well as the confidence
limits on those parameters. 
We carry out the fit for quasar-pixel pair separations $r \le 50 \hmpc$,
although our conclusions are insensitive to this value.

The results are plotted in Figure \ref{ampcontam}.
For the $\Delta_{\rm fiber}=1$ dataset, we find a total
$\chi^{2}$ for the best fit of 1526 for 988 datapoints, which shows that the
light contamination which dominates the fit is fairly well modelled by 
the quasar stack. It is perhaps not surprising the fit to the contamination is
not perfect, as the contamination model is extremely simple.
We have plotted the 
1,2 and 3 $\sigma$ confidence contours on the parameters $a_{\rm CDM}$
and $a_{\rm contam}$ in Figure \ref{ampcontam}. The central value of 
$a_{\rm contam}$ is $1.75 \times 10^{-3}$. The 1  $\sigma$ confidence
interval on $a_{\rm CDM}$ is consistent with unity, and also with zero. 
The results from $\Delta_{\rm fiber}=1$ are therefore consistent with
the sum of significant contamination from quasar light and
a quasar-\lya correlation at the level predicted by the CDM model
fit to the fiducial dataset.

We have carried out the same fitting to datasets with 
different restricted values of $\Delta_{\rm fiber}$ in the other panels
of Figure \ref{ampcontam}. We can see that for cases with
 $\Delta_{\rm fiber} \le 4$ there is a significant detection of a
contamination contribution. The value of $a_{\rm contam}$ is lower
for $\Delta_{\rm fiber} 2-4$ than for $\Delta_{\rm fiber} =1$ 
at the factor of $10-4$ level. 
 In  all cases the measurements
are also consistent with an additional quasar-\lya correlation at
the level of $a_{\rm CDM}=1$.  
 For $\Delta_{\rm fiber}=5-10$ and $\Delta_{\rm fiber} \ge 6$, 
the contamination is not detectable.

The results of the fitting shown in Figure \ref{ampcontam} are a good sign
that the signal and contamination are behaving in a particular way which 
can be accounted for by excising pairs of pixels from close fibers. It 
is useful however to examine the results for individual datapoints in
more detail. Of particular interest is  
the  measured quasar-\lya surface brightness as a function of $r_{\perp}$.
As mentioned above, we expect contamination at the CCD level for a given
fiber separation
to be independent of $r_{\perp}$. We have taken the measured 
$\xi_{{\rm q}\alpha}(r_{\perp},r_{\parallel})$ values for different
datasets limited by $\Delta_{\rm fiber}$ and averaged 
them between $r_{\parallel}=\pm 50 \hmpc$ .We plot this average,
$\langle \xi_{{\rm q}\alpha} \rangle$ as a function of $r_{\perp}$
in Figure \ref{vsrperp}. In each panel we also plot a
dashed  horizonal line
showing a level of contamination that is independent of $r_{\perp}$,
a dotted line showing the level of the signal from quasar-\lya
clustering (we plot $\langle \xi_{{\rm q}\alpha} \rangle  (r_{\perp})$
computed from the model shown in Figure\ref{sigmapi}, right panel),
and a solid line which is the sum of the two.

We can see that in the $\Delta_{\rm fiber}=1$ panel there is the strong
contamination signal which is consistent with being independent
of $r_{\perp}$. The small relative contribution of the actual
signal makes little difference in this panel. In the 
 $\Delta_{\rm fiber}=2$ and  $\Delta_{\rm fiber}=3$ panels the level
of contamination is lower (note that the y-axes are different in 
the different panels), but it still dominates over the expected signal,
with again no sign of a dependence on $r_{\perp}$. For the fiducial
panel (bottom right), we can see that the measured 
$\langle \xi_{{\rm q}\alpha} \rangle  (r_{\perp}$ does have significant
dependence on $r_{\perp}$, and this has a similar form and amplitude
to the best fit model to the data in Section 3.2. There is no sign
of a contaminating (independent of $r_{\perp}$)
component to this measurement.

Having carried out these tests on contamination from the 
\lya emission line from nearby quasars, it is pretty clear that the
signal we are seeing in our fiducial dataset cannot be caused by 
straightforward scattered light coming from nearby fibers.
There remains the possibility however that there is an additional 
contaminating component from some other mechanism
that has a dependence on $r_{\perp}$. This is difficult to imagine,
but one can construct tests for this based on other spectral features
than the \lya line. In the next two subsections we do this, first centered
on the quasar CIV emission line and then using stars and the H$\alpha$
line.

\begin{figure}
\centerline{
\psfig{file=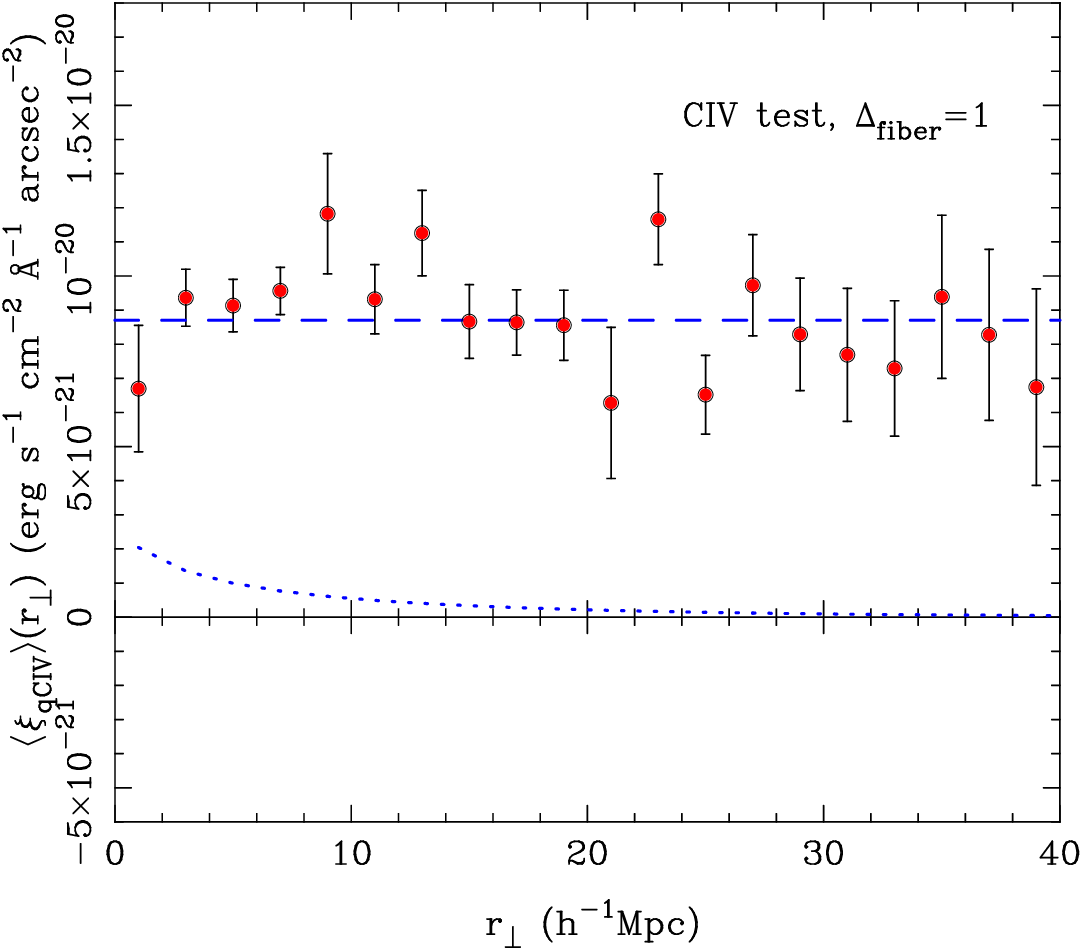,angle=-90.,width=7.0truecm}}
\centerline{
\psfig{file=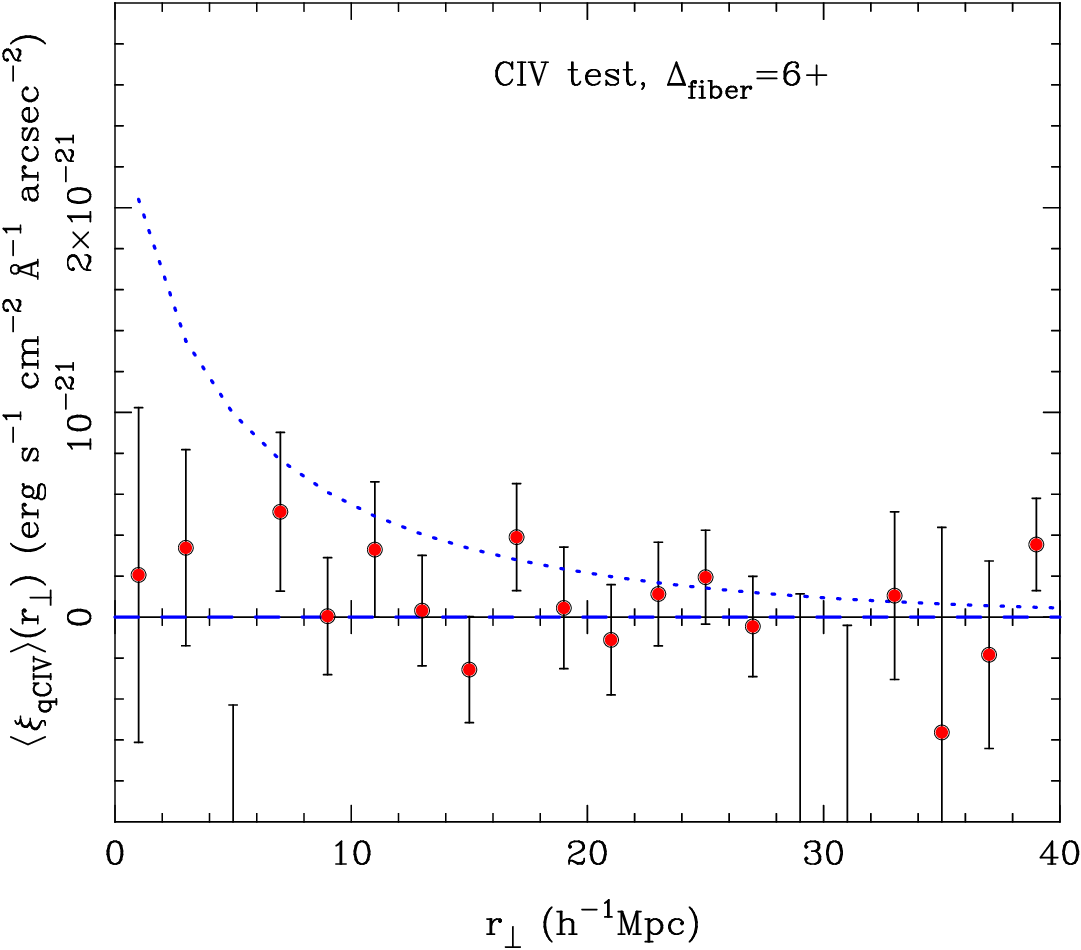,angle=-90.,width=7.0truecm}
}
\caption{
A null test using Carbon IV-quasar cross-correlations. 
Here we show 
 $\xi_{\rm qCIV}$ measured from BOSS data averaged between 
$r_{\parallel}=\pm 50  \hmpc$
and plotted as function of $r_{\perp}$. We show results for
2 different values of $\Delta_{\rm fiber}$ as indicated in each panel.
The points with error bars show the observational results. To
aid comparison with previous plots, the prediction
of the theoretical model from Equation 11 for the quasar-\lya 
cross correlation is shown as a dotted line. A
dashed line indicates scattered light contamination of the
sort discussed in Section 3.2 which is independent of $r_{\perp}$.
\label{vsrperpciv}}
\end{figure}

\subsection*{A.3 Tests with CIV}
As a test of scattered light contamination, we carry out a different
correlation between galaxy spectrum pixels and quasar positions. The only
difference from our standard analysis (Section 3) is that we compute
the cross-correlation at the wavelength of CIV
at the redshift of the quasar. We use a rest
wavelength of 1550 \AA\ for CIV  (close to the center of CIV doublet).
 If our interpretation of the
contamination from nearby quasars from the previous subsection is correct,
we would expect to see strong contamination from the CIV line for
low $\Delta_{\rm fiber}$ and  
 then no signal $\xi_{{\rm q}{CIV}}$ for 
large $\Delta_{\rm fiber}$ (because there should be no strong intergalactic
CIV line emission).

\begin{figure}
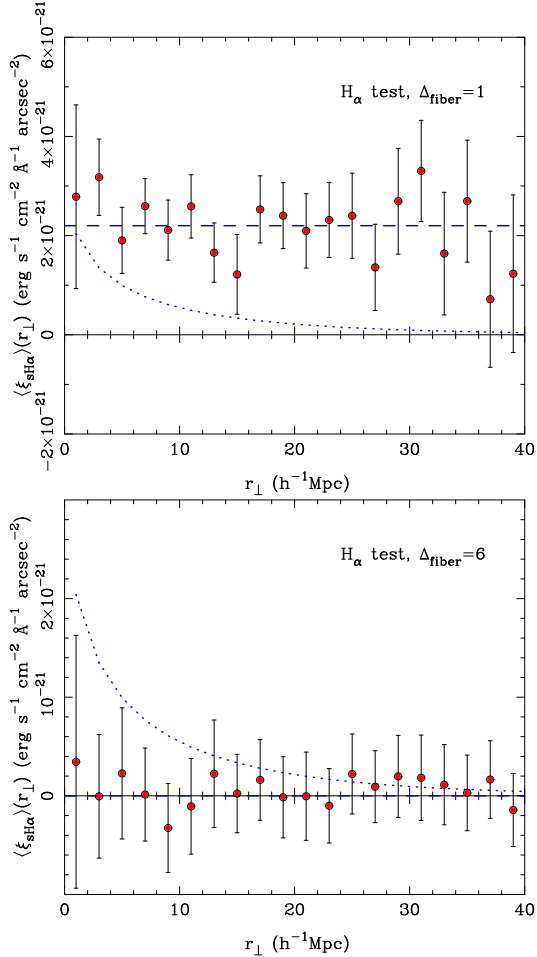

\centerline{
\psfig{file=ha_rperp_df1.ps,angle=-90.,width=7.0truecm}
}
\centerline{
\psfig{file=ha_rperp_fid.ps,angle=-90.,width=7.0truecm}
}
\caption{
A null test using H$\alpha\ $ - star cross-correlations. 
Here we show 
 $\xi_{\rm sH}\alpha$ measured from BOSS data averaged between 
$r_{\parallel}=\pm 50  \hmpc$
and plotted as function of $r_{\perp}$. We show results for
2 different values of $\Delta_{\rm fiber}$ as indicated in each panel.
The points with error bars show the observational results. To
aid comparison with previous plots, the prediction
of the theoretical model from Equation 11 for the quasar-\lya 
cross correlation is shown as a dotted line. A
dashed line indicates scattered light contamination of the
sort discussed in Section 3.2 which is independent of $r_{\perp}$.
\label{vsrperpha}}
\end{figure}

As with Figure \ref{vsrperp}, we plot 
$\langle \xi_{{\rm q}\alpha} \rangle  (r_{\perp})$, in Figure \ref{vsrperpciv}.
We show results for $\Delta_{\rm fiber}=1$ and for $\Delta_{\rm fiber}\ge 6$.
In the case of $\Delta_{\rm fiber}=1$ there is strong contamination from 
the quasar, which does not depend on $r_{\perp}$. The amplitude of the 
contamination is  approximately three times smaller than that from \lya,
which is the ratio one would expect from the relative strengths of
the two lines in Figure \ref{stackqso}. For the $\Delta_{\rm fiber}\ge 6$
we see that there is no evidence for contamination, and, although the 
error bars are relatively large,
there is no sign of any $\xi_{{\rm q}{CIV}}$ signal. We have plotted
the fiducial  $\xi_{{\rm q}{\alpha}}$ model on both panels with a dotted line.
That the contamination behaves the same way as the \lya results
in the previous subsection but that there is no $\xi_{{\rm q}{CIV}}$ signal
is further evidence for the reality of the  $\xi_{{\rm q}{\alpha}}$
measurement.

\subsection*{A.4 Tests with stars}

A further test of stray light contamination which 
we can carry out is to cross-correlate the positions
of stars with the galaxy pixels.
We use all stars in the SDSS DR10 catalog (Ahn et al. 2014), which is 171612
objects in total.
 In this case we choose the
H$\alpha$ line, with rest wavelength 6562.8 \AA\ and cross-correlate
star positions with
galaxy pixels centered on this wavelength. We measure
the resulting star-H$\alpha$
cross-correlation, 
$\langle \xi_{{\rm sH}\alpha} \rangle  (r_{\perp}$ is
for two values of 
$\Delta_{\rm fiber}$, 1 and $\ge 6$. In this test, all the stars are
obviously  at the
same rest wavelength, which effectively means that in this measurement
we are stacking galaxy spectra together. This test is therefore
likely to be more sensitive to the CIV test to residual 
artifacts in photocalibration of the spectrometer (see e.g. Figure 4 
of Busca \etal 2013). Nevertheless it is useful to see whether 
any contamination has an $r_{\perp}$ dependence.

We plot the results for $\langle \xi_{{\rm sH}\alpha} \rangle  (r_{\perp}$
in  Figure \ref{vsrperpha}, where we can see that there is again 
a sign of stellar light contamination when $\Delta_{\rm fiber}=1$
and that this contamination appears to be independent of  $r_{\perp}$.
For $\Delta_{\rm fiber}\ge6$, the measurement is consistent with zero,
again a sign that the cut in  $\Delta_{\rm fiber}$ removes 
scattered light contamination.

The results from the tests in Appendix A therefore  suggest
that although light contamination from quasars can be readily
seen in the data, it can be eliminated to a high degree
by  excluding close pairs of fibers. When this is done,
the resulting measured quasar-\lya surface brightness correlation seems
to be real. If instead it is produced by some as yet unknown systematic
this effect would
have to reproduce the $r_{\perp}$ and
$r_{\parallel}$ dependence expected from a cosmological signal, be
measurable only in the \lya wavelength range and not around
other emission lines, 
and be independent of the position of the spectra on the detector.

Having searched for and as far as we can tell eliminated the effects of
strong light contamination, we now turn to other potential
systematic effects in the next 2 appendices.

\section*{Appendix B: Large-scale surface brightness correction}

\begin{figure}
\centerline{
\psfig{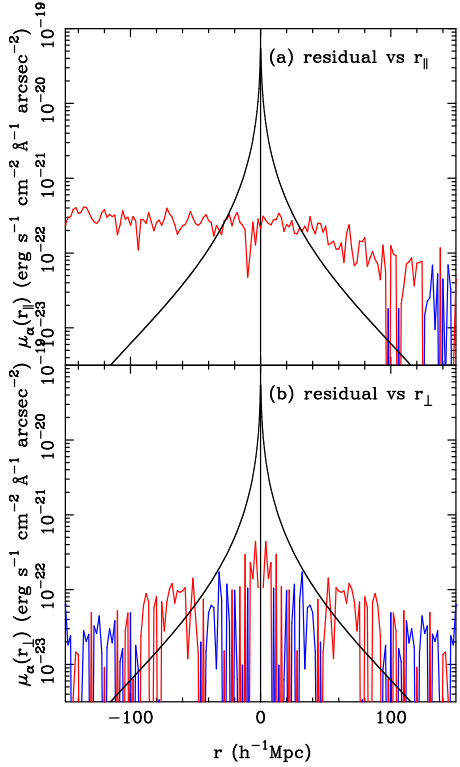}
}
\caption{ 
Large scale residual flux (Equations \ref{rppcorr} and
\ref{rppcorr2}. Negative vales are shown in blue and 
positive values in red.
\label{rpp}}
\end{figure}

A potential systematic error on our measurement is any obscuration (such
as that produced by Galactic dust) that may affect both quasar selection
and \lya surface brightness, thus producing a spurious cross-correlation
between the two on the plane of the sky. One can correct for  such
a cross-correlation by searching for an   
\xiqa\ $(r_{\perp},r_{\parallel})$ 
signal which is non-zero for large values of $r$, where
no physical (3D) correlation would be expected. This correlation
which will be a function of $r_{\perp}$ only
(in the parallel line-of-sight approximation)
 can then be subtracted from the \xiqa\ signal.
 We have done this, computing
the surface brightness correction $\mu_{\alpha}$ given by 

\begin{equation}
\mu_{\alpha}(r_{\perp})=
\frac{1}{2(x_{a}-x_{b})}\left[
\int_{-x_{a}}^{-x_{b}} \xi_{{\rm q}\alpha}(r_{\parallel},r_{\perp})
dr_{\parallel}+
\int_{x_{b}}^{x_{a}} \xi_{{\rm q}\alpha}(r_{\parallel},r_{\perp})
dr_{\parallel}\right]
\label{rppcorr}
\end{equation}

where $x_{a}= 400 \hmpc$ and $x_{b}=80 \hmpc$ and 
\xiqa$(r_{\perp},r_{\parallel})$
is the qso-\lya emission cross-correlation. The value of 
$\mu_{\alpha}(r_{\perp})$ is shown in the bottom panel of figure \ref{rpp}.
We can see that there is no coherent structure, as $\mu_{\alpha}(r_{\perp})$
becomes both positive and negative as $r_{\perp}$ varies.
We have also plotted the best fit linear CDM model (as a function
of $r$) for the qso-\lya
cross-correlation (from Section \ref{modfit}), 
and we can see that this dominates
over the $\mu_{\alpha}(r_{\perp})$ correction on small scales
$r_{\perp} < 40 \hmpc$, as we would hope. 

There is also another analogous correction, but one that applies in the 
orthogonal direction. Any systematic effect which affects the \lya emission
in the line-of-sight direction (including redshift evolution in 
the \lya surface brightness), and which is correlated with evolution
in the quasar population with redshift (or at least evolution in the efficiency
of quasar selection with redshift) could produce a spurious qso-\lya
emission cross-correlation. We compute how \xiqa\
varies as a function of $r_{\parallel}$ for large $r_{\perp}$ values where
there should be minimal physical clustering. In this case the 
the surface brightness correction $\mu_{\alpha}$ is given by

\begin{equation}
\mu_{\alpha}(r_{\parallel})=\frac{2}{(x_{a}^{2}-x_{b}^{2})}
\int_{x_{b}}^{x_{a}} \xi_{{\rm q}\alpha}(r_{\perp},r_{\parallel})
r_{\perp}dr_{\perp}
\label{rppcorr2}
\end{equation}
where we use  the same values of $x_{a}$ and $x_{b}$ as in Equation
\ref{rppcorr}.
The results for $\mu_{\alpha}(r_{\parallel})$ are shown in the top
panel of Figure \ref{rpp}. We can see that in this case there is a
significant trend in the surface brightness correction with 
 $\mu_{\alpha}(r_{\parallel})$ gradually decreasing from positive
 values of $\sim 4 \times 10^{-22}$ 
erg s$^{-1}$
cm$^{-2}$ \AA $^{-1}$ arcsec$^{-2}$
for $r_{\parallel}=-150 \hmpc$ to $\sim -10^{-23}$ 
erg s$^{-1}$
cm$^{-2}$ \AA $^{-1}$ arcsec$^{-2}$. 
for $r_{\parallel}=+150 \hmpc$. As with the previous ($r_{\perp}$)
correction, the CDM model dominates on scales $r_{\perp}< 40 \hmpc$,
where most of the clustering signal is located.

In our analyses in the main body of the paper, we have applied these
$\mu_{\alpha}(r_{\parallel})$ and $\mu_{\alpha}(r_{\perp})$ corrections
to the computation of the qso-\lya emission cross-correlation 
\xiqa\ . We have done this on a quasar-pixel pair basis,
computing $r_{\parallel}$ and $r_{\perp}$ for each pair, and then 
subtracting the appropriate value of $\mu_{\alpha}$. In all cases, the effect
of the correction is small, as we would expect given the 
much large amplitude of the clustering signal compared to the surface
brightness corrections that can be seen in Figure \ref{rpp}. 
For example,
in our fiducial case, without applying the large-scale surface-brightness
corrections, we find a best fitting values of the 
shape and amplitude parameters of Section 3.2 of
\ampqa\ $= 3.5$  erg s$^{-1}$
cm$^{-2}$ \AA $^{-1}$ arcsec$^{-2}$. 
and $\Omega_{\rm m}=0.26$, which represent 
differences of $0.5 \sigma$ and $0.5 \sigma$
in the parameters respectively.

\begin{figure}
\centerline{
\psfig{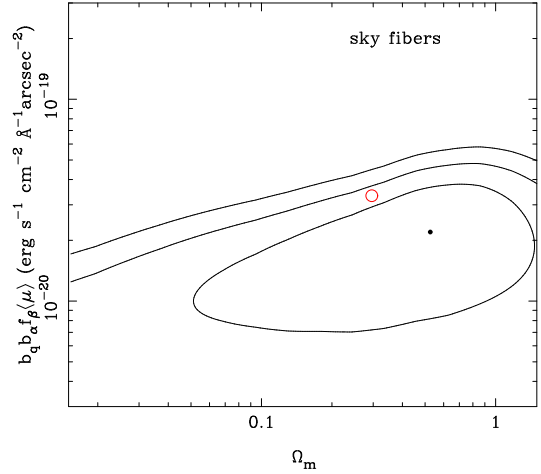}
}
\caption{ Fit parameters for the CDM model fit (as in Figure \ref{gacont}) to
the quasar-\lya cross-correlation function (Equation \ref{model}) using sky 
fibers.
 The
point shows the best fit values of the amplitude (\ampqa\ )
and shape ($\Omega_{\rm m}$) and the contours show the $1, 2 $ and $3 \sigma$
confidence intervals. The open circle shows the best fit values of
the fit parameters for the fiducial sample.  
\label{gaskyf}}
\end{figure}

\section*{Appendix C: Sample tests}

In this appendix we report on three consistency tests of our cross-correlation
results. One test uses sky fibers
to search for \lya emission instead of galaxy fibers, and the second looks at
how the luminosity of the orginally targeted galaxy  affects the quasar-\lya 
cross-correlation. The third test investigates how quasar luminosity affects
the cross-correlation.

\subsection*{C.1 Sky fibers}

\begin{figure}
\centerline{
\psfig{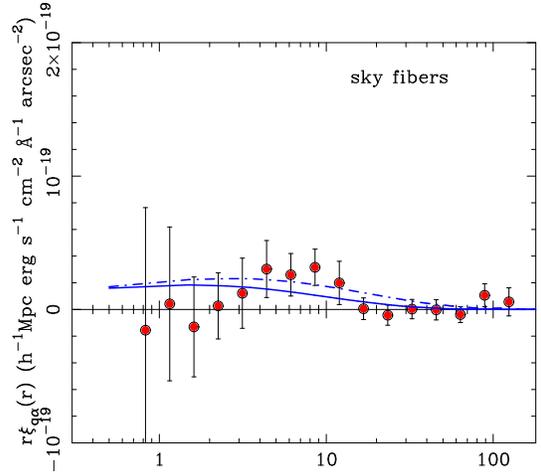}
}
\caption{ The quasar-\lya cross correlation function \xiqar\
(Equation \ref{xieq}) computed using sky fibers instead of galaxy-subtracted
galaxy fiber. The solid line shows a fit to the CDM model and
the dash-dotted line shows the fit from our fiducial sample (Figure 
\ref{gacont}),
which used galaxy-subtracted galaxy fibers.
\label{cdmskyf}}
\end{figure}

Ideally one would like to be able to use fibers positioned on random areas
of the sky to carry out intensity mapping, without needing to worry about
subtracting a foreground galaxy. This approach is being carried out 
for example by HETDEX (Blanc \etal 2011).
In our case, however, there are a number
of such random fibers which were obtained for each plate, to use in 
sky-subtraction. The number of fibers available for use to use is 146065,
approximately 15\% as many as there are galaxy fibers in our fiducial dataset.
We have taken these sky fibers and carried out the quasar-\lya
cross-correlation of Equation \ref{xieq}. The calculation
was the same as in the fiducial case, including  subtraction
of a background from fiber (Figure \ref{resflux}).

 We show the results for \xiqar\
for the sky fibers
in Figure \ref{cdmskyf}. We can see that the datapoints appear to be consistent
with the trend delineated by the best fit to the fiducial computation
(also shown, as a dash-dotted line), although the measurement is much noisier.
We fit the same CDM linear correlation function as was carried out 
in Section \ref{modfit} 
and show the best fit parameters and confidence contours in 
Figure \ref{gaskyf}.
We find \ampqa\ $=2.2^{+1.0}_{-1.0}\times10^{-20}$ 
erg s$^{-1}$
cm$^{-2}$ \AA $^{-1}$ arcsec$^{-2}$
and $\Omega_{\rm m}=0.52^{+0.48}_{-0.36}$.
 Here we can see that there that the confidence contours
are very wide, and that the clustering signal is present at only the
 $2 \sigma$ confidence level. The results are  also consistent
with the results from the fiducial sample at the 1.5 $\sigma$ level. 
At present therefore, this shows
that the use of galaxy and sky fibers in computing the cross-correlation is
approximately equivalent, although the uncertainty is obviously large. 

In the future, it would arguably be best to use randomly distributed fibers to
carry out the cross-correlation, to avoid any potential selection biases.
In the current sample case these biases are below the level of detectability,
but one can imagine two types of bias, related to the fact that galaxy
fibers are selected in regions of above average sky brightness and sky fibers
selected to be in regions of below average sky brightness. Both effects
 could in principle
bias  the measurement of \lya surface brightness
in opposite directions. We now carry out a further test to constrain
this effect, by splitting the 
spectra into two halves bases on target LRG luminosity.

\begin{figure}
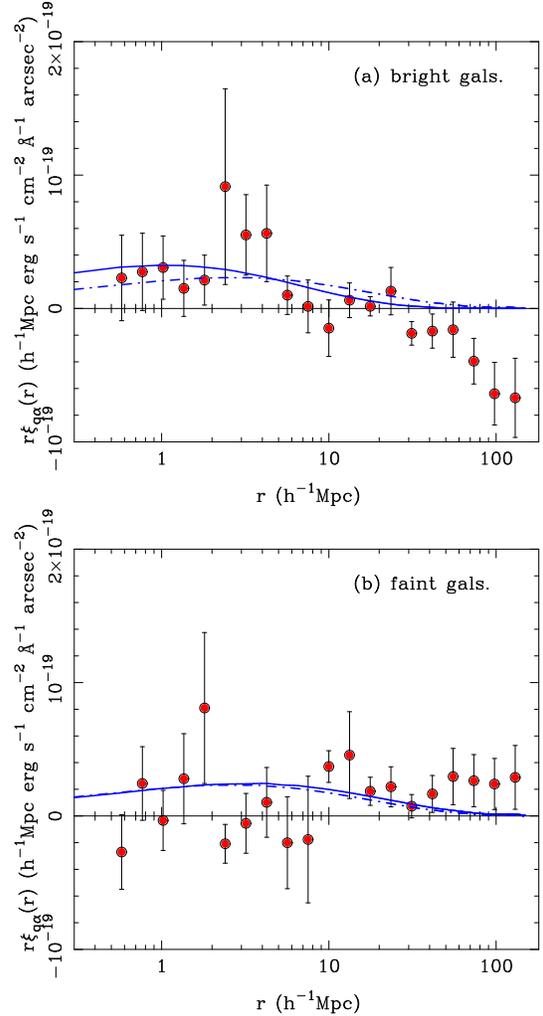

\centerline{
\psfig{file=cdm_toplum.ps,angle=-90.,width=7.0truecm}
}
\centerline{
\psfig{file=cdm_botlum.ps,angle=-90.,width=7.0truecm}
}
\caption{
 Quasar-\lya cross-correlation \xiqar\ for two
subsamples of spectra originally targeted at LRGs with (a) luminosity 
above the median value and (b) below the median value. The solid curve 
shows the best fit CDM model and the dash-dotted line the CDM fit
to the fiducial sample (all LRGs).
\label{cdmlum}}
\end{figure}

\subsection*{C.2 Sample split by galaxy luminosity}

We  divide the galaxy spectrum sample into two halves, based on the 
measured SDSS $r$ band luminosity (no k-correction was applied). One half
consists of galaxies above the median luminosity of the whole dataset
($5.2\times 10^{40}$ erg s$^{-1}$), and the other half those below it. The
median r band luminosities of the halves are $6.9\times 10^{40}$ erg s$^{-1}$
and $3.8\times 10^{40}$ erg s$^{-1}$

\begin{figure}
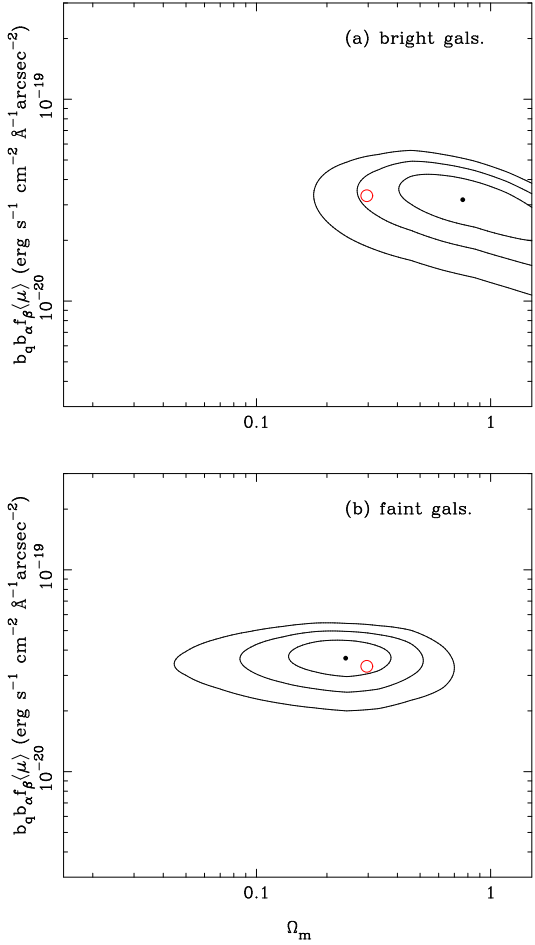

\centerline{
\psfig{file=gamma_amp_toplum.ps,angle=-90.,width=7.0truecm}
}
\centerline{
\psfig{file=gamma_amp_botlum.ps,angle=-90.,width=7.0truecm}
}
\caption{
CDM amplitude and shape
parameters fitted to the  Quasar-\lya cross-correlation \xiqar\ for two
subsamples of spectra originally targeted at LRGs with (a) luminosity 
above the median value and (b) below the median value. 
 The
points shows the best fit values of the amplitude (\ampqa\ )
and shape ($\Omega_{\rm m}$) and the contours show the $1, 2 $ and $3 \sigma$
confidence intervals. The open circles show the best fit values of
the fit parameters for the fiducial sample.  
\label{galum}}
\end{figure}

We note that the LRG galaxy surface brightnesses measured from BOSS spectra
are of order $10^{-17}$ erg s$^{-1}$
cm$^{-2}$ \AA $^{-1}$ arcsec$^{-2}$, approximately $2-3$ orders of magnitude
brighter than the mean \lya surface brightness $\langle \mu_{\alpha} 
\rangle$
that we have measured in Section \ref{sfrd}.
We therefore expect that it is unlikely
that regions of excess background 
\lya surface brightness could have caused certain
LRGs to be preferentially selected and therefore bias our measurements.

\begin{figure}
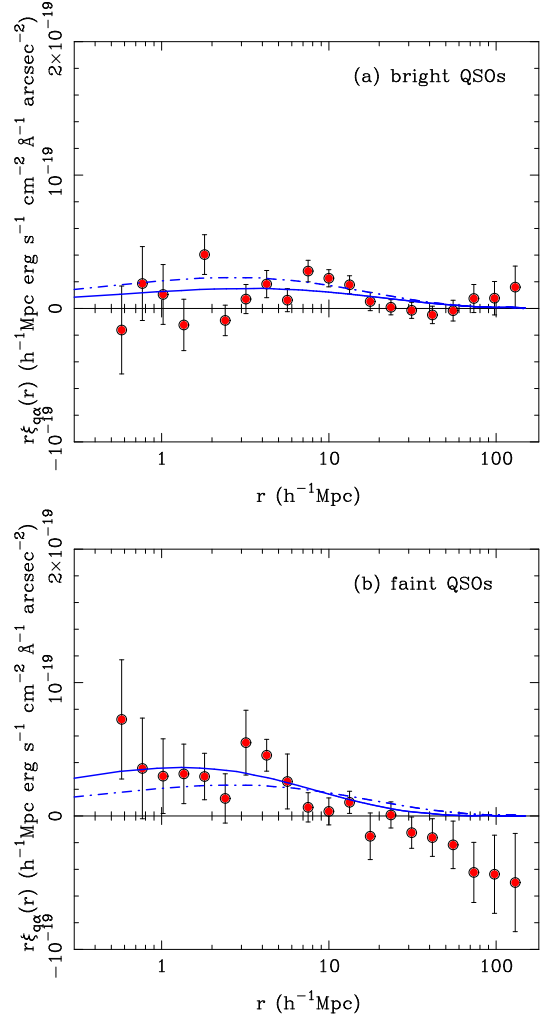

\centerline{
\psfig{file=cdm_qtoplum.ps,angle=-90.,width=7.0truecm}
}
\centerline{
\psfig{file=cdm_qbotlum.ps,angle=-90.,width=7.0truecm}
}
\caption{ 
 Quasar-\lya cross-correlation \xiqar for two
subsamples of quasar data with (a) luminosity 
above the median value and (b) below the median value. The solid curve 
shows the best fit CDM model and the dash-dotted line the CDM fit
to the fiducial sample (all quasars).
\label{cdmqlum}}
\end{figure}

We measure the quasar-\lya cross correlation \xiqar\ for the two samples
of spectra and show the results in Figure \ref{cdmlum}.
Both subsamples show evidence of clustering that is consistent with
the CDM model shape and the fiducial amplitude.
This can be seen in a quantitative manner from Figure
\ref{galum}, where we show the fit 
parameters.
We find 
\ampqa\ =$3.1^{+0.6}_{-0.7}\times10^{-20}$ 
erg s$^{-1}$ cm$^{-2}$ \AA $^{-1}$ arcsec$^{-2}$, 
for the bright galaxy spectra and 
\ampqa\ =$(3.8\pm 0.5)\times10^{-20}$ erg s$^{-1}$ cm$^{-2}$ 
\AA $^{-1}$ arcsec$^{-2}$, 
for the faint galaxy spectra.
 Both samples are consistent with the amplitude of the fiducial result at the 
$1 \sigma$ level 
(the fiducial sample fractional error bar is 
bar $^{+12\%}_{-11\%}$ for fixed $\Omega_{\rm m}=0.30$).
This indicates that any bias present in the mean
surface brightness $\langle \mu_{\alpha} \rangle$ due to
fiber target selection is likely to be at the $\sim 10\%$
level or less.

\subsection*{C.3 Sample split by quasar luminosity}

\begin{figure}
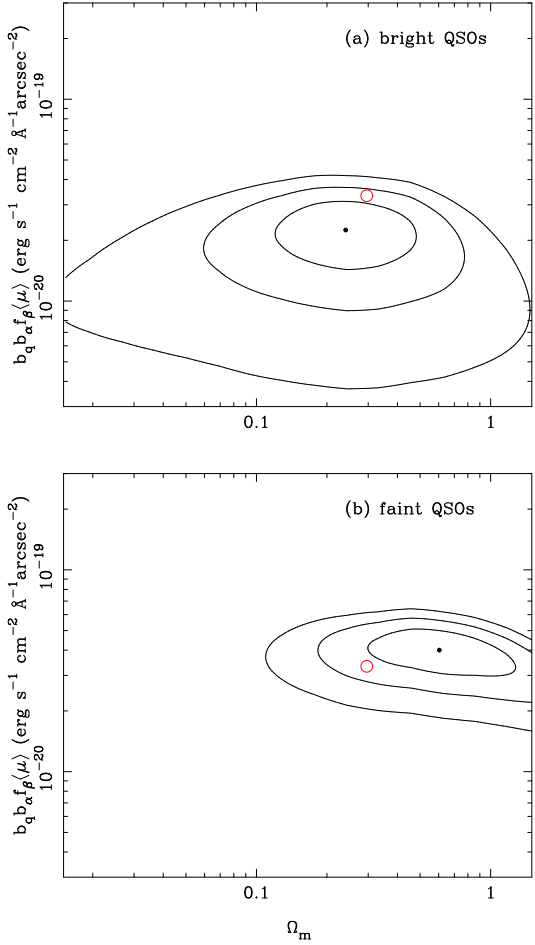

\centerline{
\psfig{file=gamma_amp_qtoplum.ps,angle=-90.,width=7.0truecm}
}
\centerline{
\psfig{file=gamma_amp_qbotlum.ps,angle=-90.,width=7.0truecm}
}
\caption{ 
CDM amplitude and shape
parameters fitted to the  Quasar-\lya cross-correlation
\xiqar\
 for two
subsamples of quasars with (a) luminosity 
above the median value and (b) below the median value. 
 The
points shows the best fit values of the amplitude (\ampqa\ )
and shape ($\Omega_{\rm m}$) and the contours show the $1, 2 $ and $3 \sigma$
confidence intervals. The open circles show the best fit values of
the fit parameters for the fiducial sample.  
\label{gaqlum}}
\end{figure}

A potential source of \lya emission clustered with quasars is
recombination radiation from dense IGM systems illuminated by the 
quasars themselves (e.g., Kollmeier \etal 2010).
We have made an estimate of the 
amplitude of this signal in Section \ref{disc} and find
it likely that it is much smaller than the signal
from star forming galaxies clustered with the quasar. One way of testing
this directly is by splitting the sample into high and low
luminosity quasars and then  measuring the \xiqar\ signal for each.
We note that this test can also function as a diagnostic for
stray light contamination from quasars, as any stray light should be 
correlated with quasar luminosity.

In this section we do this, measuring \xiqar\ the subsample of
quasars with SDSS r band luminosity above the median, and with the
subsample with luminosities below the median. The median luminosity
of the bright subsample is 3.45 times the median luminosity of
the faint sample, and the median redshifts of the
quasars in each are $z=2.66$ and $z=2.44$ respectively.

We can see in Figures \ref{cdmqlum} that the \xiqar\ results for two
subsamples do not look very different. In particular if the quasar
luminosities were causing significant fluorescent emission from
nearby IGM material one might expect the brighter subsample
to exhibit a steeper \xiqar\  on small scales, which is not the 
case. From Figure \ref{gaqlum} we can see that the fit
parameters for the bright subsample are within $1\sigma$ of the
fiducial result and the faint subsample within $2 \sigma$.  The amplitude
of parameter (\ampqa\ ) (for
Planck $\Omega_{\rm m}=0.30$) is actually 
slightly lower for the bright subsample, 
being 
$(2.3\pm 0.6)\times10^{-20}$ erg s$^{-1}$ cm$^{-2}$ \AA $^{-1}$ arcsec$^{-2}$, 
whereas for the faint subsample it 
is $3.9^{+0.7}_{-0.6}\times10^{-20}$ erg s$^{-1}$
cm$^{-2}$ \AA $^{-1}$ arcsec$^{-1}$. Given that the quasar subsamples
are different in luminosities by a factor of 3.45 one would expect there
to be a significant difference the 
\xiqar\ results in the opposite direction if quasar properties
were significantly affecting the large-scale \lya intensity
around them. As mentioned above, this lack of dependence
on quasar luminosity is also strong evidence that
stray light from quasars scattered in the spectrograph is not
contaminating our measured cross-correlation signal.

\subsection*{C.4 Other: sample split by fiber position in 
spectrograph}

It is possible to imagine other sample tests based on various
observational parameters, which 
could help to identify sources of systematic error. 
One such test is to split the sample into
two subsets based on position of fibers in the spectrograph. Of the
1000 fibers,  those  labelled with numbers 100-400 and 600-900 are 
positioned towards the center
of the spectrograph and the rest are positioned more towards the edge.
The reason for such a test is that the optical quality of the
instrument cameras degrades towards the edge, and this might affect
the signal to noise ratio and/or cause other problems.

\begin{figure}
\psfig{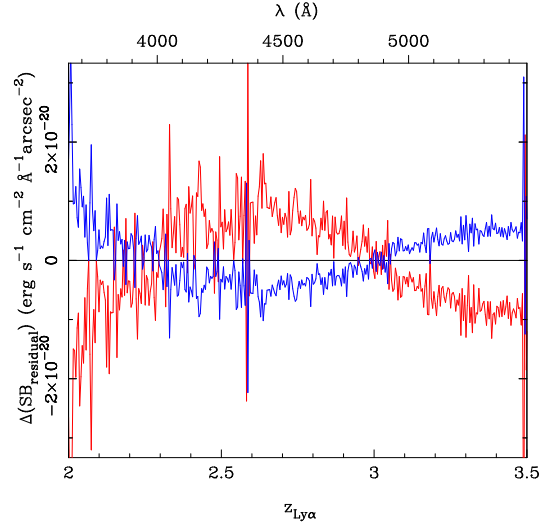}
\caption{ 
The difference in average residual flux for two subsamples of
 LRG spectra selected to be close to the center of
of the spectrograph camera (blue line) or edge of spectrograph
camera (red line). The residual flux is that left after subtraction 
of the best fit galaxy model from each spectrum,
and the difference plotted here  is
the difference between each 
subsample and the fiducial result plotted as a black 
line in Figure \ref{resflux}.
\label{edgevscen}}
\end{figure}

We carry out this split, and find that the average residual flux after 
subtraction of the model LRG spectra is somewhat different for the two 
subsamples. The residual flux for the fiducial sample was plotted in 
Figure \ref{resflux}, where we saw that it fluctuated between 
approximately
$\pm3\times10^{20} \ergs
\cm^{-2} \angs^{-1}\asec^{-2}$ for most of the relevant redshift range.
In Figure \ref{edgevscen} we plot for each
subsample (edge and center) $\Delta({\rm SB_{residual}})$,
the residual flux for the subsample minus the residual flux
for the fiducial sample. We can see that there are indeed 
differences- the residual flux in fibers close to the edge of the cameras
is systematically higher than the fiducial sample (and the center of
camera sample) by $\sim 10^{-20} \ergs               
\cm^{-2} \angs^{-1}\asec^{-2}$
 over the wavelength range $\sim 4000-5000$ \AA . As
this covers the important \lya redshift range $z=2.3-3.0$, this could indeed
have consequences for our measurements.

Using the appropriate mean residual flux for each subsample, we 
compute the quasar-\lya emission cross-correlation \xiqa .
It should be noted that the subsamples are not quite the same size
(there are 400 ``edge'' fibers per plate vs 600 ``center'' fibers).
We  fit the usual  CDM model  parameters
and  find 
\ampqa\ =$3.9^{+0.5}_{-0.5}\times10^{-20}$ 
erg s$^{-1}$ cm$^{-2}$ \AA $^{-1}$ arcsec$^{-2}$, and 
$\Omega_{\rm m}=0.31^{+0.14}_{-0.10}$ 
for the central fibers  and 
\ampqa\ =$(1.5 \pm0.7)\times10^{-20}$ erg s$^{-1}$ cm$^{-2}$ 
\AA $^{-1}$ arcsec$^{-2}$, and $\Omega_{\rm m}=0.44^{+0.96}_{-0.33}$
for the edge fibers. The central fiber results are within 1 $\sigma$ of the
fiducial results, but the edge fiber parameters have larger errors and
are  2.5 $\sigma$ below the fiducial results. 
We have also tried averaging the \xiqa results from the two subsamples
before fitting and find fit
parameters consistent at the $1 \sigma$ level with the fiducial 
results.

This test appears to show
that the degradation of the optical quality of the spectrograph cameras
close to 
the edge of their field of view does affect our ability to measure \xiqa\ .
Given this degradation,
it is reasonable to expect that the error bars for the edge sample are
underestimated and that this may account for the 2.5 $\sigma$ difference
with the fiducial sample. The fibers close to the center of camera sample 
appears to be
responsible for most of our ability to determine \xiqa .
This information should be useful for future measurements of this type, but
we do not use it to effect any changes to the fiducial analysis in this
paper.

\end{document}